\documentclass[journal, 12pt, a4wide]{article}

\usepackage{graphics} 
\usepackage{graphicx}    
\usepackage{epsfig} 
\usepackage{amsmath} 
\usepackage{amssymb}  
\usepackage{psfrag}
\usepackage{array}
\usepackage{multirow}
\usepackage[table]{xcolor}
\usepackage{enumerate}
\usepackage{multicol}
\usepackage{mleftright}

\usepackage[mathscr]{eucal}
\usepackage{bm}
\usepackage{bbm}
\usepackage{dsfont}

\usepackage{subfigure}
\usepackage{theorem}
\usepackage{algorithm}

\usepackage{enumitem}  
\setlist[description]{leftmargin=\parindent}

{\theorembodyfont{\upshape}

}

\newcommand{\beq}{\begin{equation}}
\newcommand{\eeq}{\end{equation}}
\newcommand{\beqa}{\begin{eqnarray}}
\newcommand{\eeqa}{\end{eqnarray}}
\newcommand{\beqan}{\begin{eqnarray*}}
\newcommand{\eeqan}{\end{eqnarray*}}

\newcommand{\bite}{\begin{itemize}}
\newcommand{\eite}{\end{itemize}}
\newcommand{\benu}{\begin{enumerate}}
\newcommand{\eenu}{\end{enumerate}}

\pdfoutput=1

 \title{\LARGE \bf Minimum energy control for complex networks}

\author{Gustav Lindmark and Claudio Altafini%
\thanks{Corresponding author: C. Altafini. Email: {\tt\small claudio.altafini@liu.se}} \\
Division of Automatic Control, Dept. of Electrical Engineering, \\
Link\"oping University, SE-58183, Link\"oping, Sweden. }

\begin{document}

\maketitle

\begin{abstract}
The aim of this paper is to shed light on the problem of controlling a complex network with minimal control energy.
We show first that the control energy depends on the time constant of the modes of the network, 
and that the closer the eigenvalues are to the imaginary axis of the complex plane, the less energy is required for complete controllability.
In the limit case of networks having all purely imaginary eigenvalues (e.g. networks of coupled harmonic oscillators), several constructive algorithms for minimum control energy driver node selection are developed. 
A general heuristic principle valid for any directed network is also proposed: the overall cost of controlling a network is reduced when the controls are concentrated on the nodes with highest ratio of weighted outdegree vs indegree. 
\end{abstract}

\paragraph{Significance statement}
Controlling a complex network, i.e., steering the state variables associated to the nodes of the networks from a configuration to another, can cost a lot of energy. The problem studied in the paper is how to choose the controls so as to minimize the overall cost of a state transfer. 
It turns out that 
the optimal strategy for minimum energy control of a complex network consists in placing the control inputs on the nodes that have the highest skewness in their degree distributions, i.e., the highest ratio between their weighted outdegree and indegree.

\section{Introduction}

Understanding the basic principles that allow to control a complex network is a key prerequisite in order to move from a passive observation of its functioning to the active enforcement of a desired behavior. 
Such an understanding has grown considerably in recent years. 
For instance the authors of \cite{liu2011controllability} have used classical control-theoretical notions like structural controllability to determine a minimal number of driver nodes (i.e., nodes of the network which must be endowed with control authority) that guarantee controllability of a network.
Several works have explored the topological properties underlying such notions of controllability 
\cite{Cornelius2013Realistic,gao2014target,Liu2015Control,Nepusz2012Controlling,Olshevsky2014Minimal,Ruths2014Control}, or have suggested to use other alternative controllability conditions 
\cite{Ding2013Studies,Nacher2014Analysis,Yuan2013Exact}.
Several of these approaches are constructive, in the sense that they provide receipts on how to identify a subset of driver nodes that guarantees controllability.
However, as observed for instance in \cite{Yan2012Controlling}, controllability is intrinsically a yes/no concept that does not take into account the effort needed to control a network.
A consequence is that even if a network is controllable with a certain set of driver nodes, the control energy that those nodes require may result unrealistically large. 
Achieving ``controllability in practice'' i.e., with a limited control effort, is a more difficult task, little understood in terms of the underlying system dynamics of a network.  
In addition, in spite of the numerous attempts \cite{Bof2015Role,Chen2016Energy,Li2015Minimum,Nacher2014Analysis,Olshevsky2015Eigenvalue,Pasqualetti2014Controllability,Summers2016Submolularity,Sun2013Controllability,Tzoumas2016Minimal,Yan2012Controlling,Yan2015Spectrum}, no clear strategy has yet emerged for the related problem of selecting the driver nodes so as to minimize the control energy. 

The aim of this paper is to tackle exactly these two issues, namely: i) to shed light on what are the dynamical properties of a network that make its controllability costly; and ii) to develop driver node placement strategies requiring minimum control energy. 
We show in the paper that for linear dynamics the natural time constants of the modes of the system are key factors in determining how much energy a control must use. 
Since the time constants of a linear system are inversely proportional to the real part of its eigenvalues, systems that have eigenvalues near the imaginary axis (i.e., nearly oscillatory behavior) are easier to control than systems having fast dynamics (i.e., eigenvalues with large real part). 
For networks of coupled harmonic oscillators, which have purely imaginary eigenvalues, we show that it is possible to obtain explicit criteria for minimum energy driver node placement. 
One of these criteria ranks nodes according to the ratio between weighted outdegree and weighted indegree. We show that for any given network such criterion systematically outperforms a random driver node assignment even by orders of magnitude, regardless of the metric used to quantify the control energy.

\section{Methods}

\paragraph{Reachability v.s. Controllability to 0.}

A linear system 
\beq
\dot x = A x + B u
\label{eq:system1_main}
\eeq
is controllable if there exists an input $ u(t)$ that transfers the $n$-dimensional state vector $ x(t)$ from any point $ x_o $ to any other point $ x_f $ in $ \mathbb{R}^n$. The Kalman rank condition for controllability, $ {\rm rank}([B \, AB \, A^2 B \, \ldots A^k B ])=n $ for $k$ sufficiently large, only provides a yes/no answer but does not quantifies what is the cost, in term of input effort, of such state transfer. 
A possible approach to investigate ``controllability in practice'' consists in quantifying the least energy that a control requires to accomplish the state transfer, i.e., in computing $ u(t) $ mapping $ x_o $ in $ x_f $  in a certain time $ t_f $ while minimizing $ \mathcal{E}(t_f) = \int_0^{t_f} \|  u(\tau)\|^2 d \tau $.
For linear systems like \eqref{eq:system1_main}, a closed form solution to this problem exists and the resulting cost is 
\beq
\mathcal{E}(t_f) = ( x_f - e^{At_f} x_o )^T W_r^{-1} (t_f) ( x_f - e^{At_f} x_o ),
\label{eq:energy_all_main}
\eeq
where the matrix $ W_r(t_f) = \int_0^{t_f} e^{A\tau} B B^T e^{A^T \tau} d \tau $ is called the {\em reachability Gramian} \cite{antsaklis2005linear}. 
The control that achieves the state transfer $ x_o \; \to \; x_f $ with minimal cost can be computed explicitly:
\beq
u(t) = B^T e^{A^T(t_f-t)} W_r^{-1} (t_f) (x_f - E^{At_f} x_o ) , \qquad t \in [0, \, t_f] .
\label{eq:control_gramian_main}
\eeq
Various metrics have been proposed to quantify the difficulty of the state transfer based on the Gramian, like its minimum eigenvalue $ \lambda_{\min}(W_r) $, its trace $ {\rm tr}(W_r) $, or the trace of its inverse $ {\rm tr}(W_r^{-1}) $, see \cite{Muller1972Analysis} and also SI for a more detailed description.

We would like now to describe how \eqref{eq:energy_all_main} depends on the eigenvalues of $A$. 
In order to do that, one must observe that \eqref{eq:energy_all_main} is the sum of contributions originating from two distinct problems: 1): {\em controllablity-from-0} (or {\em reachability}, as it is normally called in control theory \cite{antsaklis2005linear}) and 2): {\em controllablity-to-0}. 
The first problem consists in choosing $ x_o =0$, in which case \eqref{eq:energy_all_main} reduces to $ \mathcal{E}_r(t_f) = x_f ^T W_r^{-1} (t_f) x_f $, while in the second $ x_f =0 $ leads to $\mathcal{E}_c(t_f) =  x_o ^T   W_c^{-1} (t_f)  x_o $ where $ W_c (t_f)  = e^{-A^T t_f}  W_r (t_f)  e^{-A t_f} $ is a second Gramian, called the {\em controllability Gramian}. 
The two problems are characterized by different types of difficulties when doing a state transfer, all related to the stability of the eigenvalues of $A$. 
For instance the reachability problem is difficult along the stable eigendirections of $A$ because the control has to win the natural decay of the unforced system to 0, while the unstable eigenvalues help the system escaping from 0 by amplifying any small input on the unstable eigenspaces, see Fig.~\ref{fig:reach_contr_mixed} for a graphical explanation. 
The surfaces of $ \mathcal{E}_r (t_f) $ shown in Fig.~\ref{fig:reach_contr_mixed} (a) reflect these qualitative differences. 
On the contrary, the influence of the eigenvalues of $A$ is the opposite for the controllability-to-0 problem shown in Fig.~\ref{fig:reach_contr_mixed} (b). 
Hence if we want to evaluate the worst-case cost of a transfer between any $ x_o $ and any $ x_f $ (problem sometimes referred to as {\em complete controllability} \cite{Sontag:1998:Mathematical}), we have to combine the difficult cases of the two situations just described. 
This can be done combining the two Gramians into a ``mixed'' Gramian $ W_m$ obtained splitting $A$ into its stable and antistable parts and forming a reachability subGramian for the former and a controllability subGramian for the latter, see SI for the details.
Such Gramian can be computed in closed form only when the time of the transfer tends to infinity. 
In the infinite time horizon, in fact, both $ W_r $ and $ W_c $ diverge, but their inverses are well-posed and depend only on the stable modes the former and the unstable modes the latter. 
These are the parts constituting the inverse of $ W_m $, see Fig.~\ref{fig:reach_contr_mixed} (c). 
A finite-horizon version of $ W_m $ (and $ W_m^{-1}$) can be deduced from the infinite horizon ones.

\paragraph{Eigenvalues of random matrices.} 
The so-called {\em circular law} states that a matrix $ A $ of entries $ a_{ij}/\sqrt{n} $ where $ a_{ij} $  are i.i.d. random variables with zero-mean and unit variance has spectral distribution which converges to the uniform distribution on the unit disk as $ n \to \infty $, regardless of the probability distribution from which the $ a_{ij} $ are drawn \cite{Allesina2015Stability}. 
A numerical example is shown in Fig.~\ref{fig:eigenval}(a) (top left panel). 
By suitably changing the diagonal entries, the unit disk can be shifted horizontally at will, for instance rendering the entire spectrum stable (Fig.~\ref{fig:eigenval}(a), top mid panel) or antistable (Fig.~\ref{fig:eigenval}(a), top right panel). 
A random matrix is typically a full matrix, meaning that the underlying graph of interactions is fully connected. 
The circular law is however valid also for sparse matrices, for instance for Erd\H{o}s-R\'enyi (ER) topologies. 
If $ p$ is the edge probability, then $A = (a_{ij})/\sqrt{p\cdot n} $ still has eigenvalues distributed uniformly on the unit disk, see Fig.~\ref{fig:metrics_sparse}(a). 

A generalization of the circular law is the {\em elliptic law}, in which the unit disk containing the eigenvalues of  $A$ is squeezed in one of the two axes. 
To do so, the pairs of entries $ \{ a_{ij} , \, a_{ji} \}$ of $A$ have to be drawn from a bivariate distribution with zero marginal means and covariance matrix expressing the compression of one of the two axes, see \cite{Allesina2015Stability}. 
Various examples of elliptic laws are shown in the lower panels of Fig.~\ref{fig:eigenval} (a). 
Also elliptic laws generalize to sparse matrices, see Fig.~\ref{fig:metrics_sparse}(a). 


\section{Results}

\begin{figure}[h!]
\begin{center}
\subfigure[]{
\includegraphics[angle=0, trim=1cm 5cm 1cm 0cm, clip=true,width=12cm]{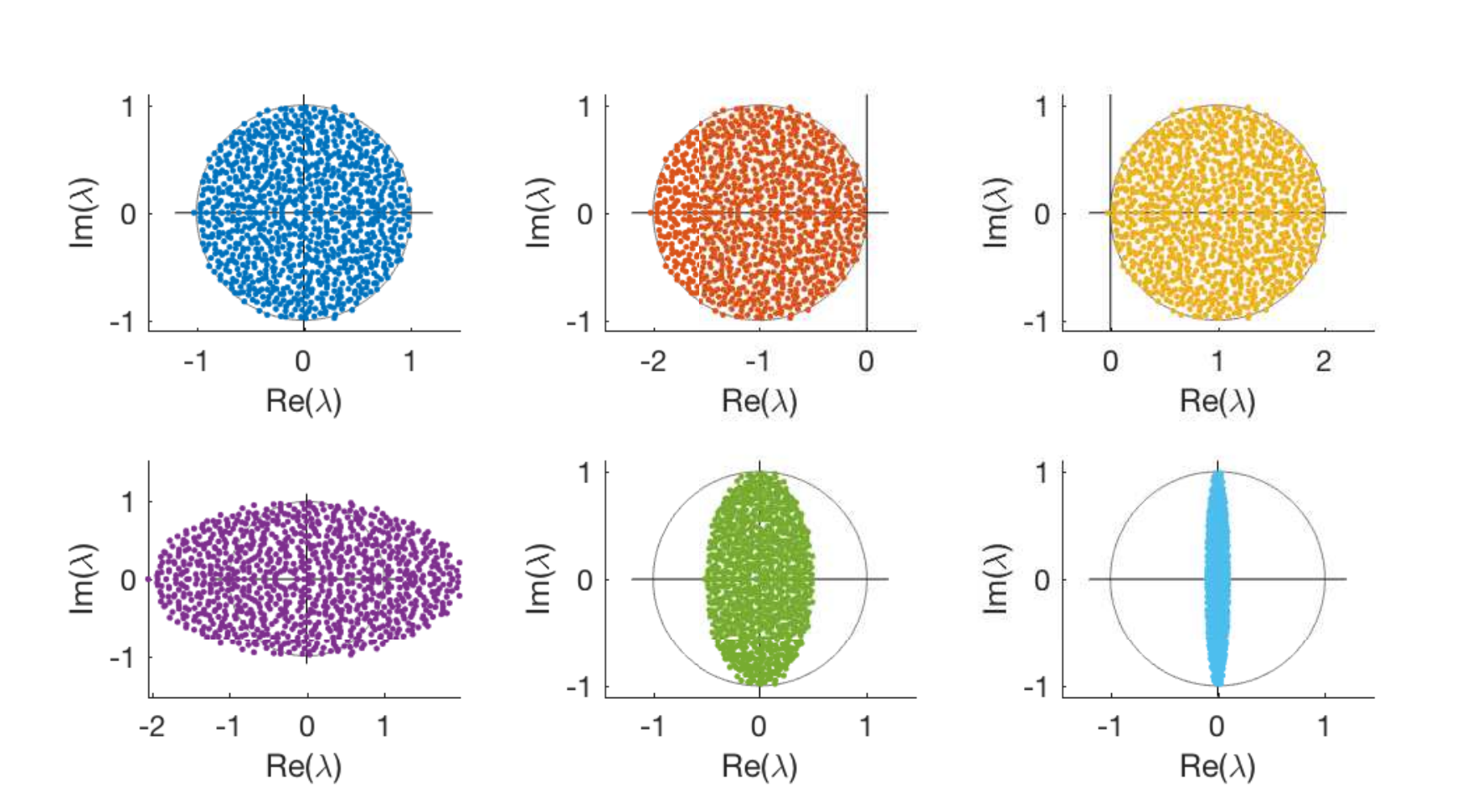}} \\
\subfigure[]{
\includegraphics[width=4cm]{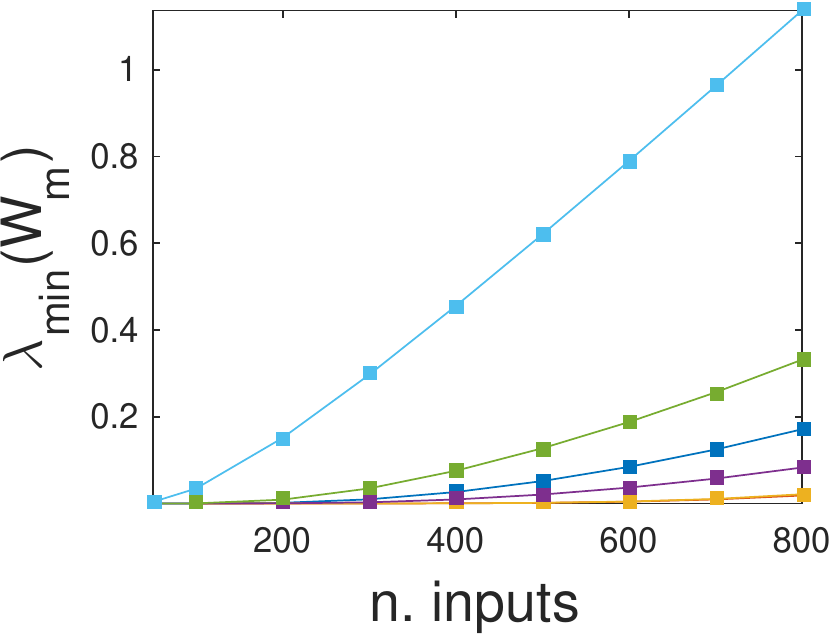}
\includegraphics[width=4cm]{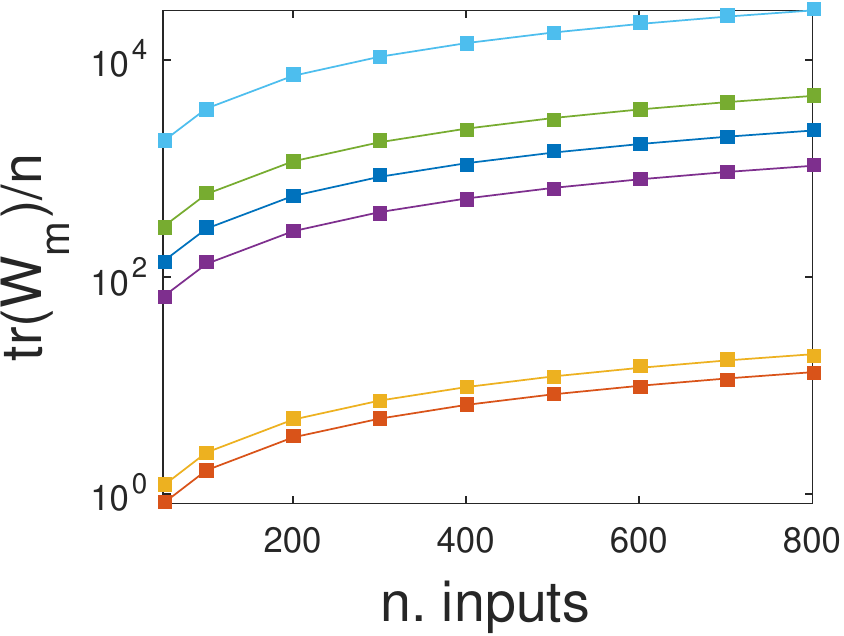}
\includegraphics[width=4cm]{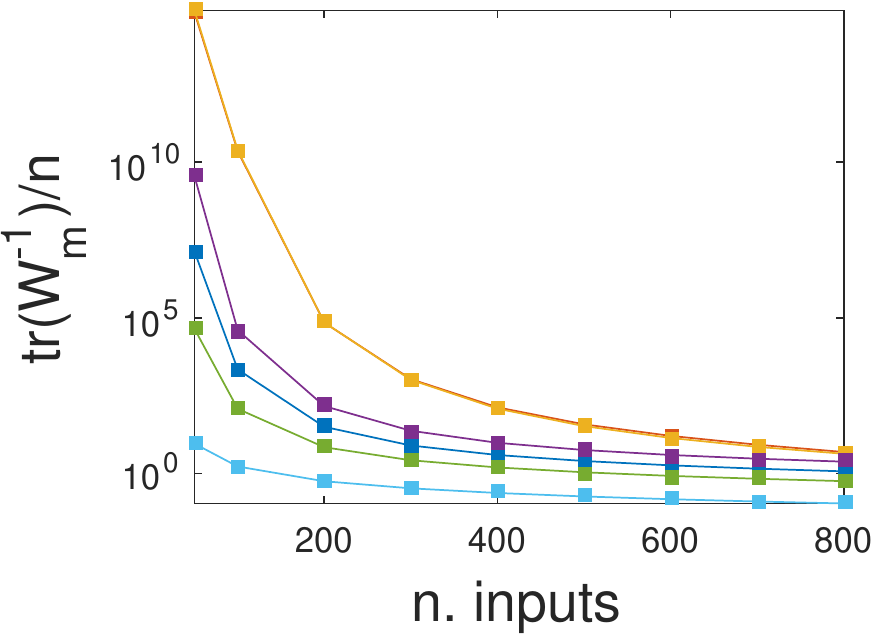}}
\caption[]{\small (a): Circular law and eigenvalue location. For a random matrix, the circular law allows to obtain state matrices $A$ with eigenvalues in prescribed locations, for instance in the unit disk (blue) or in the shifted unit disk (red and yellow) by altering the diagonal of $A$. The elliptic law allows to squeeze the eigenvalue location along one of the two axes of the complex plane (violet, green and cyan).
(b): Control energy for various metrics when the number of (randomly chosen) inputs grows. The data show a mean over 100 realizations of dimension $ n=1000$ (for each realization 100 different edge weights assignments are considered). The color code is as in (a). For all three metrics used to measure the control energy ($\lambda_{\min}(W_m) $, $ {\rm tr}(W_m) $ which should both be maximized, and $ {\rm tr}(W_m^{-1} )$ which should be minimized), the performances are strictly a function of the position of the eigenvalues of $A$. The minimum of the control energy is achieved when the eigenvalues have very small real part (cyan) and worsen with growing real part, following the order: cyan, green, blue, violet.}
\label{fig:eigenval}
\end{center}
\end{figure}

\paragraph{Control energy as a function of the real part of the eigenvalues of $A$.} 
 In a driver node placement problem, the inputs affect a single node, hence the columns of $ B $ are elementary vectors, i.e., vectors having one entry equal to 1 and the rest equal to 0. 
When $A$ is a random matrix, the underlying graph is generically fully connected, hence issues like selection of the number of driver nodes based on the connectivity of the graph become irrelevant. 
Having disentangled the problem from topological aspects, the dependence of the control effort from other factors, like the spectrum of $A$, becomes more evident and easier to investigate. 
If for instance we place driver nodes at random and use the mixed Gramian $ W_m $ to form the various energy measures mentioned above for quantifying the control effort, then we have the results shown in Fig.~\ref{fig:eigenval}(b). 
As expected, all indicators improve with the number of inputs.
What is more interesting is that when we repeat the computation for the various spectral distributions of Fig.~\ref{fig:eigenval}(a), the result is that the cost of controllability decreases when the (absolute value of the) real part of the eigenvalues of $A$ decreases.
All measures give an unanimous answer on this dependence, regardless of the number of inputs considered. 
In particular, when $A$ has eigenvalues which are very close to the imaginary axis (lower right panel of Fig.~\ref{fig:eigenval}(a) and cyan curves in Fig.~\ref{fig:eigenval}(b)) then the worst-case controllability direction is easiest to control (i.e., $\lambda_{\min}(W_m) $ is bigger), but also the average energy needed for controllability on all directions decreases (i.e., $ {\rm tr}(W_m) $ increases and $ {\rm tr}(W_m^{-1} ) $ decreases).

Recall that in a linear unforced dynamical system the real part of the eigenvalues of $A$ determines how fast/slow a system converges to the origin (stable eigenvalues, when real part of $ \lambda(A)$ is negative) or diverges to $ \infty$ (unstable eigenvalues, when real part of $ \lambda(A) $ is positive). 
Such convergence/divergence speed grows with the absolute value of the real part of $ \lambda(A)$. 
In the complete controllability problem, both stable and unstable modes of $A$ are gathered together, and all have to be ``dominated'' by the controls to achieve controllability. 
When the modes of the system are all slow, like when they are very close to the imaginary axis, then the energy needed to dominate them all is lower than when some of them are fast (i.e., the eigenvalues have large real part). 

An identical result is valid also for sparse matrices. In particular, for ER graphs with edge probability $ p=0.05$ and coefficients from a bivariate normal distribution (yielding elliptic laws as in Fig.~\ref{fig:metrics_sparse}(a)), 
the various norms used to quantify input energy are shown in Fig.~\ref{fig:metrics_sparse}(b). 
Their pattern is identical to the full graph case of Fig.~\ref{fig:eigenval}(b).

The computations shown in Fig.~\ref{fig:eigenval}(b) are performed with the infinite-horizon mixed Gramian $ W_m$ described in the SI, because such $ W_m$ can be easily computed in closed form.
A finite-horizon $ W_m (t_f) $ can be approximately obtained from it, but the arbitrarity of $ t_f $ makes it hard to set up an unbiased comparison of the various spectral distributions of $A$ of Fig.~\ref{fig:eigenval}(a), which are characterized by widely different time constants (inversely correlated to the amplitude of the real part of $ \lambda(A)$). 
Observe in Fig.~\ref{fig:metrics_time} 
how the various measures of controllability computed with a finite-time $ W_m(t_f) $ tend all to the infinite-time $ W_m$ but with different speeds.

\begin{figure}[h]
\begin{center}
\subfigure[]{
\includegraphics[width=5cm]{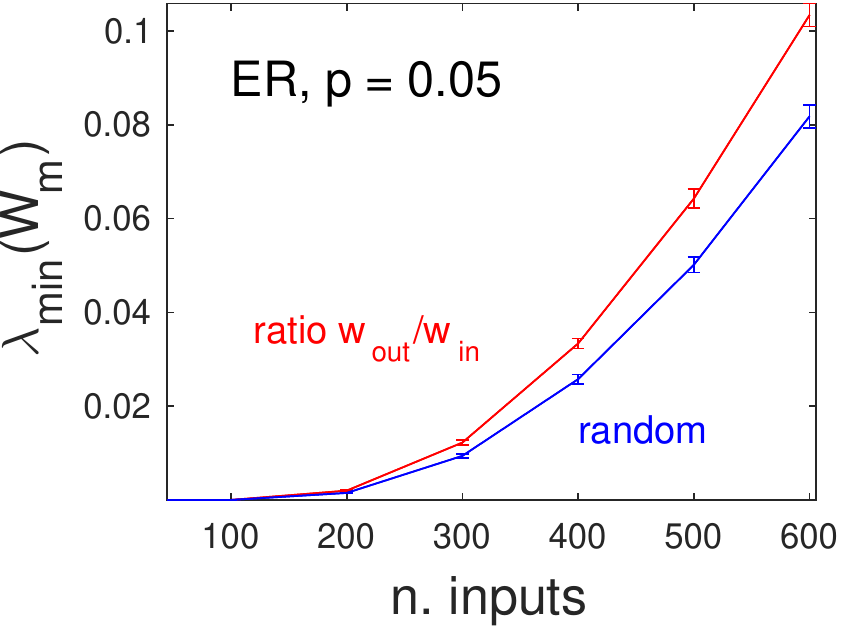}}
\subfigure[]{
\includegraphics[width=6cm]{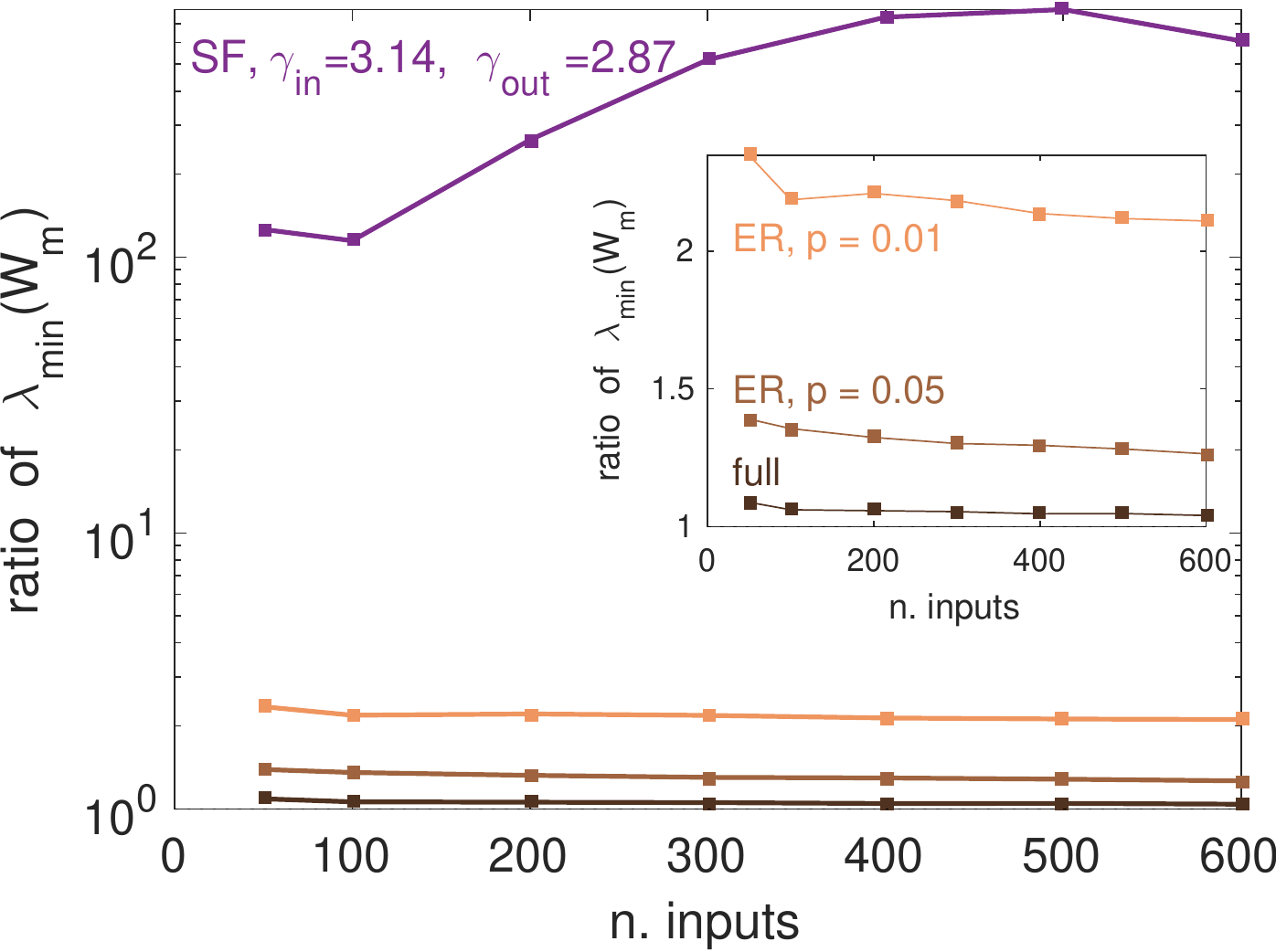}}
\subfigure[]{
\includegraphics[angle=0, trim=0cm 0cm 0cm 6cm, clip=true, width=10cm]{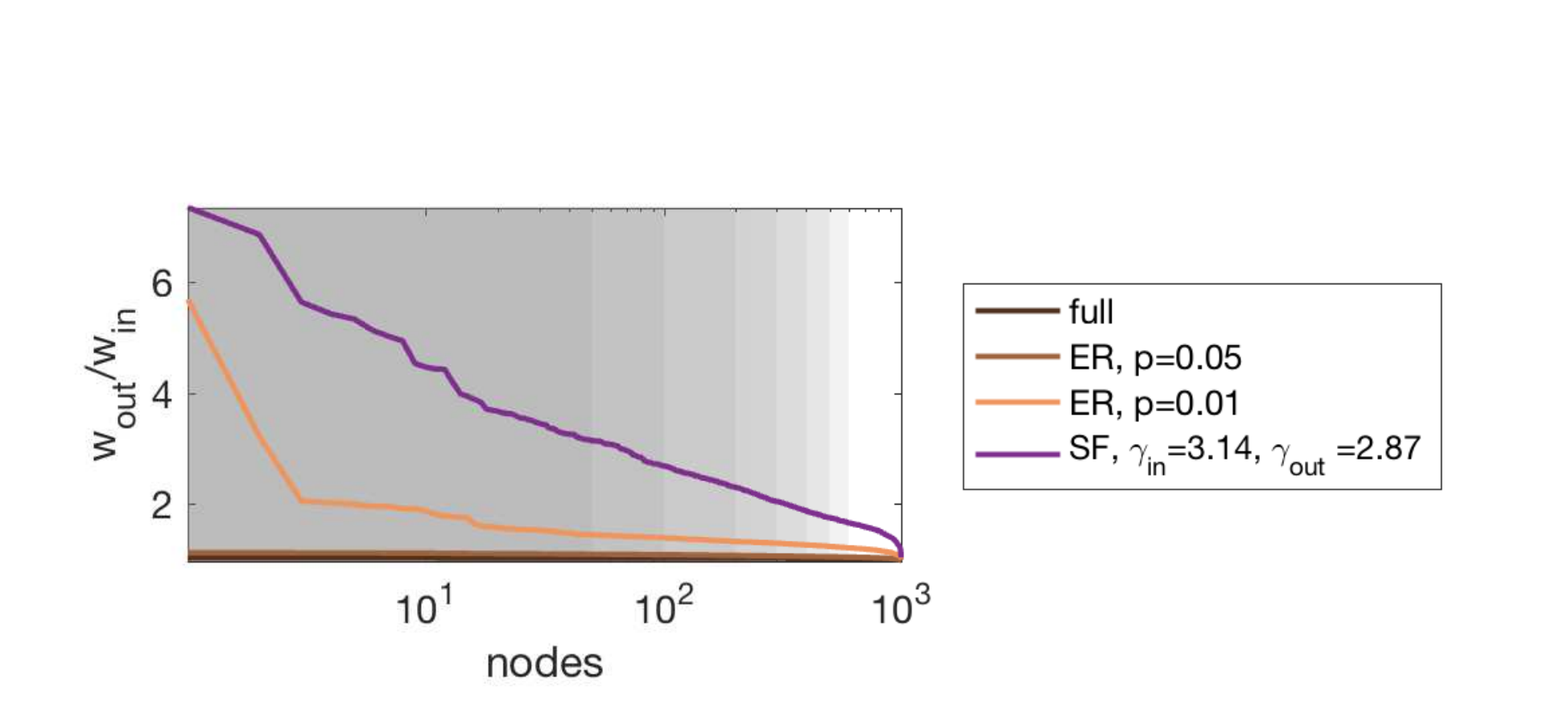}}
\caption[]{\small Driver node placement strategy: ranking according to $ r_w= w_{\rm out}/w_{\rm in}$. 
(a): When the driver nodes are chosen according to the ratio $ r_w $, then all the control energy measures improve with respect to a random node selection. Here $ \lambda_{\min} (W_m)$ is shown, the other energy measures are in Fig.~S3. 
Measures are means (and st. dev.) over 100 realizations of size $ n=1000$; for each realization 100 edge weight assignments are tested. 
(b): For ER networks, the improvement in $ \lambda_{\min} (W_m)$ increases with the sparsity of the graph (inset: zoomed comparison in linear scale). For other topologies, like SF directed graphs with indegree exponent $ \gamma_{\rm in} = 3.14 $ and outdegree exponent $ \gamma_{\rm out} =2.87 $, the improvement is remarkably more significant (two orders of magnitude, violet curve, see also Fig.~S5 
for more details). 
(c): the ratio $r_w $ of  ranked nodes is shown. For SF networks, the fraction of nodes having high $ r_w $ is much bigger than that of ER networks, and this leads to the much better performances in terms of control energy. 
The shaded areas represent the values of $m$ tested in our computations. }
\label{fig:driver_node_placement}
\end{center}
\end{figure}

\paragraph{Driver node placement based on weighted connectivity.} 
In the analysis carried out so far the driver nodes were chosen randomly.
A topic that has raised a remarkable interest in recent times (and which is still open in our knowledge) is devising driver node placement strategies that are efficient in terms of input energy \cite{Bof2015Role,Chen2016Energy,Li2015Minimum,Olshevsky2015Eigenvalue,Pasqualetti2014Controllability,Summers2016Submolularity,Tzoumas2016Minimal,Yan2015Spectrum}. 
If we consider as weighted indegree and outdegree at node $ i$ the sum of the weights in absolute value of all incoming or outgoing edges, i.e., $ w_{\rm in} (i) = \sum_{j=1}^n |a_{ij}| $ and $ w_{\rm out}(i) = \sum_{j=1}^n |a_{ji}| $ (a normalization factor such as $ \sqrt{p \cdot n} $ can be neglected), then a strategy that systematically beats random input assignment consists in ranking the nodes according to the ratio $ r_w(i) = w_{\rm out} (i) / w_{\rm in}(i)$ and placing inputs on the nodes with highest $ r_w$. 
In Fig.~\ref{fig:driver_node_placement}(a) the $ \lambda_{\min} (W_m) $ of this driver node placement strategy is compared with  a random selection.
If for full graphs the improvement is minimal, as the graphs become sparser it increases, and for ER networks with $p=0.01$ the $ \lambda_{\min} (W_m) $ obtained by controlling nodes with high $ r_w $ is more than twice that of the random choice of controls, see Fig.~\ref{fig:driver_node_placement}(b). 
As can be seen in Fig.~\ref{fig:driver_node_placement2}, 
all measures of input energy show a qualitatively similar improvement. 
When the topology of the network renders the values of $ r_w $ more extreme, like when direct scale-free (SF) graphs are chosen, with indegree exponent bigger than outdegree exponent \cite{Bollobas:2003:Directed}, see Fig.~\ref{fig:driver_node_placement4}, 
then the improvement in choosing driver nodes with high $ r_w $ becomes much more substantial, even of orders of magnitude bigger than a random selection, see Fig.~\ref{fig:driver_node_placement}(b) and Fig.~\ref{fig:driver_node_placement2_SF} 
for more details.

What we deduce from such results is that once the technical issues associated with minimal controllability can be neglected, a general criterion for controlling a network with a limited input cost is to drive the nodes  having the maximal disembalance between their weighted outdegree and indegree. 
Notice that our computation of weighted out/indegrees considers the total sum of weights in absolute value. 
When signs are taken into account in computing $ w_{\rm in} $ and $ w_{\rm out}$, then no significant improvement over random input placement is noticeable.
This is connected to the quadratic nature of the Gramian. 

\begin{figure}[h]
\begin{center}
\subfigure[]{
\includegraphics[angle=0, trim=8cm 20cm 40cm 30cm, clip=true, width=6cm]{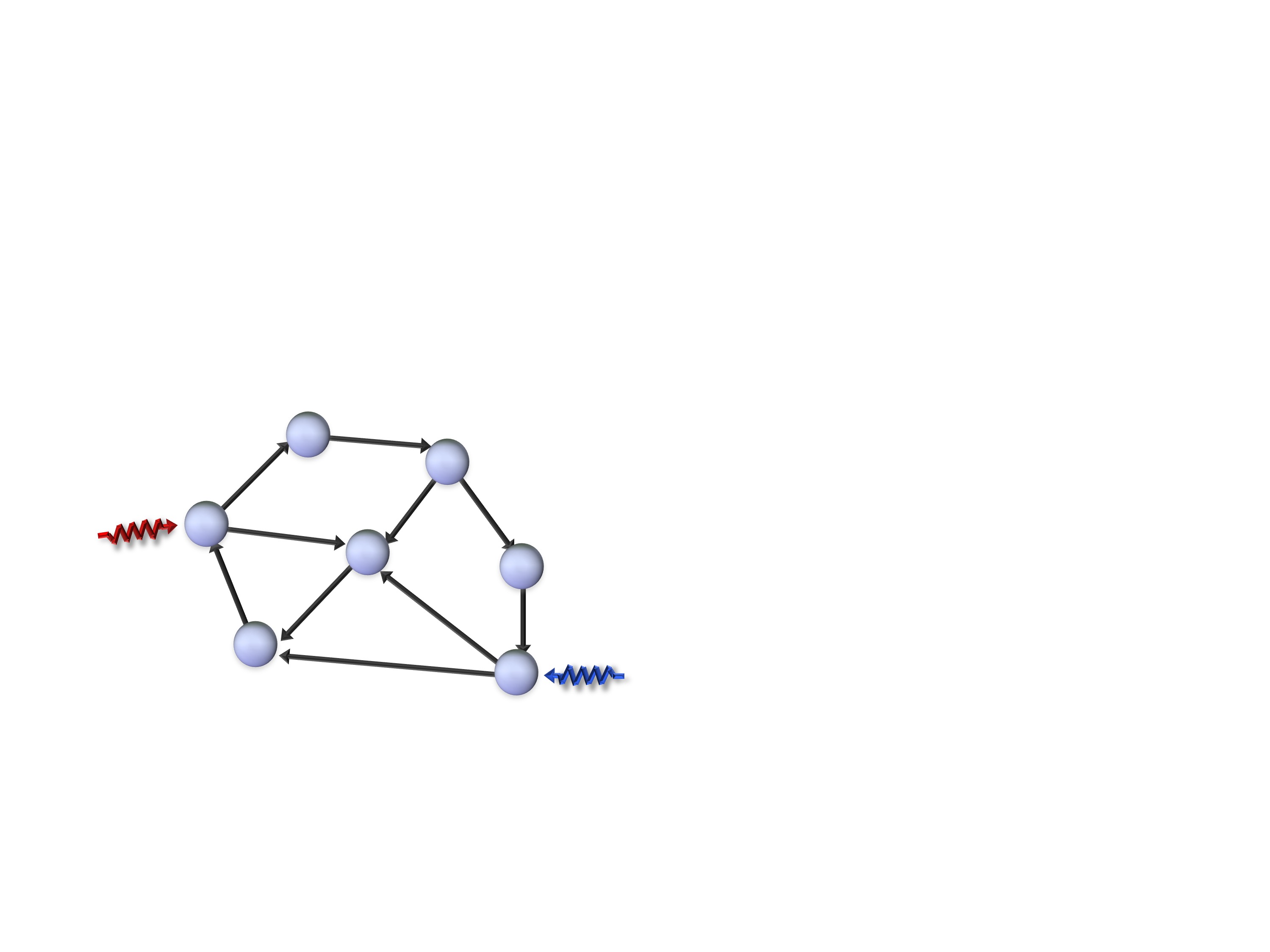}
\includegraphics[angle=0, trim=4cm 20cm 40cm 30cm, clip=true, width=6cm]{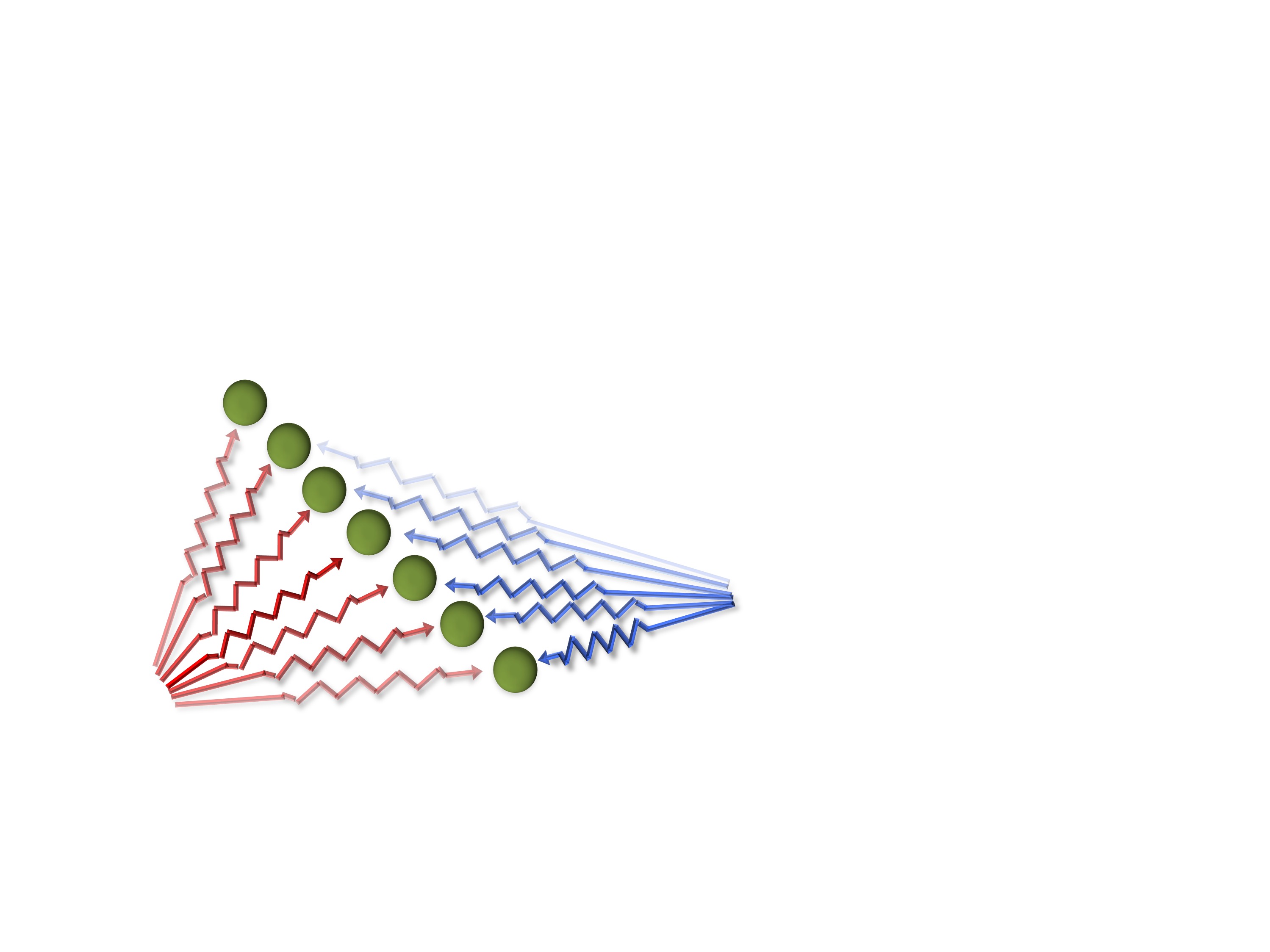}}
\subfigure[]{
\includegraphics[width=4cm]{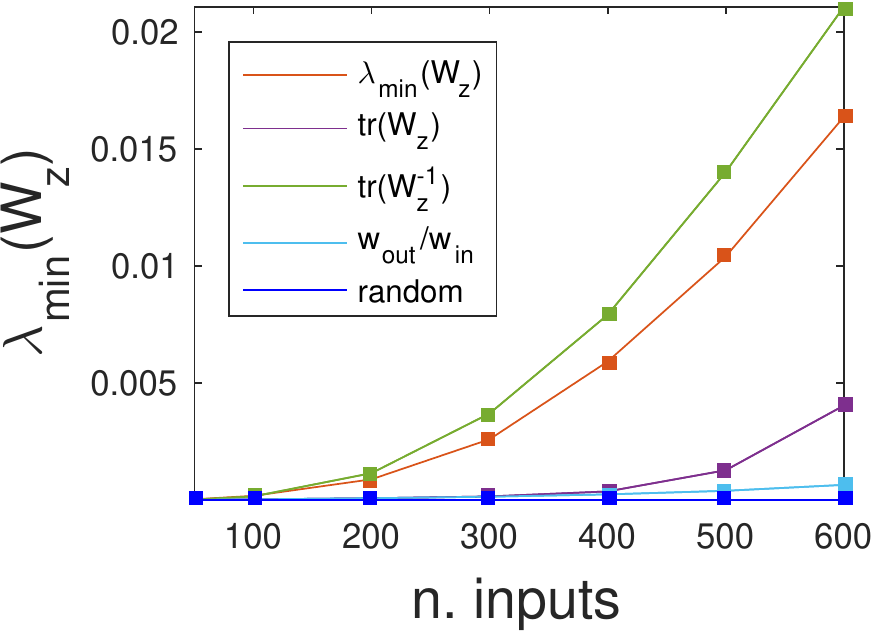}
\includegraphics[width=4cm]{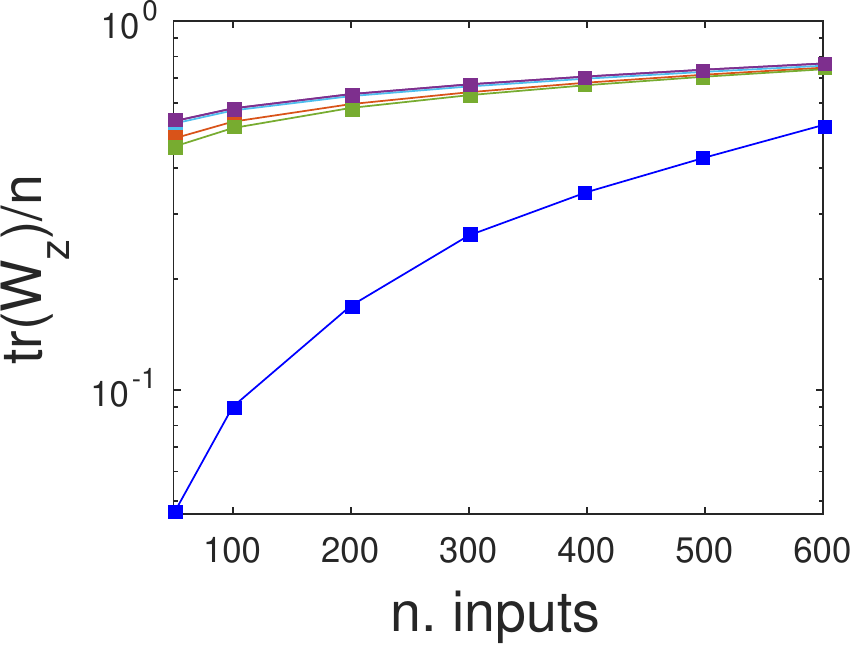}
\includegraphics[width=4cm]{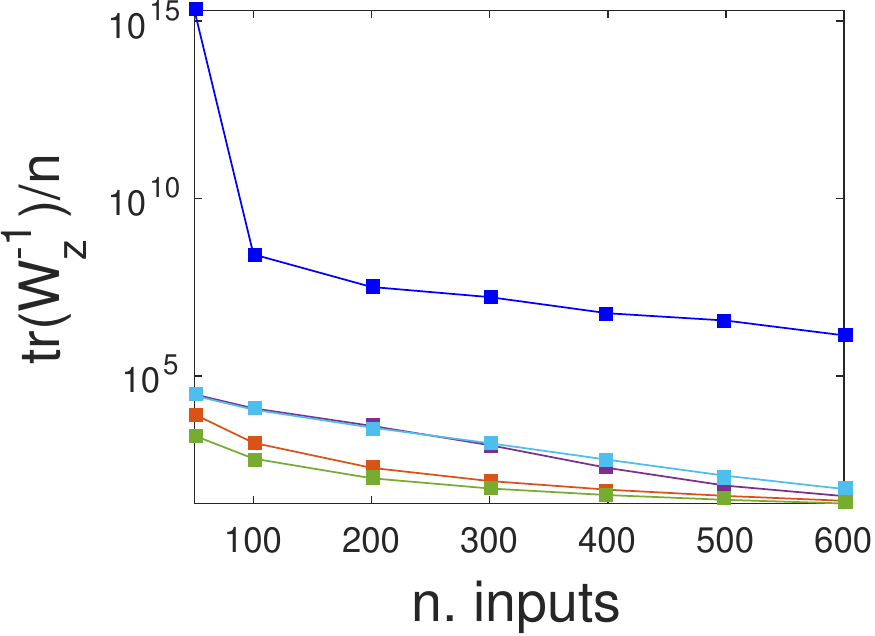}} 
\subfigure[]{
\includegraphics[width=9cm]{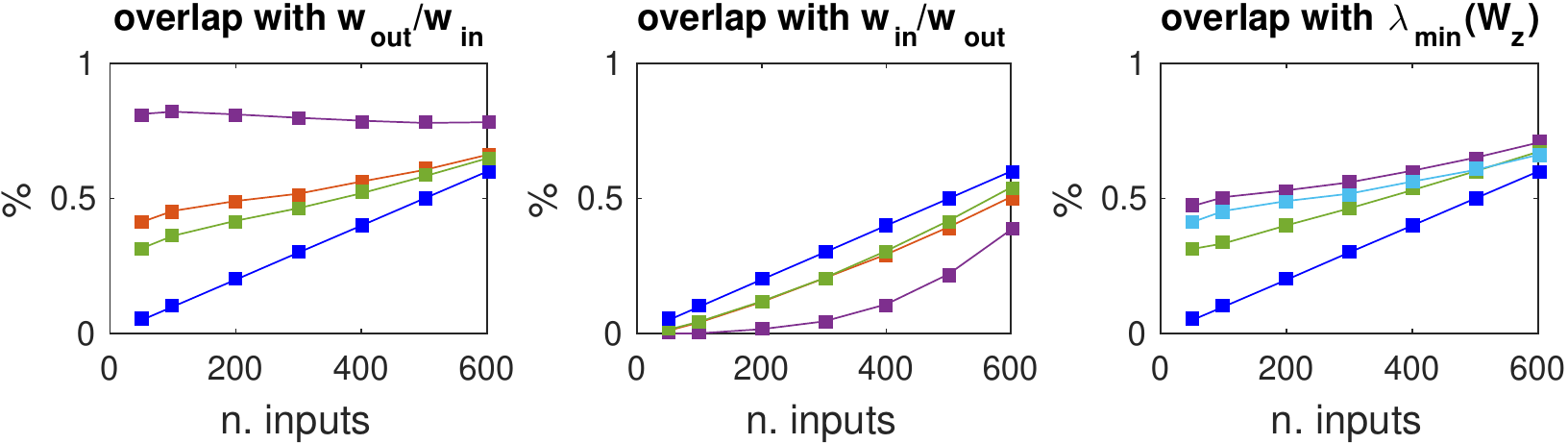}}
\subfigure[]{
\includegraphics[width=3cm]{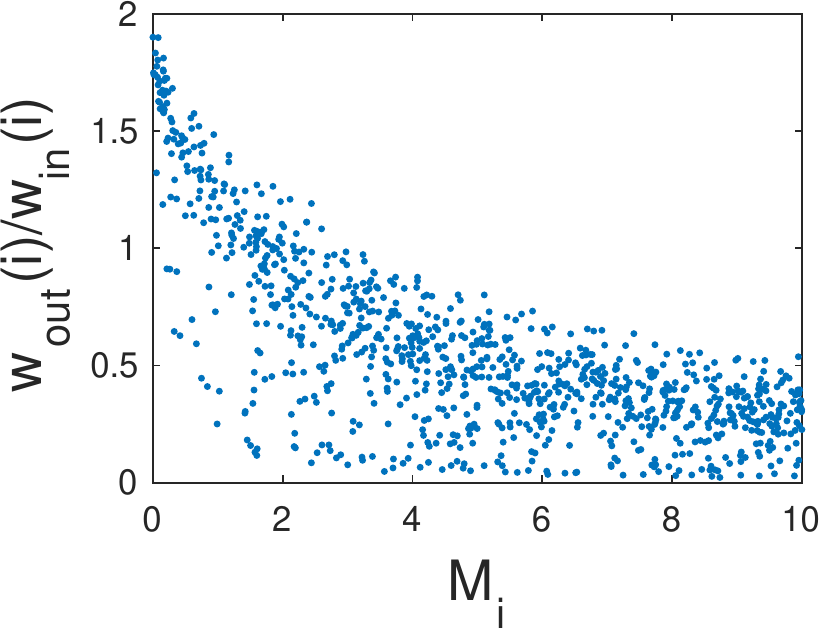}}
\caption[]{\small Driver node placement strategies for a network of coupled harmonic oscillators.
(a): The concept of driver node is basis dependent: when the basis changes in state space (for instance we pass from \eqref{eq:oscill_main3} to \eqref{eq:oscill_main4}), the control inputs no longer target a single node, but become spread across the entire state space (now decoupled into non-interacting modes). 
(b): Comparison of different driver node placement strategies for $ n=1000$ coupled harmonic oscillators. Shown are means over 100 realizations (with 100 edge weights samples taken for each realization). Red: driver node placement based on $ \lambda_{\min}(W_{z})$.  Violet: placement based on $ {\rm tr}(W_{z})$. Green: placement based on $ {\rm tr}(W^{-1}_{z})$. Cyan: placement based on $ w_{\rm out}/w_{\rm in} $. Blue: random input assignment. 
All driver node placement strategies always beat a random assignment, often by orders of magnitude. The green and red curves give similar performances and so do the cyan and violet. Notice that for $ {\rm tr}(W_{z})$ the violet curve gives the exact optimum. 
(c): Overlap in the node ranking of the different driver node placement strategies. Color code is the same as in (b). The only highly significant overlap is between $ w_{\rm out}/w_{\rm in} $ and $ {\rm tr}(W_z)$, while $ \lambda_{\min}(W_{z})$ and $ {\rm tr}(W^{-1}_{z})$ correspond to different node ranking patterns. Notice that none of the strategies orders nodes according to $ w_{\rm in}/w_{\rm out} $ (mid panel). (d) Inverse correlation between $ M_i $ and $ w_{\rm out}/w_{\rm in} $ (correlation coefficient around $-0.75$ in average).}
\label{fig:driver_node_placement_oscill}
\end{center}
\end{figure}

It is worth emphasizing that for a dynamical system the concept of driver node is not intrinsic, but basis-dependent. 
In fact, just like the idea of adjacency matrix of a graph is not invariant to a change of basis in state space, so inputs associated to single nodes (i.e., to single state variables) in the original basis become scattered to all variables in another representation of the same system, see Fig.~\ref{fig:driver_node_placement_oscill}(a). 
If we take a special basis in which the modes are decoupled (for instance the Jordan basis), then the contribution of the nodes to the modes (i.e., the eigenvectors of $A$) provide useful information for the investigation of minimum input energy problems. 
The topic is closely related to the so-called participation factors analysis in power networks \cite{Perez1982Selective}. 
Also quantities like $ w_{\rm in} $ and $ w_{\rm out}$ are basis-dependent and become nearly equal for instance if in \eqref{eq:system1_main} we pass to a Jordan basis. 
On the contrary, the eigenvalues of $A$ are invariant to a change of basis. 
Hence as a general rule, the control energy considerations that are consequence of the time constants of the system (like the dependence on the real part of the eigenvalues illustrated in Fig.~\ref{fig:eigenval}) are ``more intrinsic'' than those that follow from the particular basis representation we have available for a network. 

\begin{figure}[h]
\begin{center}
\subfigure[]{
\includegraphics[width=5cm]{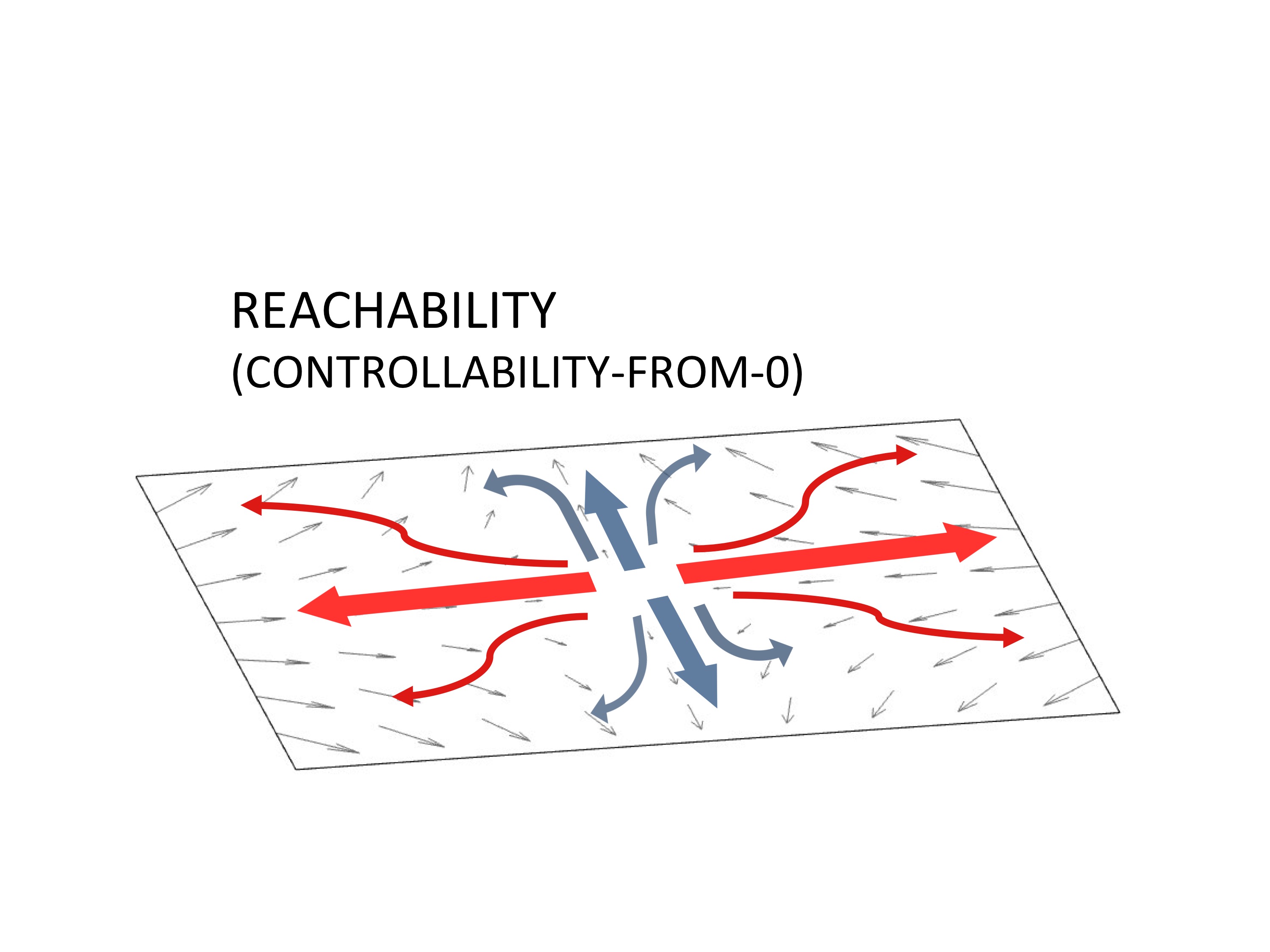}
\includegraphics[width=8cm]{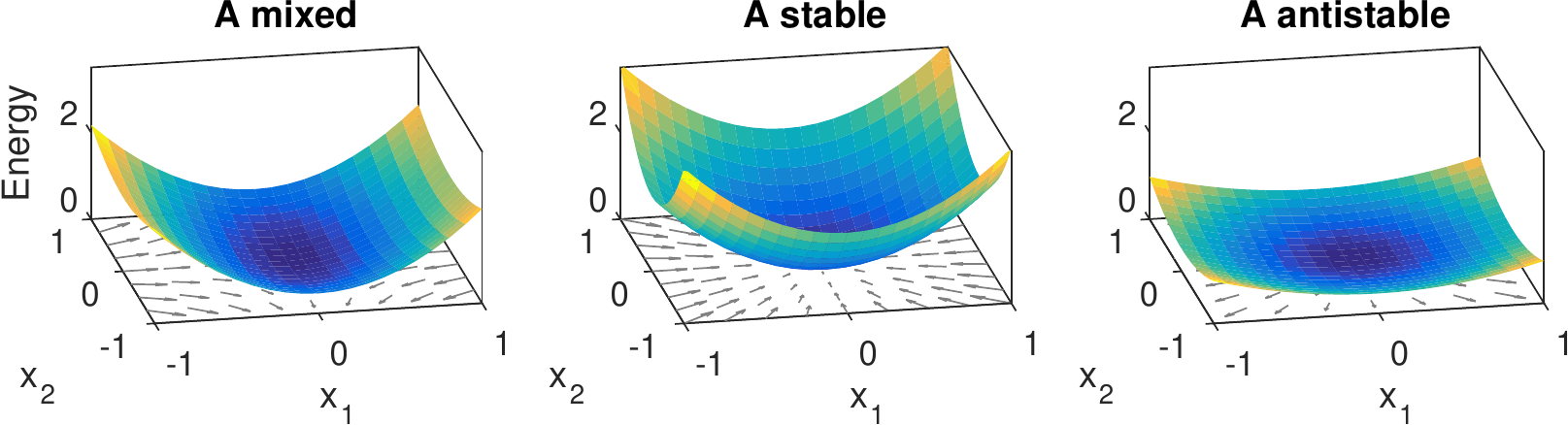}}
\subfigure[]{
\includegraphics[width=5cm]{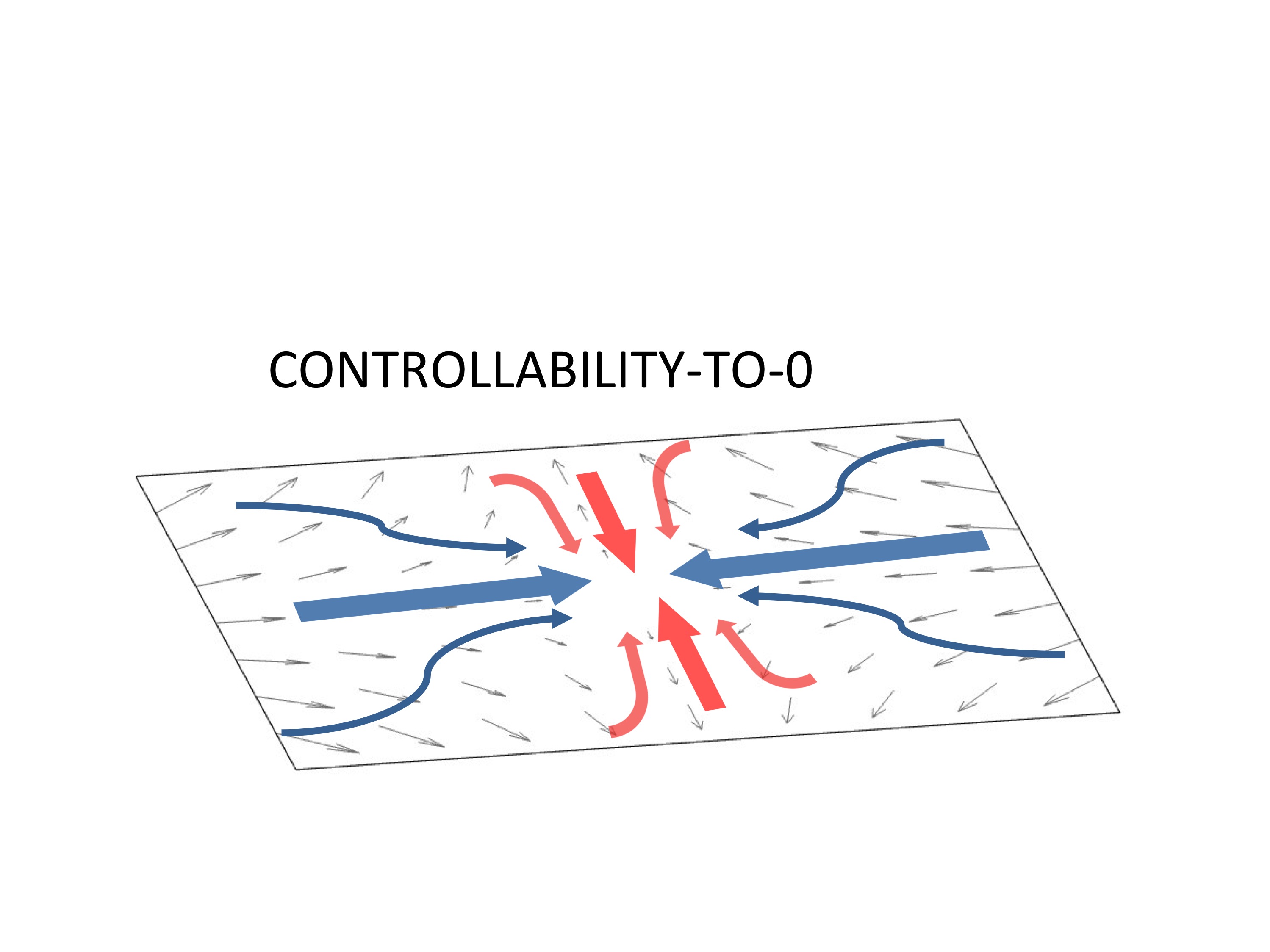}
\includegraphics[width=8cm]{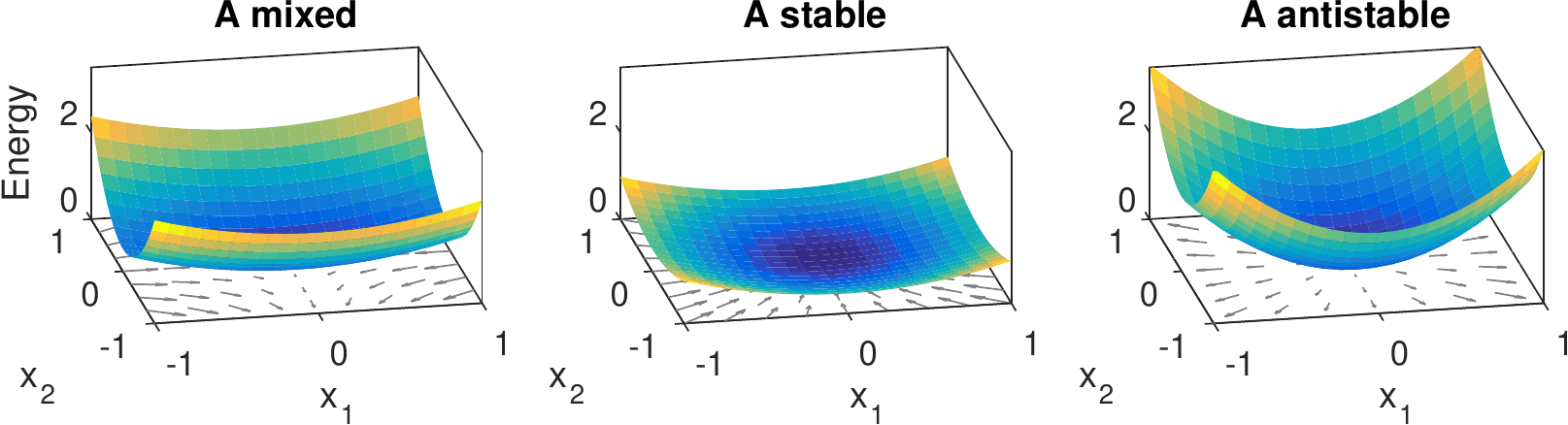}}
\subfigure[]{
\includegraphics[width=5cm]{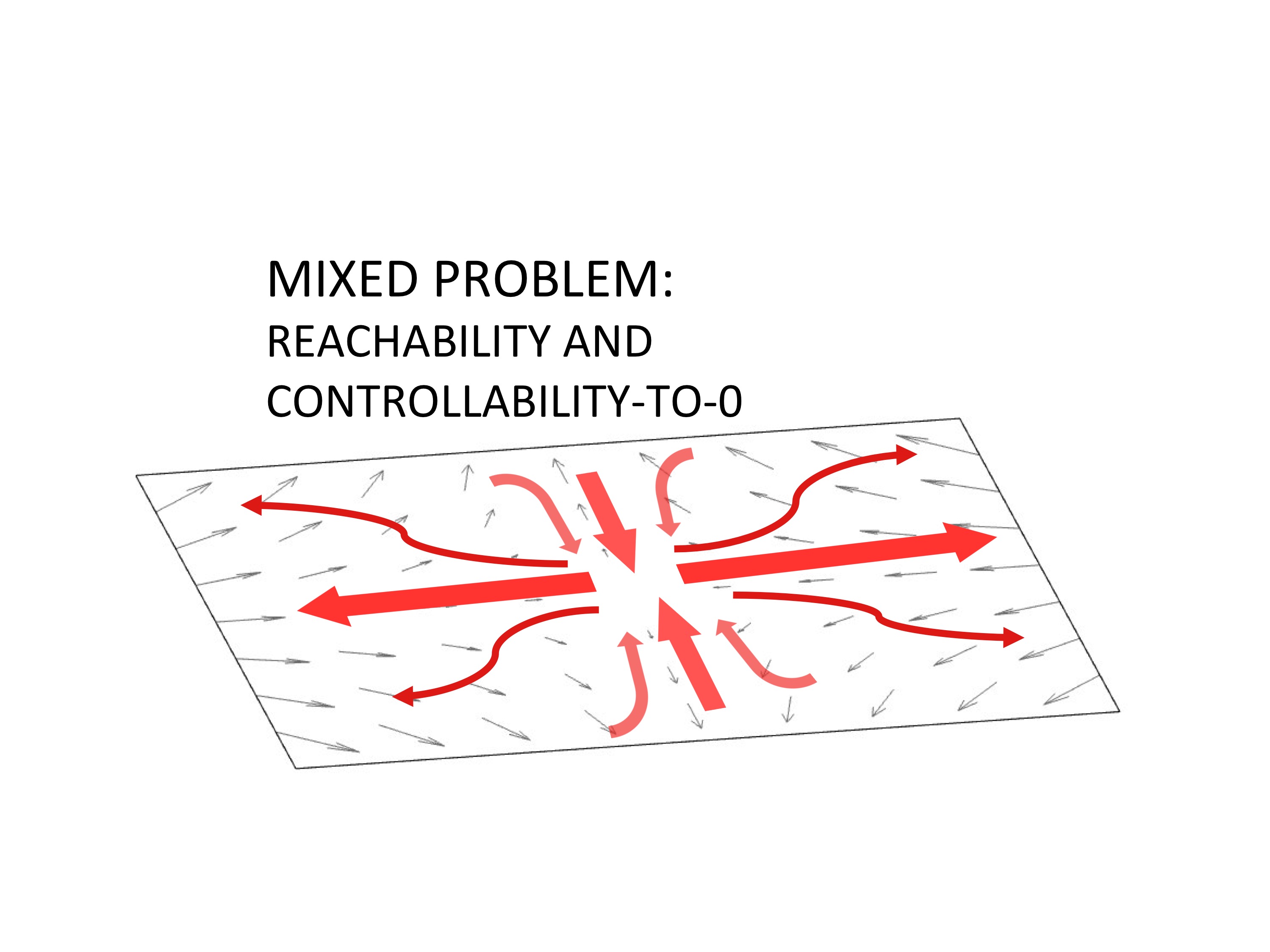}
\includegraphics[width=8cm]{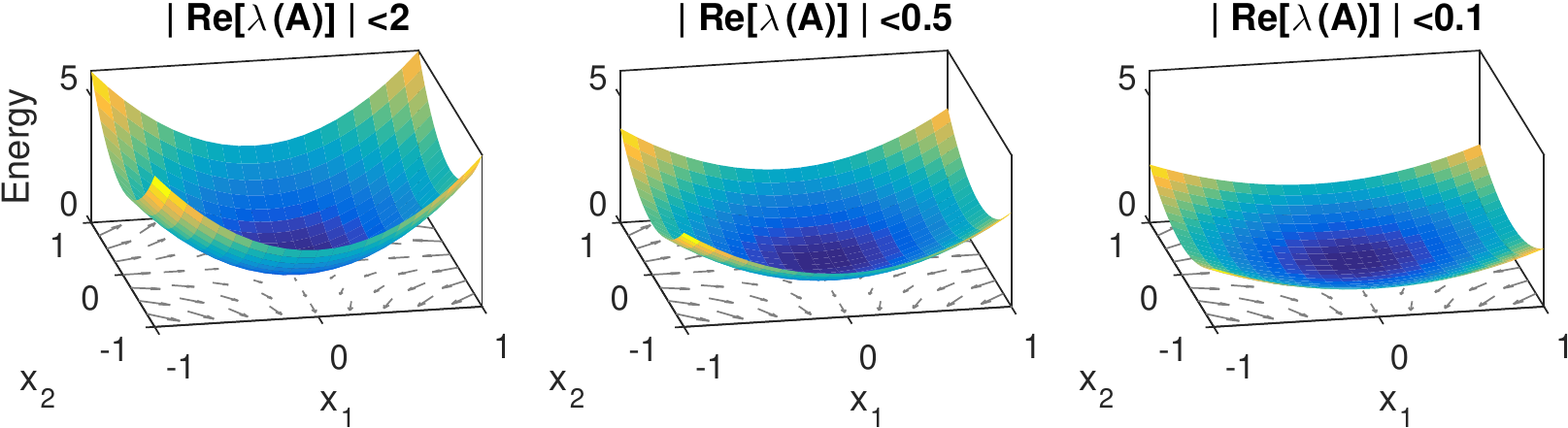}}
\caption[]{\small Reachability and Controllability-to-0 problems. 
(a): The reachability (or controllability-from-0) problem is difficult along the stable eigendirections of $A$ (red curves in the leftmost panel) and easy along the unstable ones (blue). This is reflected in the surfaces of $ \mathcal{E}_r(t_f) = x_f^T W_r^{-1} (t_f) x_f$ shown in the 3 rightmost panels. 
In particular, the reachability problem requires limited control energy when $A$ is antistable (rightmost panel).
(b):  The controllability-to-0 problem is difficult along the unstable eigendirections of $A$ (red) and easy along the stable ones (blue). The input energy surfaces, $ \mathcal{E}_c(t_f) = x_o^T W_c^{-1} (t_f) x_o$, reflect these properties. 
The case of $A$ stable requires the least control energy.
(c): The problem studied in this paper is a mixture of the two cases, collecting the worst-case of both. 
When the real part of the eigenvalues of $A$ is squeezed towards the imaginary axis as in the lower right panels of Fig.~\ref{fig:eigenval}(a), the input energy 
reduces accordingly.}
\label{fig:reach_contr_mixed}
\end{center}
\end{figure}

\paragraph{Real part of the eigenvalues and controllability with bounded controls.}
From what we have seen above, the control energy is least when the real part of the eigenvalues tends to vanish.
In the limit case of all eigenvalues of $A$ being purely imaginary, we recover a known result from control theory affirming that controllability from any $ x_o $ to any $ x_f $ can be achieved in finite time by means of control inputs of bounded amplitude. 
As a matter of fact, an alternative approach used in control theory to take into account the control cost of a state transfer is to impose that the amplitude of the input stays bounded for all times (rather than the total energy), and to seek for conditions that guarantees controllability with such bounded controls \cite{brammer1972controllability,jakobson1977,lee1986foundations}.
Assume $ u \in \Omega $, with $ \Omega $ a compact set containing the origin, for instance $ \Omega = [ -1, \, 1 ]^m$, where $ m $ is the number of control inputs. 
The constraint $ u \in \Omega $ guarantees that we are using at all times a control which has an energy compatible with the physical constraints of our network.
The consequence is, however, that reaching any point in $ \mathbb{R}^n $ may require a longer time, or become unfeasible. 
In particular a necessary and sufficient condition for any point to be reachable from 0 in finite time when $ u \in \Omega $ is that no eigenvalue of $A$ has a negative real part, see SI.
This is clearly connected with our previous considerations on the reachability problem without bounds on $ u$: when all modes of $A$ are unstable then the input energy required to reach any state from 0 is low (Fig.~\ref{fig:reach_contr_mixed}(a)) and becomes negligible for sufficiently long time horizons.
On the contrary, transferring any state to 0 in finite time with $ u \in \Omega $ is possible if and only if no eigenvalue of $A$ has a positive real part. Also in this case the extra constraints on the input amplitude reflects the qualitative reasoning stated above and shown in Fig.~\ref{fig:reach_contr_mixed}(b).
Also in the bounded control case, considering a generic transfer from any state $ x_o $ to any other state $ x_f $ means combining the two scenarios just described: formally a system is completely controllable from any $ x_o $ to any $ x_f $ in finite time and with bounded control amplitude $ u \in \Omega $ if and only if all eigenvalues of $A$ have zero real part, see SI for the details.
The findings discussed above for $ u $ unbounded are completely coherent with this alternative approach to ``practical controllability''.


\paragraph{Systems with purely imaginary eigenvalues: the case of coupled harmonic oscillators.}
A special case of linear system with purely imaginary eigenvalues is a network of $n$ coupled harmonic oscillators, represented by a system of second order ODEs
\beq
M \ddot q + K q = B u 
\label{eq:oscilla_main1}
\eeq
where $ M= M^T >0 $ is the inertia matrix, $ K = K^T \geqslant 0 $ is the stiffness matrix, typically of the form $ K = K_d + L $, with $K_d \geqslant 0 $ diagonal and $ L $ a Laplacian matrix representing the couplings. 
In \eqref{eq:oscilla_main1} the controls are forces, and the input matrix $B$ contains elementary vectors in correspondence of the controlled nodes. 
The state space representation of \eqref{eq:oscilla_main1} is
 \beq
 \dot x = A_o x + B_o u 
\label{eq:oscill_main3}
\eeq
with 
\[
x=  \begin{bmatrix} M q \\ M \dot q \end{bmatrix} \in \mathbb{R}^{2n}, \qquad 
A_o = 
\mleft[ \begin{array}{c|c}
0 & I \\
\hline
-  KM^{-1} & 0 
\end{array}\mright] , \quad \text{ and  } \quad 
B_o =
\mleft[ \begin{array}{c}
0  \\
\hline
  B 
\end{array}\mright].
\]
The system \eqref{eq:oscill_main3} has purely imaginary eigenvalues equal to $ \pm i \omega_j$, $ j=1, \ldots, n$, where $ \omega_j $ are the natural frequencies of the oscillators. 
If $ \Omega^2 = {\rm diag} ( \omega_1^2 , \ldots, \omega_n^2 )$ and $ \Psi =\begin{bmatrix} \psi^1 \ldots \psi^n \end{bmatrix} $ is the matrix of corresponding eigenvectors, then in the so-called modal basis the oscillators are decoupled and one gets the state space representation
\beq
\begin{split} \dot z & = A_1z + B_1 u  \\
& = \mleft[ \begin{array}{c|c}
0 & I \\
\hline
- \Omega^2 & 0 
\end{array}\mright] z + 
\mleft[ \begin{array}{c}
0  \\
\hline
 \Psi^{T}  M^{-1} B 
\end{array}\mright] u
\end{split} 
\label{eq:oscill_main4}
\eeq
where $ z = \mleft[ \begin{array}{c|c}
\Psi^{-1}  & 0 \\
\hline
0 & \Psi^{-1} 
\end{array}\mright] x $.
See SI for the details.
When a system has purely imaginary eigenvalues, the finite time Gramian diverges as $ t_f \to \infty$.
However, in the modal basis \eqref{eq:oscill_main4} the Gramian is diagonally dominant and linear in $t_f$, hence as $ t_f $ grows it can be approximated by a diagonal matrix which can be computed explicitly \cite{Arbel1981Controllability}:
\beq
 W_{z} (t_f) \approx  \sum_{j=1}^n \frac{\beta_j}{2 M_j^2 }  \begin{bmatrix}  
\frac{(\psi^1_j )^2 }{\omega_1^2 } \\ 
 & \ddots \\
& &  \frac{(\psi^n_j )^2 }{ \omega_n^2} \\ 
& & & (\psi^1_j )^2\\ 
& & & & \ddots \\
& & & & &  (\psi^n_j )^2
\end{bmatrix} t_f ,
\label{eq:oscill_main5}
\eeq
where $ \beta_j =1 $ if the $ j$-th input is present and 0 otherwise. 
Using \eqref{eq:oscill_main5}, the three measures of control energy adopted in this paper give rise to simple strategies for minimum energy driver nodes placement, which in some cases can be computed exactly for any $ n$ (for instance for the metric $ {\rm tr}(W_z)$, see SI). 
Fig.~\ref{fig:driver_node_placement_oscill} shows that such strategies are always beating a random driver node placement, often by orders of magnitude.

Also $ w_{\rm out}/w_{\rm in} $ is still a good heuristic for driver node placement strategy.
This can be understood by observing that the model \eqref{eq:oscilla_main1} is symmetric hence for it in- and out-degrees are identical. However, since $ A_o $ has rows rescaled by $M^{-1}$, $ w_{\rm out}$ is affected directly: when the inertia $ M_{ii}$ is big, the corresponding $ w_{\rm out}(i) = \sum_{j=1}^n K_{ji}/M_{ii}$ is small and viceversa. 
No specific effect is instead induced on $ w_{\rm in}$. 
In the representation \eqref{eq:oscill_main3}, selecting nodes according to $  w_{\rm out} /   w_{\rm in} $ means placing control inputs on the lighter masses, see Fig.~\ref{fig:driver_node_placement_oscill} (d). 
When the harmonic oscillators are decoupled ($ L=0$) then $ m< n $ means controllability is lost, but nevertheless the least control energy of the $m$ inputs is indeed obtained when driving the oscillators of least inertia. 
A weak (and sparse) coupling allows to recover controllability, while the least inertia as optimal driver node strategy becomes suboptimal. 
When the coupling becomes stronger (for instance when the coupling graph is more connected) then the inertia of an oscillator is less significant as a criterion for selection of driver nodes: the modes of the system are now spread throughout the network and no longer localized on the individual nodes. 
As shown in Fig.~\ref{fig:driver_node_placement_oscill_full}, 
in a fully connected network of harmonic oscillators, driver node strategies based on $ w_{\rm out}/w_{\rm in} $ and on $ {\rm tr}(W_{z})$ tend to perform considerably worse for the other measures ($ \lambda_{\min}(W_{z})$ and ${\rm tr}(W_{z}^{-1})$), while for a sparse graph (here ER graphs with $ p=0.05$), of the three explicit optimal driver node placement strategies available in this case, $ {\rm tr}(W_{z})$ has a high overlap with $ w_{\rm out} /  w_{\rm in} $, see Fig.~\ref{fig:driver_node_placement_oscill} (b), while the other two tend to rank controls in somewhat different ways.
Given that in this case we have three strategies that are (near) optimal for the chosen measure of control energy, the dissimilarity of the node rankings of these three strategies means that the driver node placement problem is heavily dependent on the way control energy is quantified.


\paragraph{Controlling vibrations of polyatomic molecules.}
Coupled harmonic oscillators are used in several applicative contexts, for instance in the active control of mechanical vibrations \cite{Gawronski1998Dynamics,Leleu2001Piezoelectric} or in that of flexible multi-degree structures (aircrafts, spacecrafts, see \cite{junkins1993introduction}) where our controllability results can be applied straightforwardly (and compared with analogous/alternative approaches described for instance in \cite{Arbel1981Controllability,Hamdan1989Measures,junkins1993introduction,Leleu2001Piezoelectric,Shaker2012Optimal,Tarokh1992Measures,vandeWal2001Review}). 
Another context in which they are used is in controlling molecular vibrations of polyatomic molecules \cite{Warren1,Zewail1980Laser}. 
The assumption of harmonicity is valid in the regime of small oscillations near equilibrium, in which the potential energy is approximated well by the parabola $ V(q) = \frac{1}{2} q^T K q$ (here $ q$ is the quantum expectation value of the displacement from the equilibrium position and only vibrational degrees of freedom are considered\footnote{For a molecule of $n$ atoms the number of independent degrees of freedom is $ 3n - 5$ i.e., 3 for each atom, minus the coordinates of the center of mass. All masses and stiffness constants should be rescaled relative to it.}). 
The node-driving setting described by \eqref{eq:oscill_main3} corresponds here to controlling the vibrations of single bonds, as in mode-selective chemistry \cite{Dian2002Conformational,Staanum2010Rotational,Zewail1980Laser}. 
In the modal basis \eqref{eq:oscill_main4}, the free evolution of the modes (i.e., $A_1$) is decoupled, but not the input matrix $ B_1$, see Fig.~\ref{fig:driver_node_placement_oscill}(a). 
The $B_1 $ terms in \eqref{eq:oscill_main4} specify how the energy exciting a specific vibrational bond propagates through the entire molecule and affects all its modal frequencies. 
The methods developed above for quantifying the control energy are applicable also to this context. 
In particular the Gramian can be used to estimate the energy spreading of a monochromatic input field: for the $ j$-th bond it is proportional to the $j$-th row of $ \Psi$ (i.e., it consists of the $ j$-th components of all eigenvectors $ \psi^1, \ldots, \psi^n$). 

The basic principle we have adopted so far (minimize the input energy, here the fluence of the pumping field) implicitly favours the selection of inputs having a broad ``coverage" in the modal basis, or, said otherwise, favours the intramolecular spreading of vibrational energy to the entire molecule. 
This is clearly not the main task of selective-mode laser chemistry, which on the contrary aims at keeping  as much energy as possible concentrated on specific bonds or modes.
Given that the power of the laser field is not a limiting factor, a control problem that can be formulated using the tools developed in this paper is for instance to choose $m$ monochromatic excitations selectively tuned on the stiffness constants of $m$  bonds (i.e. for us certain columns of $ B_o $) so as i) to guarantee controllability; ii) to maximize the power at a certain mode $ \omega_k$ (representing for example a conformational change that one wants to achieve in the molecule \cite{Dian2002Conformational}). 
For the diagonal Gramian \eqref{eq:oscill_main5}, this amounts to choosing the indexes $ j_1, \ldots, j_m \in \{ 1, \ldots, n \} $ such that $ \sum_{\ell=1}^m ( \psi^k_{j_\ell} )^2 $ is maximized, a problem which can be solved exactly once $ \Psi$ is known. 
Notice that even when energy spreads through the bonds because of the couplings, it is in principle possible to ``refocus'' it towards a single bond using the dynamics of the system. 
In the linear regime, a formula like \eqref{eq:control_gramian_main} can be used to compute explicitly the controls needed to refocus it on a specific bond, corresponding for instance to $ q_f $ having a single non-zero component. 
This does not require to solve an optimal control problem, as proposed for instance in \cite{Shi1988Optimal,Shi1990Optimal}. 

\begin{figure}[h]
\begin{center}
\subfigure[]{
\includegraphics[angle=0, trim=0cm 7cm 0cm 0cm, clip=true, width=10cm]{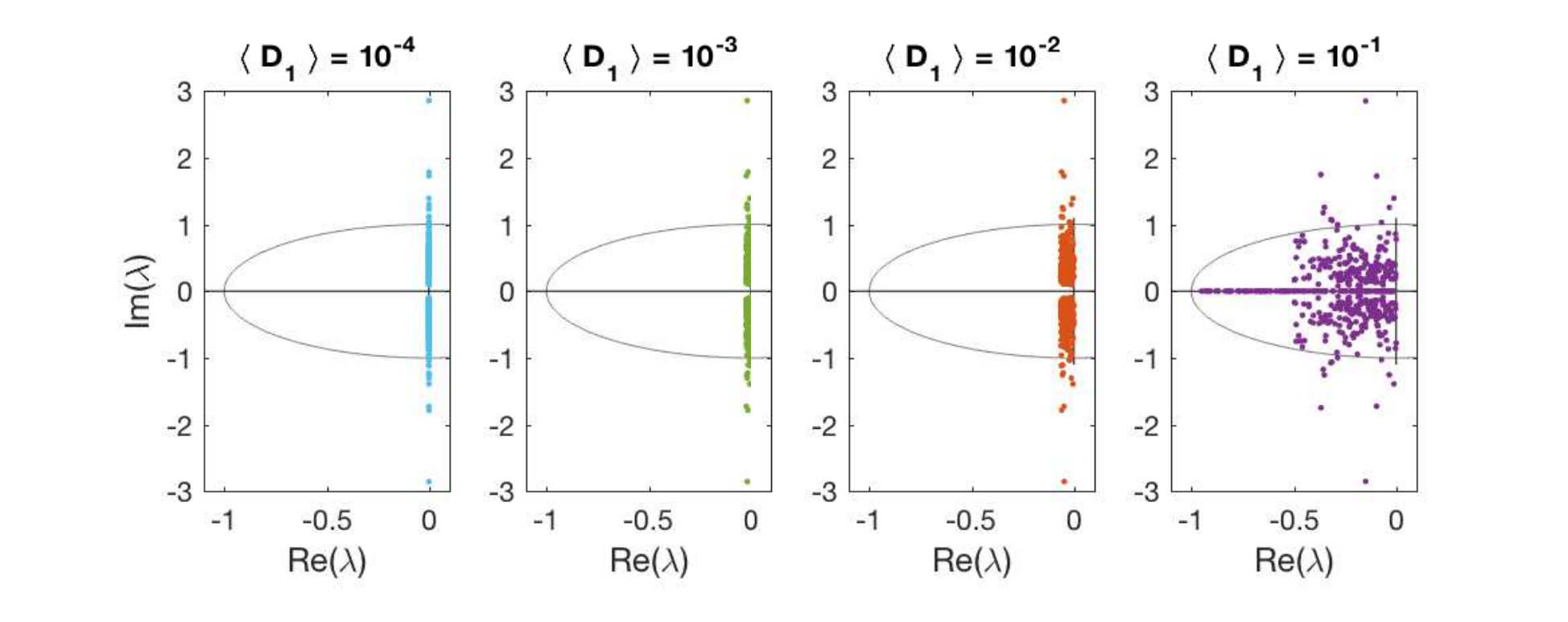}} \\
\subfigure[]{
\includegraphics[angle=0, trim=0cm 00cm 0cm 0cm, clip=true, width=4cm]{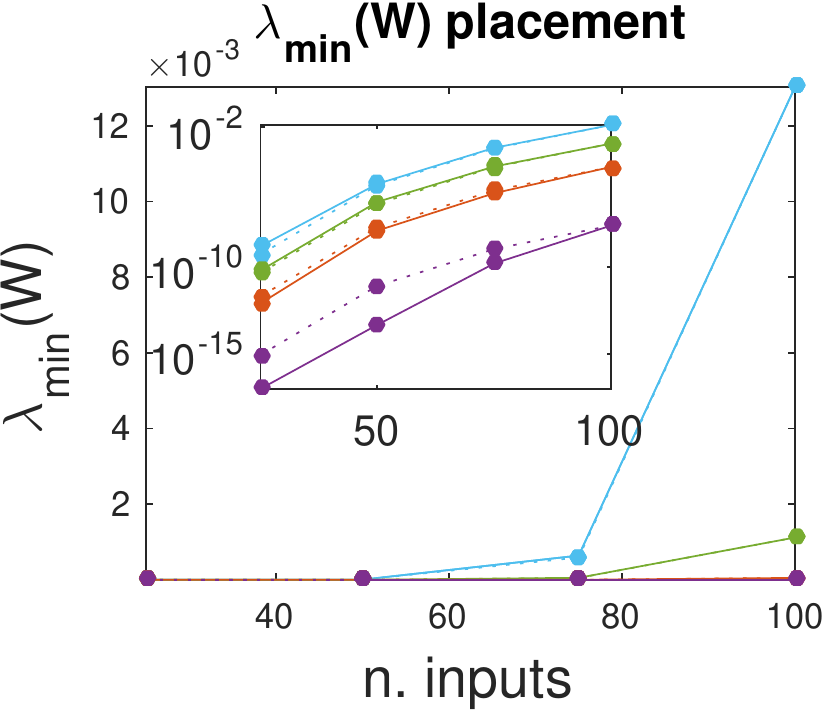}
\includegraphics[angle=0, trim=0cm 0cm 0cm 0cm, clip=true, width=4cm]{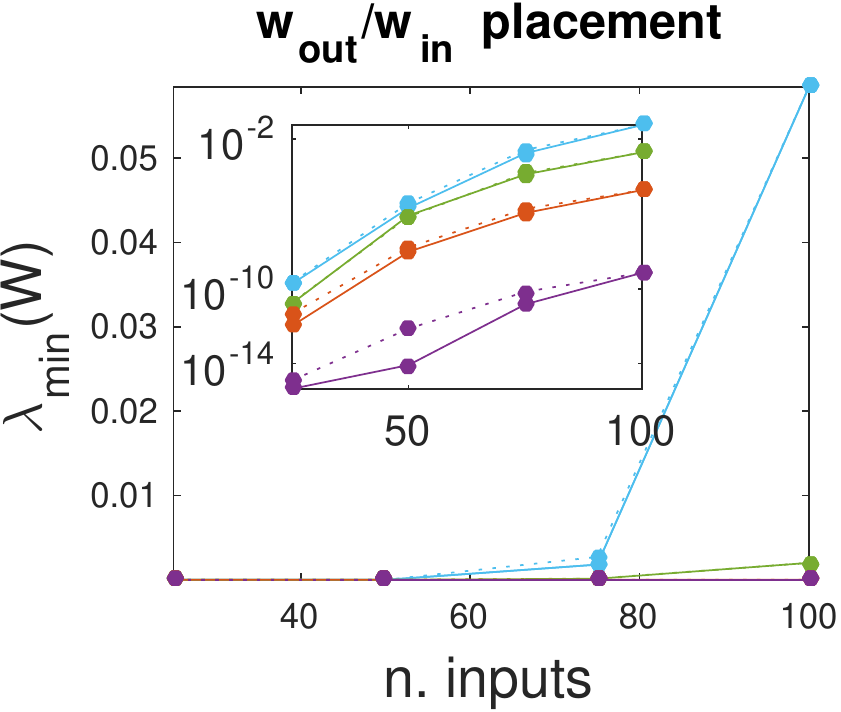}
\includegraphics[angle=0, trim=0cm 0cm 0cm 0cm, clip=true, width=4cm]{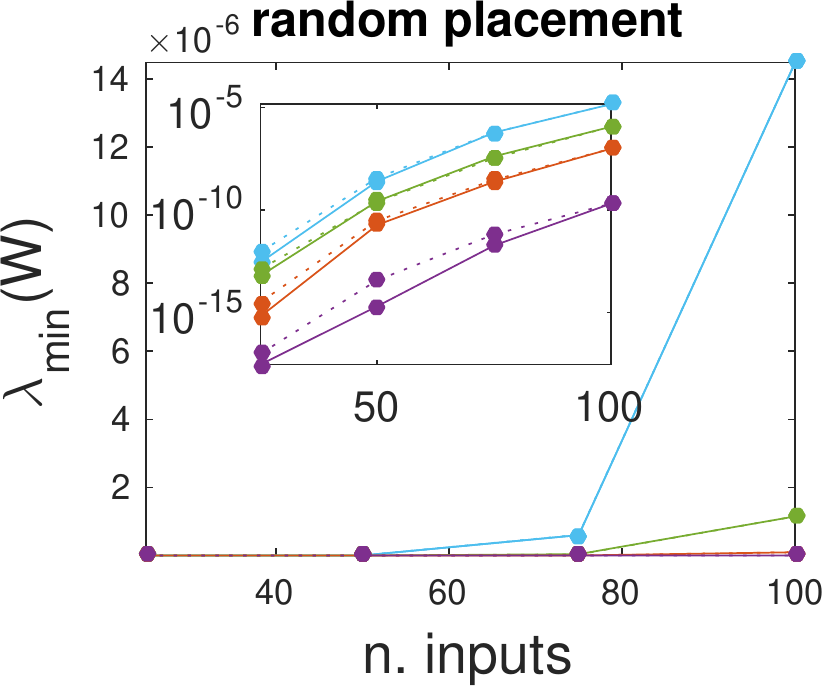}}
\caption[]{\small Minimum energy control of power grids for varying damping coefficients.
(a): The eigenvalues of the state space system \eqref{eq:damped_main2} for the North EU power grid \cite{Menck2014How} with uniformly distributed masses $ (\langle M_i \rangle = 10$) and damping coefficients that vary across 4 orders of magnitude.
(b): Control energy for the metric $ \lambda_{\min} (W_r ) $ when the driver nodes are placed according to $ \lambda_{\min} (W_r) $ (left panel), $ w_{\rm out}/w_{\rm in} $ (mid panel), or randomly (right panel). The values of $ \lambda_{\min} (W_r ) $ corresponding to the 4 choices of damping made in (a) are shown in solid lines (same color code as in (a)), while in dotted lines the values of $ \lambda_{\min}(W_z ) $ are shown (suitably normalized to eliminate the explicit dependence from $ t_f$, see \eqref{eq:oscill_main5}).
The insets show the same quantities in log scale. Values are all averages over 100 realizations.
For all three driver node placement strategies, the performances worsen as the damping is increased. Comparing the three panels, $ w_{\rm out}/w_{\rm in} $ performs similarly to $ \lambda_{\min}(W_r)$, and both outperform a random placement by orders of magnitude.}
\label{fig:driver_node_placement_dampedoscill}
\end{center}
\end{figure}

\paragraph{Minimum energy control of power grids.}
In the linear regime, power grids can be modeled as networks of weakly damped coupled harmonic oscillators \cite{grigsby2007power}.
The so-called swing equation corresponds in fact to the following network of damped and coupled harmonic oscillators
\beq
M \ddot q  + D\dot q + K q = B u ,
\label{eq:damped1}
\eeq
where $ D$ is the matrix of dampings which we assume to be proportional, that is, that in the modal basis $ D_1 = \Psi^T M^{-1}  D  M^{-1}\Psi $ is diagonal. 
 In the state space representation \eqref{eq:oscill_main3}, one gets then 
 \beq
 \dot x = 
\mleft[ \begin{array}{c|c}
0 & I \\
\hline
-  K M^{-1} & - D M^{-1}  
\end{array}\mright] x +
\mleft[ \begin{array}{c}
0  \\
\hline
B 
\end{array}\mright]
u ,
\label{eq:damped_main2}
\eeq
while in the modal basis
\beq
\dot z = 
\mleft[ \begin{array}{c|c}
0 & I \\
\hline
- \Omega^2  & - D_1 
\end{array}\mright] z +
\mleft[ \begin{array}{c}
0  \\
\hline
 \Psi^T M^{-1} B 
\end{array}\mright]
u .
\label{eq:damped_main3}
\eeq
For weak damping, the driver node selection strategies illustrated above can be applied to the model \eqref{eq:damped_main2} and so can the method based on $ w_{\rm out}/w_{\rm in}$. 
We have investigated the minimum energy control of several power grids listed in Table~\ref{tab:powergrids}, 
varying the dampings across several orders of magnitude, see Fig.~\ref{fig:driver_node_placement_dampedoscill} (a).
As expected, for all of them the energy required to achieve controllability increases as the real part of the eigenvalues moves away from the imaginary axis, see Fig.~\ref{fig:driver_node_placement_dampedoscill} (b) and Figs.~\ref{fig:driver_node_placement_dampedoscill2}-\ref{fig:driver_node_placement_dampedoscill_USA}. 
All strategies still beat a random driver node placement, even those based on the Gramian \eqref{eq:oscill_main5}, formally valid only for undamped dynamics.

\renewcommand{\thefigure}{S\arabic{figure}}
\renewcommand{\thetable}{S\arabic{table}}
\renewcommand{\theequation}{S\arabic{equation}}
\setcounter{figure}{0} 

\newpage

 \title{\LARGE \bf SUPPLEMENTARY MATERIAL \\ $\;$ \\}

\maketitle

\section{Methods}

\subsection{Control energy: finite-time horizon formulation}

Consider a linear system 
\beq
\dot x = A x + B u
\label{eq:system1}
\eeq
where $ x \in \mathbb{R}^n $ is the state vector, $ A \in \mathbb{R}^{n\times n } $ is the state update matrix, $ B \in \mathbb{R}^{n\times m } $ is the input matrix, and $ u $ is the $m$-dimensional input vector. 
The {\em reachable set} of \eqref{eq:system1} in time $ t_f$ from $ x_o $ is the set
\[
\mathcal{R}_{t_f}(x_o) =\{ x\in \mathbb{R}^n \;\; \text{ s. .t.  } \; \; \exists \; u\,: [0, \, t_f] \to \Omega \; \text{ s. t. }  \; \phi(t_f, u, x_o ) = x \}
\]
where $ \phi(t, u, x_o) $ is the solution of \eqref{eq:system1} at time $ t$ with input $u$ and $ \Omega$ is the admissible set of control inputs, here $ \Omega = \mathbb{R}^m$.

The system \eqref{eq:system1} is {\em reachable} (or {\em controllable from the origin} \cite{antsaklis2005linear}) in time $t_f$ if any $ x_f \in \mathbb{R}^n $ can be reached from $0$ by some control $ u \in \Omega $ in time $t_f$, i.e. if $ \mathcal{R}_{t_f}(0) = \mathbb{R}^n$. 
It is said {\em controllable to the origin} if any $ x_o \in \mathbb{R}^n $ can be brought to $0$ by some control $ u \in \Omega$ in time $t_f$. 
The system \eqref{eq:system1} is said {\em completely controllable} in time $ t_f $ if $ \mathcal{R}_{t_f} (x_o) = \mathbb{R}^n $ for any $ x_o \in \mathbb{R}^n$. 

\paragraph{Finite-time Gramians.}
The time-$t_f$ reachability (or controllability from 0) Gramian is the symmetric matrix
\beq
W_r(t_f) = \int_0^{t_f} e^{A\tau} B B^T e^{A^T \tau} d \tau ,
\label{eq:gram_reach}
\eeq
while the time-$t_f$ controllability to 0 Gramian (normally called the controllability Gramian, \cite{antsaklis2005linear}) is 
\beq
W_c(t_f) = \int_0^{t_f} e^{-A\tau} B B^T e^{-A^T \tau} d \tau .
\label{eq:gram_contr}
\eeq
The two Gramians are positive definite whenever $ (A, \, B) $ is controllable, and are related by 
\[
W_r(t_f) = e^{At_f} W_c(t_f) e^{A^T t_f}.
\]
Similarly, for their inverses,
\[
W_r^{-1} (t_f) = e^{- A^T t_f} W_c^{-1} (t_f) e^{- A t_f}.
\]

\paragraph{Finite-time control energy for state transfer.}
The transfer of the state from any $ x_o $ into any other $ x_f $ in time $t_f$ can be accomplished by many controls. 
In order to quantify how costly a state transfer is on a system, one can choose to consider the control input that minimizes the input energy, i.e., the functional 
\beq
\mathcal{E}(t_f) =\int_0^{t_f} \| u(\tau)\|^2 d \tau .
\label{eq:energy-cost}
\eeq
Such control can be computed explicitly \cite{antsaklis2005linear} as
\beq
u(t) = B^T e^{A^T(t_f-t)} W_r^{-1} (t_f) (x_f - E^{At_f} x_o ) , \qquad t \in [0, \, t_f]
\label{eq:control_gramian}
\eeq
and the corresponding transfer cost as
\beq
\mathcal{E}(t_f) = ( x_f - e^{At_f} x_o )^T W_r^{-1} (t_f) ( x_f - e^{At_f} x_o ) .
\label{eq:energy_all}
\eeq

The various metrics that have been proposed in the recent and old literature to quantify the energy needed for state transfer between any two states $ x_o $ and $ x_f $ are in fact all based on the Gramian \cite{Muller1972Analysis}:
\begin{enumerate}
\item $ \lambda_{\min} (W_r ) = \lambda_{\max} (W_r^{-1} ) $: the min eigenvalue of the Gramian (equal to the max eigenvalue of $ W_r^{-1}$) is a worst-case metric, estimating the energy required to move along the direction which is most difficult to control.
\item $ {\rm tr}(W_r ) $: the trace of the Gramian is inversely proportional to the average energy required to control a system.
\item $ {\rm tr}(W_r^{-1} )$: the trace of the inverse of the Gramian is proportional to the average energy needed to control the system.
\end{enumerate}

Minimizing the control energy means maximizing the first and second measure or minimizing the third. 
To be more precise on how the control energy is formed, we have to split the state transfer energy \eqref{eq:energy_all} into subtasks:
\begin{enumerate}
\item \label{enum:item1}
$ x_o =0 $ (reachability problem) 
\[
\Longrightarrow \quad \mathcal{E}_r(t_f) = x_f^T W_r^{-1} (t_f) x_f ;
\]
\item $ x_f =0 $ (controllability to 0 problem) 
\[
\Longrightarrow \quad \mathcal{E}_c(t_f) = x_o ^T e^{A^Tt_f}  W_r^{-1} (t_f) e^{At_f} x_o 
=  x_o ^T   W_c^{-1} (t_f)  x_o .
\]
\end{enumerate}
In particular, both $ W_r (t_f) $ and $ W_c (t_f) $ enter into the cost function. In particular,  to quantify the amount of control energy of these problems we need to compute the inverse of $ W_r(t_f)$ and $ W_c(t_f) $.

Let us look at how the stability/instability of the eigenvalues influences the two costs $ \mathcal{E}_r (t_f) $ and $ \mathcal{E}_c(t_f)$.  
\begin{itemize}
\item If $A$ is stable (i.e., $ {\rm Re}[\lambda(A)]<0$), then ``escaping from 0'' (i.e., the reachability problem) requires more energy than transferring to 0, (i.e., the controllability to 0 problem) because the modes of $A$ naturally tend to converge to 0. 
\item If $A$ is antistable (i.e., $ {\rm Re}[\lambda(A)]>0$, $ -A $ is stable), then the opposite considerations are valid: the modes of $A$ tend to amplify the magnitude of the state, simplifying the reachability problem but complicating the controllability to 0 problem.
\item If $A$ has eigenvalues with both negative and positive real part, the two situations coexist. 
\end{itemize}
Hence computing a plausible measure of control energy for a generic state transfer $ x_o \to x_f $ requires to take into account the ``difficult'' directions of both cases.


\subsection{Control energy: infinite-time horizon formulation}

When $ t_f\to \infty$, then $ \mathcal{E}(t_f) $ converges (or diverges) to a quantity 
\beq
 \mathcal{E} = \int_0^\infty \| u(\tau)\|^2 d \tau, 
\label{eq:energy_infty}
\eeq 
and so do $ \mathcal{E}_r(t_f) $ and $ \mathcal{E}_c(t_f)$.

When $ t_f \to \infty $, both Gramians become infinite-time integrals, which may be convergent or divergent, depending on the modes of $A$.
If $A$ stable, then 
\beq
W_r = \int_0^\infty e^{A\tau} B B^T e^{A^T \tau} d \tau 
\label{eq:gram_reach2}
\eeq
exists finite and it is positive definite if $ (A, \, B) $ controllable. 
If instead $A$ is antistable, then it is
\beq
W_c = \int_0^\infty e^{-A\tau} B B^T e^{-A^T \tau} d \tau 
\label{eq:gram_contr2}
\eeq
to exist finite and positive definite when $ (A, \, B) $ controllable. 
In the mixed eigenvalues cases the two expression \eqref{eq:gram_reach2} and \eqref{eq:gram_contr2} both diverge.

\paragraph{Controllability to 0 in the infinite-time horizon.}
Let us observe what happens for instance to the controllability to 0 problem according to the eigenvalues of $A$.
\begin{itemize}
\item If $A$ is stable, then in correspondence of $ u=0$, $ \lim_{t\to \infty} x(t) =0$ for all $ x_o$, meaning that the controllability to 0 problem can be solved with zero energy $ \mathcal{E}_c =0$. 
Furthermore, the integral \eqref{eq:gram_reach} converges to \eqref{eq:gram_reach2}, whose value can be computed solving the following Lyapunov equation:
\beq
A W_r + W_r A^T + B B^T =0 .
\label{eq:lyapunov_gram1}
\eeq
Such a solution always exists and it is $ W_r >0 $ (positive definite) if the pair $ (A, \, B ) $ is controllable.
\item When instead $A$ is antistable, then the integral \eqref{eq:gram_reach} diverges as $ t_f\to \infty $. 
Hence the solution with $ u=0$ is no longer feasible as all modes are unstable (and diverge as soon as $ x_o\neq 0$), meaning that to find a minimizer of \eqref{eq:energy_infty} we have to proceed in some other way.
Since $ (A, \, B) $ controllable, we can determine $ u(t) $ as if we were computing a stabilizing feedback law, i.e., expressing $ u(t) $ as a function of the state $ x(t)$ so that the resulting closed loop system converges to 0 asymptotically. 
Such a feedback law can be computed solving in $P$ an algebraic Riccati equation (ARE)
\beq
P(-A) + (-A^T) P + P B B^T P =0 .
\label{eq:are1}
\eeq
Such ARE admits a positive definite solution $P$, which can in turn be computed solving in $L$ the Lyapunov equation (in $ -A$, which is stable, hence a solution $ L>0 $ always exist)
\beq
(-A)L + L(-A^T) + B B^T =0 ,
\label{eq:lyap-L}
\eeq
and then setting $ P = L^{-1}$. 
It can be verified directly that the controllability Gramian $  W_c  $ in \eqref{eq:gram_contr2} is one such solution $L$. 
Correspondingly we obtain $ P= W_c^{-1}$. 
From the theory of linear-quadratic regulators (in particular \cite{trentelman2012control}, Ch.~10) the feedback controller 
\beq
 u = - B^T P x(t) 
 \label{eq:feedb-law1}
 \eeq
guarantees stability of the closed-loop system
\[
\dot x = (A- B B^T P ) x
\]
i.e., $ A- BB^TP $ is a stable matrix. 
The feedback law \eqref{eq:feedb-law1} also minimizes the input energy \eqref{eq:energy_infty} which is equal to $ \mathcal{E}_c = x_o^T W_c^{-1} x_o$.

\item When $A$ has eigenvalues with both positive and negative real part and no purely imaginary eigenvalues, then the two situations described above occur simultaneously. 
Assume $A$ is split into two diagonal blocks, one consisting of only eigenvalues with negative real part and the second only of eigenvalues of positive real part. 
This can always be achieved through a change of basis \cite{Zhou1999Balanced}.
Split $ B $ and $ x(t) $ accordingly:
\beq
x=\begin{bmatrix} x_1 \\ x_2 \end{bmatrix}, \quad 
A=\begin{bmatrix}  A_1 & 0 \\ 0 & A_2 \end{bmatrix} , \quad B=\begin{bmatrix} B_1 \\ B_2 \end{bmatrix} , \qquad 
\begin{array}{l} \text{Re}[\lambda(A_1)] < 0 \\ \text{Re}[\lambda(A_2)] >0 
\end{array}
\label{eq:split12}
\eeq
In the infinite time horizon, the $ u=0 $ control steers optimally the $ x_1 $ subvector, while for the $ x_2 $ part a feedback controller provides the energy-minimizing solution. 
From \eqref{eq:are1} we obtain that the ARE has solution 
\[
P= \begin{bmatrix} 0 & 0 \\ 0 & P_2 \end{bmatrix} 
\]
where $ P_2 $ solves the ARE for the $ (A_2, \, B_2 ) $ subsystem. 
Hence the control input 
\[
u= - B B^T P = - B B^T  \begin{bmatrix} 0 & 0 \\ 0 & W_{2,c}^{-1} \end{bmatrix}
\]
achieves a transfer to the origin with minimal energy cost equal to $ \mathcal{E}_c = x_o^T P x_o = x_{2,o}^T W_{2,c}^{-1} x_{2,o}$.
Furthermore, combining \eqref{eq:lyapunov_gram1} and \eqref{eq:lyap-L}, we have that for \eqref{eq:split12} the following two Lyapunov equations must hold simultaneously:
\begin{subequations}
\label{eq:2ARE}
\beqa
A_1W_{1,r} + W_{1,r}A_1^T + B_1 B_1^T & =& 0  \\
(-A_2)W_{2,c} + W_{2,c}(-A_2^T) + B_2 B_2^T & = & 0  
\eeqa
\end{subequations}

\end{itemize}

\paragraph{Reachability in the infinite-time horizon.}
Let us now consider the reachability problem (i.e., controllability from 0). 
Now the roles of stable and unstable eigenvalues are exchanged. 
\begin{itemize}
\item If $A$ stable, reachability requires an active control (here a destabilizing state feedback law) in order to steer $ x(t)$ out of the origin.
The energy-optimal solution consists in choosing $ u = - B^T P x(t) $ with $ P>0 $ solution of the ARE
\beq
PA + A^T P + PB B^T P = 0
\label{eq:are_reach}
\eeq
or, equivalently, $ P= K^{-1} $ with $ K $ solution of the Lyapunov equation
\beq
AK + K A^T + BB^T =0 .
\label{eq:lyap_reach}
\eeq
$A$ is stable, hence $ K>0 $ solving \eqref{eq:lyap_reach} and $ P>0 $ solving \eqref{eq:are_reach} always exist. 
The resulting closed loop matrix $ A- BB^T P $ must be antistable. 
From \eqref{eq:lyapunov_gram1} and \eqref{eq:lyap_reach} it can also be $ K = W_r $, the reachability Gramian. 
\item If $ A$ antistable, then $ u=0 $ is the minimal energy controller (an infinitesimal amount energy at $ t=0 $ is enough to ``kick'' the system towards the right direction $ x_f $ when initialized in $ x_o=0$; this amount of energy is negligible in the infinite time horizon considered here).
Since $ -A $ stable, a Lyapunov equation like \eqref{eq:lyap-L} holds, with solution $ L= W_c $. 

\item When $A$  has eigenvalues with both positive and negative real part and no purely imaginary eigenvalues, then a decomposition like \eqref{eq:split12} can be obtained through a change of basis. 
The complete ARE has now solution 
\[
P = \begin{bmatrix} P_1 & 0 \\ 0 & 0 \end{bmatrix} = \begin{bmatrix} W_{1,r}^{-1} & 0 \\ 0 & 0 \end{bmatrix} ,
\]
and the controller achieving the transfer with minimal energy is 
\[
u = - B B^T  \begin{bmatrix} W_{1,r}^{-1} & 0 \\ 0 & 0 \end{bmatrix} ,
\]
for an amount of energy equal to 
\[
\mathcal{E}_r = x_{1,f} ^T W_{1,r}^{-1} x_{1,f}
\]
The decomposition \eqref{eq:split12} also in this case induce a pair of Lyapunov equations identical to \eqref{eq:2ARE}.
\end{itemize}


\subsection{Mixed Gramian in infinite and finite time horizon}
In order to assemble the considerations of the previous sections, it is useful to introduce a third Gramian which we call {\em mixed Gramian}, $ W_m$ and which gathers the directions difficult to control of both the reachability and the controllability to 0 problems.

\paragraph{Infinite-time horizon mixed Gramian}
Assume that the spectrum of $A$ contains $ k$, $0\leqslant k \leqslant n $, eigenvalues with negative real part, and $ n-k $ eigenvalues with positive real part (and no purely imaginary eigenvalues). 
Then, as already mentioned above, there exist a change of basis $ V$ bringing $ A$ into the form \eqref{eq:split12}:
\beq
\begin{bmatrix} \bar A_1 & 0 \\ 0 & \bar A_2 \end{bmatrix} = V A V^{-1} 
\label{eq:basisV}
\eeq
and, correspondingly,
\[
\begin{bmatrix} \bar B_1  \\ \bar B_2 \end{bmatrix} = V B
\]
with $ {\rm Re}[\lambda(\bar A_1)]<0 $ and $ {\rm Re}[\lambda(\bar A_2 ) ]>0$. 
In the new basis, the two Lyapunov equations \eqref{eq:2ARE} hold, which can be rewritten as 
\beq
\begin{bmatrix} \bar A_1 & 0 \\ 0 & - \bar A_2 \end{bmatrix}  \bar W + \bar W 
\begin{bmatrix} \bar A_1 & 0 \\ 0 & - \bar A_2 \end{bmatrix} ^T + \begin{bmatrix} \bar B_1 \bar B_1 ^T & 0  \\ 0 & \bar B_2  \bar B_2 \end{bmatrix} ^T =0
\label{eq:2lyap-diag}
\eeq
with 
\[
\bar W_m = \begin{bmatrix} \bar W_{1,r} & 0 \\ 0 & \bar W_{2,c} \end{bmatrix} 
\]
the mixed Gramian.
Following \cite{Zhou1999Balanced}, the expression of the mixed Gramian in the original basis is 
$ W_m= V^{-1} \bar W_m V^{-T} $. 
By construction, the mixed Gramian matrix $ W_m $ always exists and summarizes the infinite-horizon contribution of the stable eigenvalues to the reachability problem and of the unstable eigenvalues to the controllability to 0 problem.

\paragraph{Finite-time horizon mixed Gramian}
Using the insight given by the previous arguments, it is possible to construct also a finite-time mixed Gramian, which weights only the modes that are difficult to control in the two state transfer problems. 
In the basis in which $A$ is split into stable and antistable diagonal blocks, \eqref{eq:basisV}, this is given by 
\[
\bar W_m (t_f) = \begin{bmatrix} \bar W_{1,r} (t_f) & 0 \\ 0 & \bar W_{2,c}(t_f) \end{bmatrix} 
\]
where $  \bar W_{1,r} (t_f) $ and $ \bar W_{2,c}(t_f) $ are the equivalent of \eqref{eq:gram_reach} and \eqref{eq:gram_contr} for the two subsystems $ (\bar A_1, \bar B_1 )$ and $ (\bar A_2, \bar B_2 ) $. 
An equation like \eqref{eq:2lyap-diag} has no coupling terms between the two subsystems (i.e., terms of the form $ \bar B_1 \bar B_2 )$. These terms disappear asymptotically, but transiently they give a contribution, hence the finite-time formulation of $ \bar W_m (t_f)  $ is only an approximation. 
In the original basis, $ W_m (t_f)= V^{-1} \bar W_m (t_f) V^{-T} $, and the input energy is  
\[
\mathcal{E}_m (t_f) = \begin{bmatrix} x_{1,f}^T & x_{2, o} ^T \end{bmatrix} W_m (t_f)^{-1} \begin{bmatrix} x_{1,f} \\ x_{2, o}  \end{bmatrix} .
\]
Clearly a proxy for this quantity is obtained by simply flipping the sign the real part of the unstable eigenvalues of $A$ and considering only the reachability problem on the resulting stable system, or shifting the eigenvalues of $A$ by adding a diagonal term (see Fig.~\ref{fig:eigenval}). 
Equivalently, all eigenvalues can be made unstable, and the controllability to 0 problem considered.


\subsection{Controllability with bounded controls}
Consider the system \eqref{eq:system1}.
Assume $ u\in \Omega$, where $ \Omega $ is a compact set of $ \mathbb{R}^m $ containing the origin in its interior. 
Assume $ (A, \, B ) $ is controllable. 
Then we have the following, see \cite{lee1986foundations,brammer1972controllability} and \cite{Sontag:1998:Mathematical}, p.~122.
\begin{itemize}
\item A necessary and sufficient condition for the origin to be steered to any point of $ \mathbb{R}^n $ in finite time (i.e., the reachability problem) is that no eigenvalue of $A$ has negative real part.
\item A necessary and sufficient condition for any point of $ \mathbb{R}^n $ to be steered to the origin in finite time (i.e., the controllability to 0 problem) is that no eigenvalue of $A$ has positive real part.
\end{itemize}
Combining the two:
\begin{itemize}
\item A necessary and sufficient condition for complete controllability (from any point $ x_o $ to any point $ x_f $) in finite time is that all eigenvalues have zero real part. 
\end{itemize}


\subsection{Control of coupled harmonic oscillators}
A network of $ n$ coupled harmonic oscillators can be written as a system of second order differential equations
\beq
 M_i \ddot q_i + (k_i + \sum_{j=1}^n k_{ij} ) q_i - \sum_{j=1}^n k_{ij} q_j = \beta_i u_i, \qquad i=1, \ldots, n ,
 \label{eq:oscill1}
 \eeq
 where $ M_i > 0$ is the mass of the $ i$-th oscillator, $ k_i \geqslant 0 $ its stiffness, $ k_{ij} \geqslant 0 $ the coupling stiffness between the $i$-th and $ j$-th oscillators, and $ \beta_i \in \{ 0, \, 1 \} $ indicates the presence or absence of a forcing term in the $i$-th oscillator. 
 In matrix form, \eqref{eq:oscill1} can be rewritten as
 \beq
 M \ddot q  + K q = B u ,
 \label{eq:oscill2}
 \eeq
 where $ M= M^T = {\rm diag} (M_i) >0$ is the mass matrix, $ K = K^T \geqslant 0 $ the stiffness matrix, and $B$ is a $ n\times m$ matrix whose columns are the elementary vectors corresponding to the $ \beta_i =1$. 
 When $ u=0 $, the solutions of \eqref{eq:oscill2} have the form $ q = \phi e^{i \omega t } $ in correspondence of the pairs $ \omega_j $ and $ \phi^j $, $ j=1, \ldots, n$, that are the solutions of the generalized eigenvalues/eigenvector equation
\beq
( - \omega^2 M + K ) \phi =0 .
\label{eq:oscill4}
\eeq
The $ \omega_j $ are called the natural frequencies of \eqref{eq:oscill2}.
Denote 
\[
 \Phi = \begin{bmatrix} \phi^1 & \ldots & \phi^n \end{bmatrix} 
 \]
the matrix of eigenvectors. 
$ \Phi $  can be used to pass to a so-called modal basis, in which the oscillators are decoupled. 
In fact, it can be verified directly that in correspondence of the change of basis $ q_1 = \Phi^{-1} q $, $ M_1 = \Phi^T M \Phi $ and $ K_1 = \Phi^T K \Phi $ are both diagonal matrices, hence 
\[
M_1 \ddot q_1 + K_1 q_1 = \Phi^T B u 
\]
has decoupled dynamics (but coupled inputs). 
 
 The state space representation of \eqref{eq:oscill2} is $ 2n $ dimensional. 
 If 
 \beq
  x = \begin{bmatrix} M & 0  \\ 0 & M  \end{bmatrix} \begin{bmatrix} q \\ \dot q \end{bmatrix},
\label{eq:oscill_basis}
\eeq
then
 \beq
 \dot x = A_o x + B_o u = 
\mleft[ \begin{array}{c|c}
0 & I \\
\hline
-  K M^{-1} & 0 
\end{array}\mright] x +
\mleft[ \begin{array}{c}
0  \\
\hline
B 
\end{array}\mright]
u .
\label{eq:oscill3}
\eeq
In terms of the state space model \eqref{eq:oscill3}, the eigenvalues are $ \lambda_j = \pm i \omega_j$, $j=1, \ldots, n$, of eigenvectors 
\[
 v_j = \mleft[ \begin{array}{c} 
\psi^j \\ \hline 
\omega_j \psi^j 
\end{array} \mright]
\]
where $ \psi^j = M \phi^j$, 
from which the purely oscillatory nature of $ A_o $ is evident. 
If we denote
\[
\Omega^2=\begin{bmatrix} \omega_1^2 & \\ & \omega_2^2 \\ & & \ddots \\ & & & \omega_n^2 \end{bmatrix} ,
\]
then, from \eqref{eq:oscill4}, $ 
M^{-1} K \Phi = \Phi \Omega $ which implies 
\[
\begin{split} \Omega^2 & = \Phi^{-1} M^{-1} K \Phi \\
& = \Phi^{-1} M^{-1}\Phi^{-T} \Phi^T   K \Phi  \\
& = M_1^{-1} K_1 .
\end{split}
\]

If $ \Psi = M \Phi$, the state space representation in the modal basis 
\[
z = T x = \mleft[ \begin{array}{c|c} 
\Psi^{-1} & 0 \\ \hline 0 & \Psi^{-1} \end{array} \mright] x
\]
is given by
\beq
\dot z = A_1 z + B_1 u = 
\mleft[ \begin{array}{c|c}
0 & I \\
\hline
- \Omega^2  & 0 
\end{array}\mright] z +
\mleft[ \begin{array}{c}
0  \\
\hline
 \Psi^{-1} B 
\end{array}\mright]
u .
\label{eq:oscill6}
\eeq
From $ M_1 = \Psi^T M^{-1} \Psi $, one gets $  \Psi^{-1} B = M_1 ^{-1} \Psi^T M^{-1} B $.

A key advantage of the modal representation is that the Gramian of the pair $ (A_1, \, B_1)$ can be computed explicitly. 
As a matter of fact, when the eigenvalues are on the imaginary axis, the integral \eqref{eq:gram_reach} (or \eqref{eq:gram_contr}) diverge, and hence the infinite-time Gramian cannot be computed. 
However, in the modal basis $ z$, $ W_{z} (t_f ) $ (or more precisely $ W_{z,r}(t_f)$) is diagonally dominant, and for $ t_f $ sufficiently long it can be approximated by its diagonal terms. 
These terms are computed explicitly in \cite{Arbel1981Controllability}:
\[
\left( W_{z} (t_f) \right)_{jj} = \begin{cases}
\frac{\left( M_1^{-1} \Psi^T M^{-1} B B^T M^{-1} \Psi M_1^{-T} \right)_{jj}t_f}{2 \omega_j} & \text{ for  } 1 \leqslant j \leqslant n \\ 
\frac{\left( M_1^{-1} \Psi^T M^{-1}  B B^T M^{-1}  \Psi M_1^{-T} \right)_{jj}t_f}{2} & \text{ for  } n+ 1 \leqslant j \leqslant 2 n .\\ 
\end{cases}
\]
If we assume that the mass matrix $ M $ is diagonal, then it is always possible to choose $ \Psi $ so that $ M_1 = I $ and $ K_1 $ diagonal, by suitably rescaling the eigenvectors $ \psi^j$. 
In this case
\[
B_1 B_1 ^T = \mleft[ \begin{array}{c} 0 \\ \hline \Psi^T M^{-1} B \end{array} \mright]
 \mleft[ \begin{array}{c|c} 0 &  B^T M^{-1}  \Psi \end{array} \mright] = \mleft[ \begin{array}{c|c} 0 & 0 \\ \hline 0 &  \Psi^TM^{-1}  B  B^T M^{-1} \Psi \end{array} \mright] ,
 \]
and the Gramian is determined by the lower diagonal block.
 When the columns of $ B$ are elementary vectors as in our case, the product $ \Psi^TM^{-1}  B  B^T M^{-1} \Psi $ can be written explicitly as sum of rank-1 matrices:
 \[
\Psi^TM^{-1}  B  B^T M^{-1} \Psi= \sum_{j=1}^n \frac{\beta_j}{M_j^2}   \begin{bmatrix} \psi^1_j \\ \vdots \\ \psi^n_j \end{bmatrix} \begin{bmatrix} \psi^1_j & \ldots & \psi^n_j \end{bmatrix}
\]
(only $m$ of the $n$ factors $ \beta_j \in \{ 0, \, 1 \}$ are nonzero) and its diagonal entries are
\beq
{\rm diag} ( \Psi^TM^{-1}  B  B^T M^{-1} \Psi )  = \sum_{j=1}^n \frac{\beta_j}{M_j^2}  \begin{bmatrix} 
 (\psi^1_j)^2 \\ & \ddots \\ & & (\psi^n_j )^2 \end{bmatrix} .
 \label{eq:oscill_diag}
 \eeq
 Hence the expression for the Gramian in the modal basis is 
 \[
 W_{z} (t_f) \approx \begin{bmatrix}  
 \sum_{j=1}^n \frac{\beta_j}{M_j^2}   \frac{(\psi^1_j )^2 }{2 \omega_1^2 } \\ 
 & \ddots \\
& &  \sum_{j=1}^n \frac{\beta_j}{M_j^2}  \frac{(\psi^n_j )^2 }{2 \omega_n^2} \\ 
& & & \sum_{j=1}^n \frac{\beta_j}{M_j^2}   \frac{(\psi^1_j )^2 }{2  } \\ 
& & & & \ddots \\
& & & & &  \sum_{j=1}^n \frac{\beta_j}{M_j^2}  \frac{(\psi^n_j )^2 }{2 } 
\end{bmatrix} t_f .
\]
Notice the linearity in $ t_f $, meaning that all components diverge to $ \infty $ with the same speed when $ t_f \to \infty$. 
This expression can be used to compute the various measures of control energy we have adopted in the paper, and hence to optimize the driver node placement problem. 
For instance, selecting inputs according to $ \lambda_{\min}(W_{z}) $ amounts to solving the following MILP max-min problem:
\[
\begin{split}
\max_{\beta_j } \min_i & \sum_{j=1}^n \frac{\beta_j}{M_j^2}  ( \psi^i_j)^2  
\\
& \begin{split} \text{ subject to  } & \sum_{j=1}^n \beta_i =m \\ 
& \; \beta_j \in \{ 0, \, 1 \}
\end{split}
\end{split}
\]
which can be solved exactly only for systems of moderate size.
However, efficient heuristics can be derived for it, such as Algorithm~\ref{alg:one}. 

\begin{algorithm}[H]
\caption{Driver node placement that maximizes $ {\rm tr} (W_{z})$.}
\begin{description}
\item{Input:}
\[
 y^i = \frac{1}{M_i^2}  \begin{bmatrix}  \frac{(\psi^1_i )^2}{2 \omega_1^2 } & \ldots &   \frac{(\psi^n_i )^2}{2 \omega_n^2 } &  \frac{(\psi^1_i )^2}{2 } & \ldots &  \frac{(\psi^n_i )^2}{2 } \end{bmatrix} , \quad i=1, \ldots, n
\]
\end{description}
\begin{enumerate}
\item Choose $ \;\; \hat i = {\rm argmax}_i ( - \| y^i \|_\infty )
= {\rm argmax}_i (\min_k y^i_k ) $
\begin{itemize}
\item $y^s = y^{\hat i} $
\item $ \mathcal{I} = \{ 1, \, 2, \ldots, n \} \setminus \{ \, \hat i \, \} $
\item $ \mathcal{O} = \{ \,\hat i \, \} $
\end{itemize}
\item For $ c =2, \, 3, \ldots, m $ 
\begin{itemize}
\item compute $ \hat j = {\rm argmax}_{j \in \mathcal{I}}  ( - \| y^s + y^j \|_\infty ) $
\item $y^s = y^s + y^{\hat j} $
\item $ \mathcal{I} = \mathcal{I}  \setminus \{ \, \hat j \, \} $
\item $ \mathcal{O} = \mathcal{O} \cup \{ \,\hat j \, \} $
\end{itemize}
  \end{enumerate}
\begin{description}
\item{Output:} $ \mathcal{O}$ 
\end{description}
\label{alg:one}
\end{algorithm}

If instead we choose to maximize the $ {\rm tr}(W_{z}) $, then we get
\[
\begin{split}
\max_{\beta_j } & \sum_{i=1}^n  \sum_{j=1}^n \frac{\beta_j}{M_j^2}  \frac{( \psi^i_j)^2}{2(1+ \omega_i^2 )}  \\
& \begin{split} \text{ subject to  } & \sum_{j=1}^n \beta_j =m \\ 
& \; \beta_j \in \{ 0, \, 1 \}
\end{split}
\end{split}
\]
Since $ {\rm tr}(W_{z}) $ is linear in the $ \beta_j$, this is a linear optimization problem, hence solvable exactly and efficiently for any $n$.
Finally, also for the minimization of $ {\rm tr}(W_{z}^{-1} ) $  
\[
\begin{split}
\min_{\beta_j } & \; {\rm tr}(W_{z}^{-1} )   \\
& \begin{split} \text{ subject to  } & \sum_{j=1}^n \beta_j =m \\ 
& \; \beta_j \in \{ 0, \, 1 \}
\end{split}
\end{split}
\]
an efficient heuristic can be set, as outlined in Algorithm~\ref{alg:two}.
\begin{algorithm}[H]
\caption{Driver node placement that minimizes $ {\rm tr} (W^{-1}_{z})$.}
\begin{description}
\item{Input:}
\[
 y^i =\frac{1}{M_i^2}   \begin{bmatrix}  \frac{(\psi^1_i )^2}{2 (1+\omega_1^2 )} & \ldots &   \frac{(\psi^n_i )^2}{2(1+ \omega_n^2) } \end{bmatrix} , \quad i=1, \ldots, n
\]
\end{description}
\begin{enumerate}
\item Choose $ \;\; \hat i = {\rm argmax}_i \sum_{k=1}^n \frac{1}{y^i_k }$
\begin{itemize}
\item $y^s = y^{\hat i} $
\item $ \mathcal{I} = \{ 1, \, 2, \ldots, n \} \setminus \{ \, \hat i \, \} $
\item $ \mathcal{O} = \{ \,\hat i \, \} $
\end{itemize}
\item For $ c =2, \, 3, \ldots, m $ 
\begin{itemize}
\item compute $ \hat j = {\rm argmax}_{j \in \mathcal{I}}  \sum_{k=1}^n  \frac{1}{y^s_k + y^j_k} $
\item $y^s = y^s + y^{\hat j} $
\item $ \mathcal{I} = \mathcal{I} \setminus \{ \, \hat j \, \} $
\item $ \mathcal{O} = \mathcal{O} \cup \{ \,\hat j \, \} $
\end{itemize}
  \end{enumerate}
\begin{description}
\item{Output:} $ \mathcal{O}$ 
\end{description}
\label{alg:two}
\end{algorithm}

Looking at an expression like \eqref{eq:oscill_diag}, it is possible to understand what kind of behavior yields good controllability properties to certain driver nodes.
For instance when measuring according to $ \lambda_{\min}(W_{z}) $, from \eqref{eq:oscill_diag}, the columns of $ \Psi^T $ express how nodes in the original basis are spread among the state variables in the modal basis. 
What is needed is an eigenbasis such that the $ j$-th component of all eigenvectors has ``support'' on all the directions of the state space, i.e., all $ \psi^1_j , \ldots, \psi^n_j $ are nonvanishing and possibly all as large as possible. 
Since $ W_{z} $ is approximated well by a diagonal matrix, the ``coverage'' effect of control inputs is additive, hence choosing a pair of controls $ i$ and $ j$ for which the sum of $ \{ \psi^k_i \}_{k=1\ldots, n} $ and $ \{ \psi^k_j \}_{k=1\ldots, n} $ has all components that are as large as possible guarantees an improvement in the control cost with respect to taking only one of the two inputs.

Notice that although $ M $ and $ K $ are both symmetric, $  K M^{-1}$ need not be, hence $ \Psi$ need not be an orthogonal matrix. It can be rendered orthogonal if a slightly different modal basis is chosen, see \cite{Gawronski1998Dynamics} for the details. 
In that case $ \Psi^T = \Psi^{-1} $ i.e., the ``coverage'' discussed here is given by the left eigenvectors of \eqref{eq:oscill4}, condition sometimes considered in the literature \cite{Hamdan1989Measures}.

Another basis for the state space that can be used in place of  \eqref{eq:oscill_basis} is given by $ \tilde x =  \begin{bmatrix} q \\ \dot q \end{bmatrix}$. 
With this choice, the state space realization is 
 \beq
 \dot{\tilde{x}}  = \tilde A_o \tilde x  + \tilde B_o u = 
\mleft[ \begin{array}{c|c}
0 & I \\
\hline
- M^{-1} K & 0 
\end{array}\mright] \tilde x +
\mleft[ \begin{array}{c}
0  \\
\hline
 M^{-1} B 
\end{array}\mright]
u .
\label{eq:oscill3_x2}
\eeq
It is straightforward to verify that in this basis the roles of $ w_{\rm in } $ and $ w_{\rm out}$ are exchanged, hence a criterion for driver node selection becomes ranking according to $ w_{\rm in}/ w_{\rm out}$ (instead of $ w_{\rm out}/ w_{\rm in}$).

\section{Datasets}
The power networks used in the paper are listed in Table~\ref{tab:powergrids}. 
All nodes are treated equally, regardless of their function as generators or loads in the real grid. 

\vspace{1cm}

\begin{table}[ht]
\scriptsize
\begin{tabular}{llll}
\hline
{\textsc{\normalsize{Network} type}}  &{\normalsize nodes} &{\normalsize edges} &{\normalsize source} \\
\hline 
\multicolumn{1}{p{5cm}}{\raggedright North EU power grid} & $236$ & $320 $ & \cite{Menck2014How} \\
\multicolumn{1}{p{5cm}}{\raggedright IEEE 300 test grid} & $300 $ & $ 409$  & https://www.ee.washington.edu/research/pstca/ \\
\multicolumn{1}{p{5cm}}{\raggedright French power grid} & $1888$ & $ 2308$ & \cite{Josz2016ACPower}  \\
\multicolumn{1}{p{5cm}}{\raggedright USA power grid} & $4941$ & $ 6591$ & \cite{Watts1998Collective} \\
\hline
\end{tabular}
\caption{Power grids used in this study.}
\label{tab:powergrids}
\end{table}




\begin{figure}[h!]
\begin{center}
\subfigure[]{
\includegraphics[angle=0, trim=1cm 4.5cm 1cm 0cm, clip=true, width=12cm]{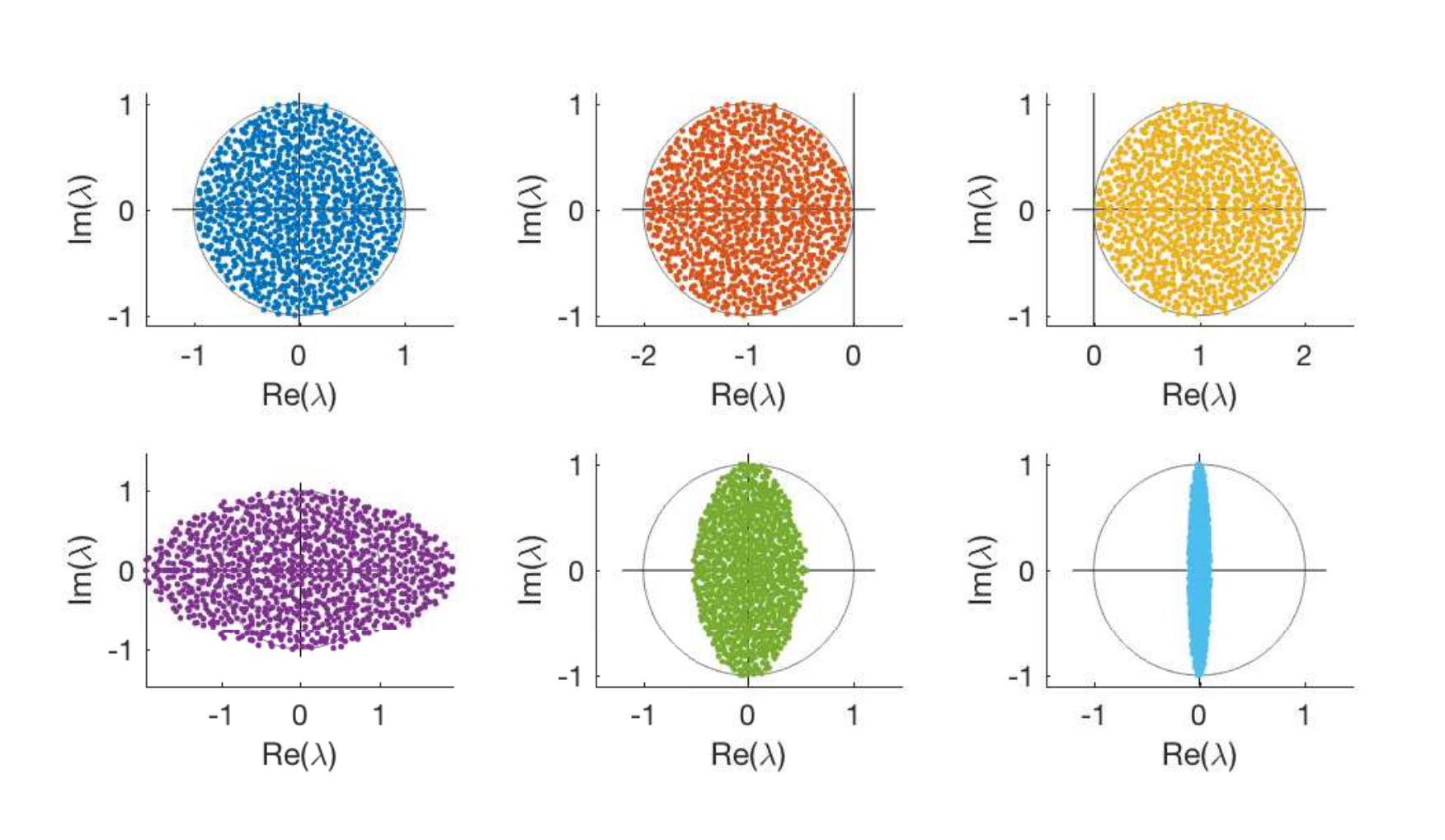}}
$\;$ 
\vspace{0.5cm}

\subfigure[]{
\includegraphics[width=4cm]{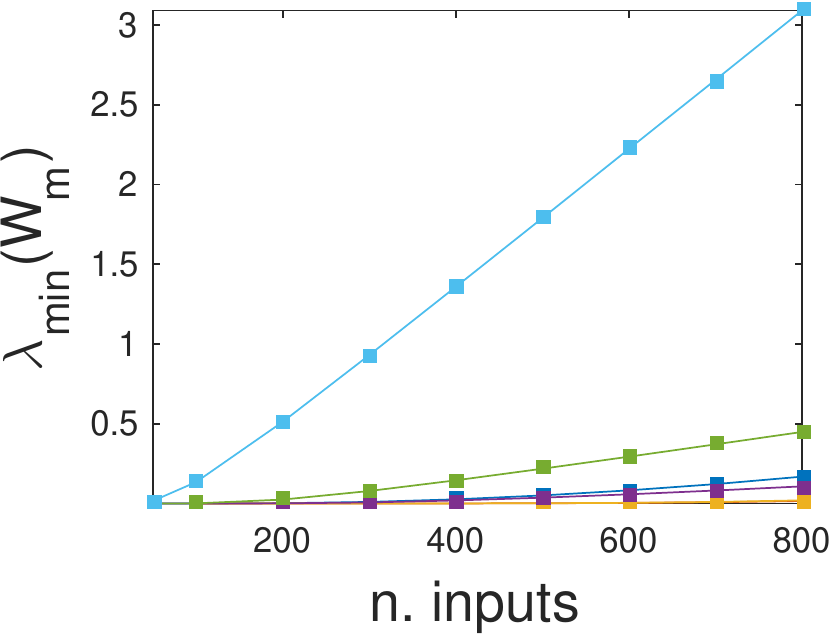}
\includegraphics[width=4cm]{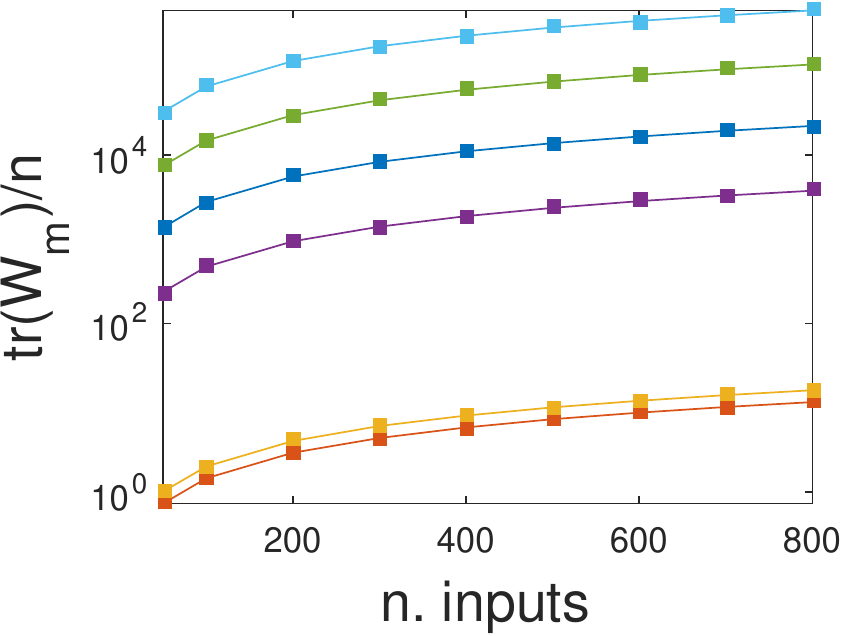}
\includegraphics[width=4cm]{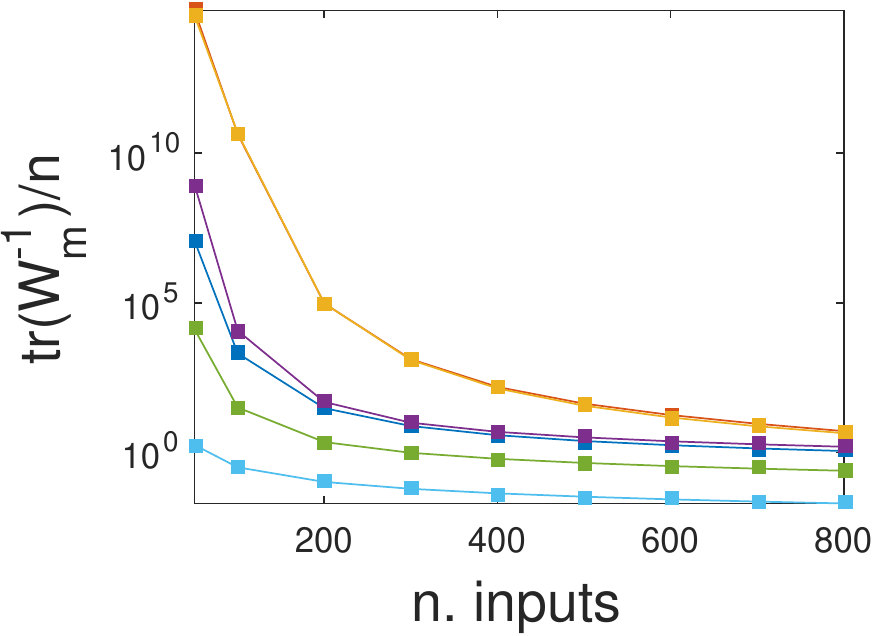}}
\caption[]{\small Analogous of Fig.~\ref{fig:eigenval}, but for networks with an ER topology with edge probability $ p=0.05$. 
(a): Six different eigenvalues locations in the complex plane for ER graphs of size $ n=1000$ and random edges weights. The circular law and the elliptic law are still valid. 
(b): Control energy for various metrics when the number of (randomly chosen) inputs grows. The data show a mean over 100 realizations of dimension $ n=1000$ (for each realization 100 different edge weights assignments are considered). The color code is as in (a). For all three metrics used to measure the control energy ($\lambda_{\min}(W_m) $, $ {\rm tr}(W_m) $ which should both be maximized, and $ {\rm tr}(W_m^{-1} )$ which should be minimized), the performances are strictly a function of the position of the eigenvalues of $A$. The minimum of the control energy is achieved when the eigenvalues have very small real part (cyan) and worsen with growing real part, following the order: cyan, green, blue, violet.}
\label{fig:metrics_sparse}
\end{center}
\end{figure}

\begin{figure}[h!]
\begin{center}\includegraphics[width=12cm]{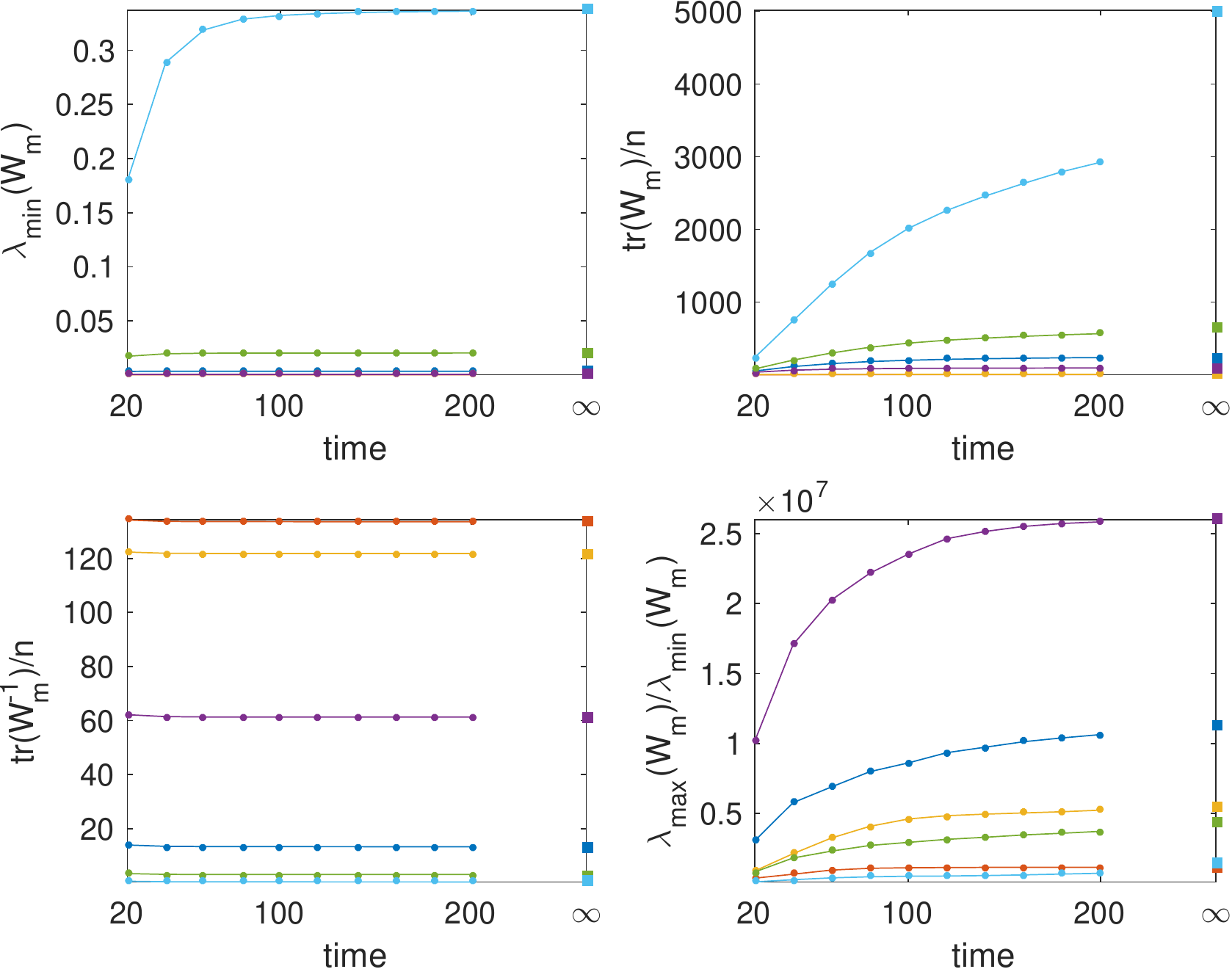}
\caption[]{\small Computing control energies in finite time and infinite time. For the various measures of control energy considered in the paper ($\lambda_{\min}(W_m) $, $ {\rm tr}(W_m) $ and $ {\rm tr}(W_m^{-1} )$), the plots show the profile in time when computations are performed using $ W_m(t_f) $, for various values of $ t_f$. The value for $ t_f =\infty $ is also shown for comparison. 
For this specific example (full network of size $ n=1000$ and $ m=400$ controls), some measures converge much faster than others.
For instance $ {\rm tr}(W_m^{-1} )$ achieves its infinite-time value extremely quickly, while $ {\rm tr}(W_m) $ converges very slow.
Also the condition number of $ W_m $ (i.e., $\lambda_{\max} (W_m)/\lambda_{\min} (W_m) $, lower right panel) converges to its asymptotic value.}
\label{fig:metrics_time}
\end{center}
\end{figure}

\begin{figure}[h]
\begin{center}
\subfigure[]{
\includegraphics[width=10cm]{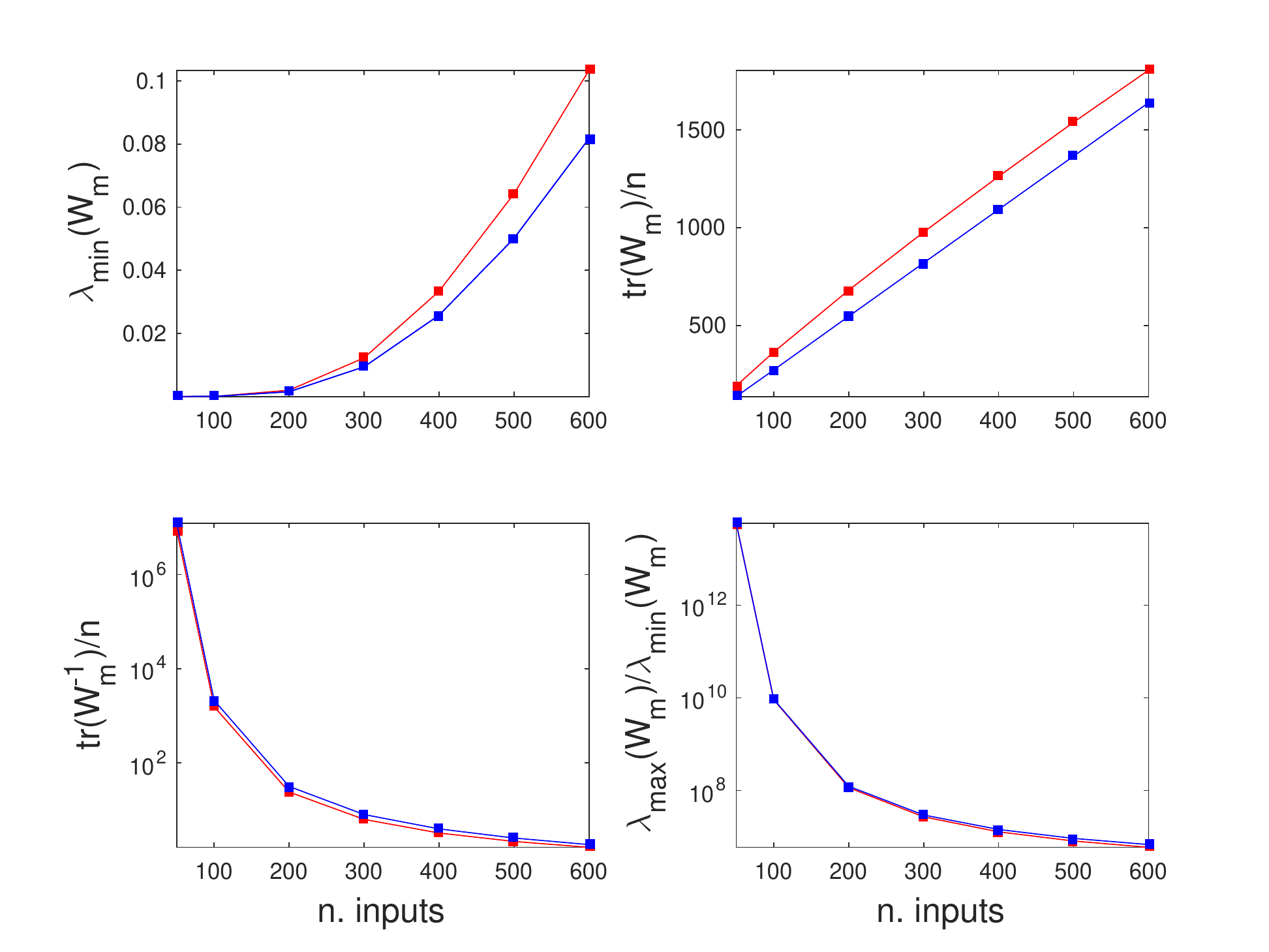}}
\\
$\;$ 
\vspace{0.3cm}

\subfigure[]{
\includegraphics[width=9cm]{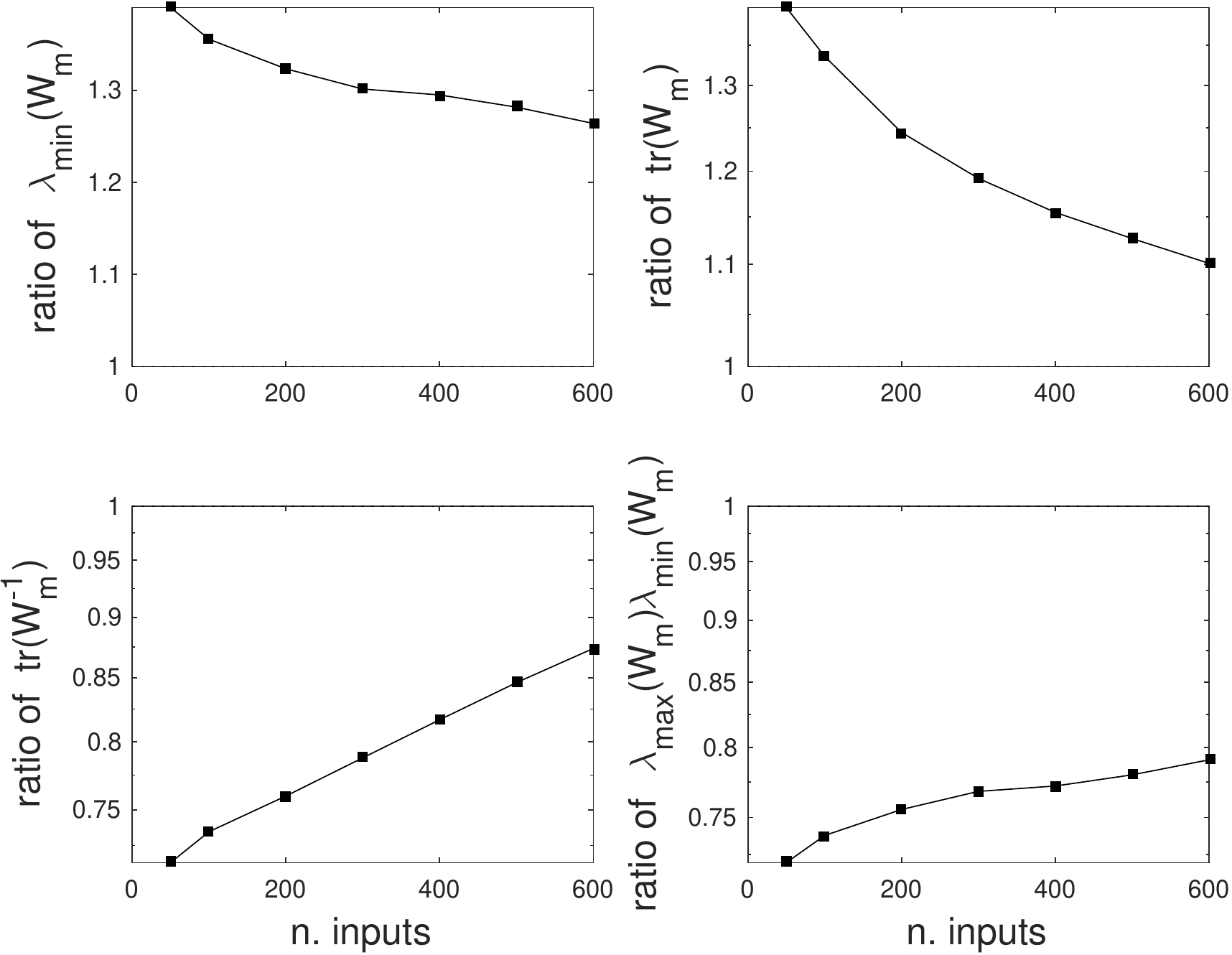}}
\caption[]{\small Driver node placement for ER networks with $ p=0.05$. 
(a): Comparison between the value of the various measures of control energy obtained for driver node placement strategies based on $ r_w = w_{\rm out}/w_{\rm in}$ (red) and the same measure for random driver node assignments (blue). 
As can be seen on the ratios shown in (b), all measures improve, especially when $ m$ is low.
Notice that also the condition number of $ W_m $ (i.e., $\lambda_{\max} (W_m)/\lambda_{\min} (W_m) $, lower right panel) improves.}
\label{fig:driver_node_placement2}
\end{center}
\end{figure}

\begin{figure}[h]
\begin{center}
\subfigure[]{
\includegraphics[width=12cm]{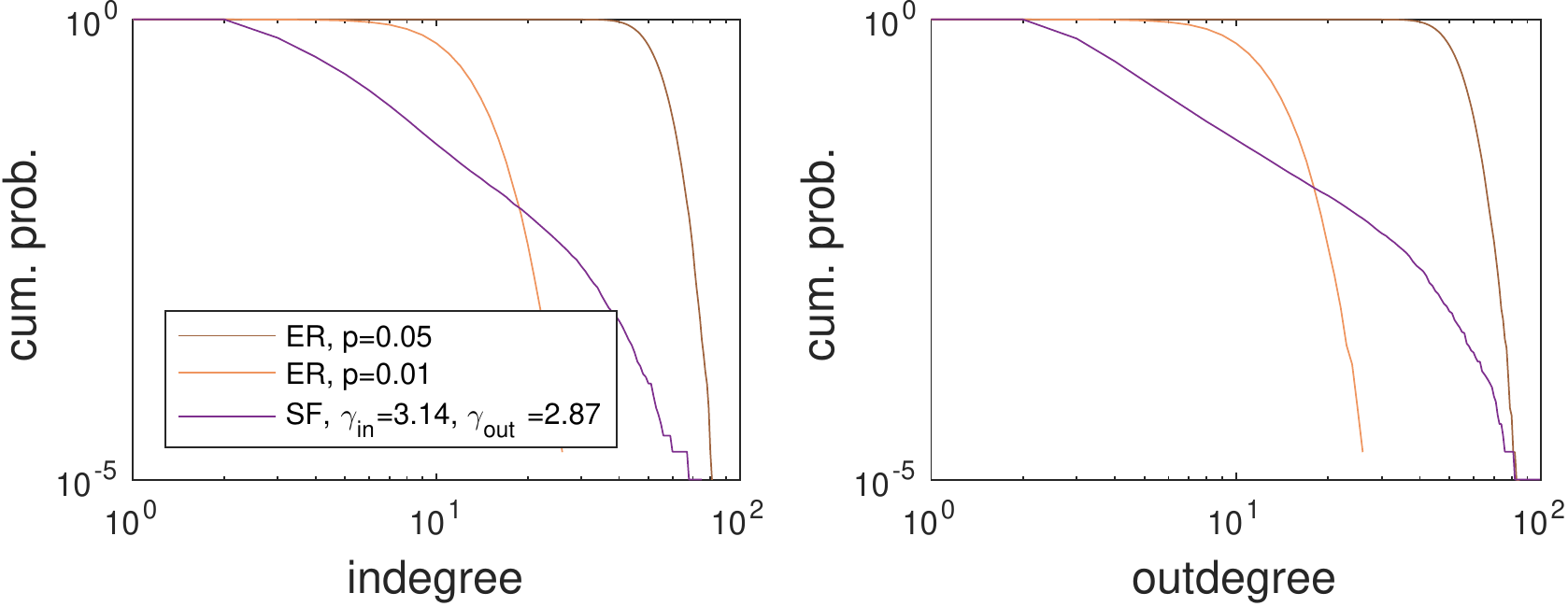}} 
\\
$\;$ 
\vspace{0.5cm}

\subfigure[]{
\includegraphics[width=10cm]{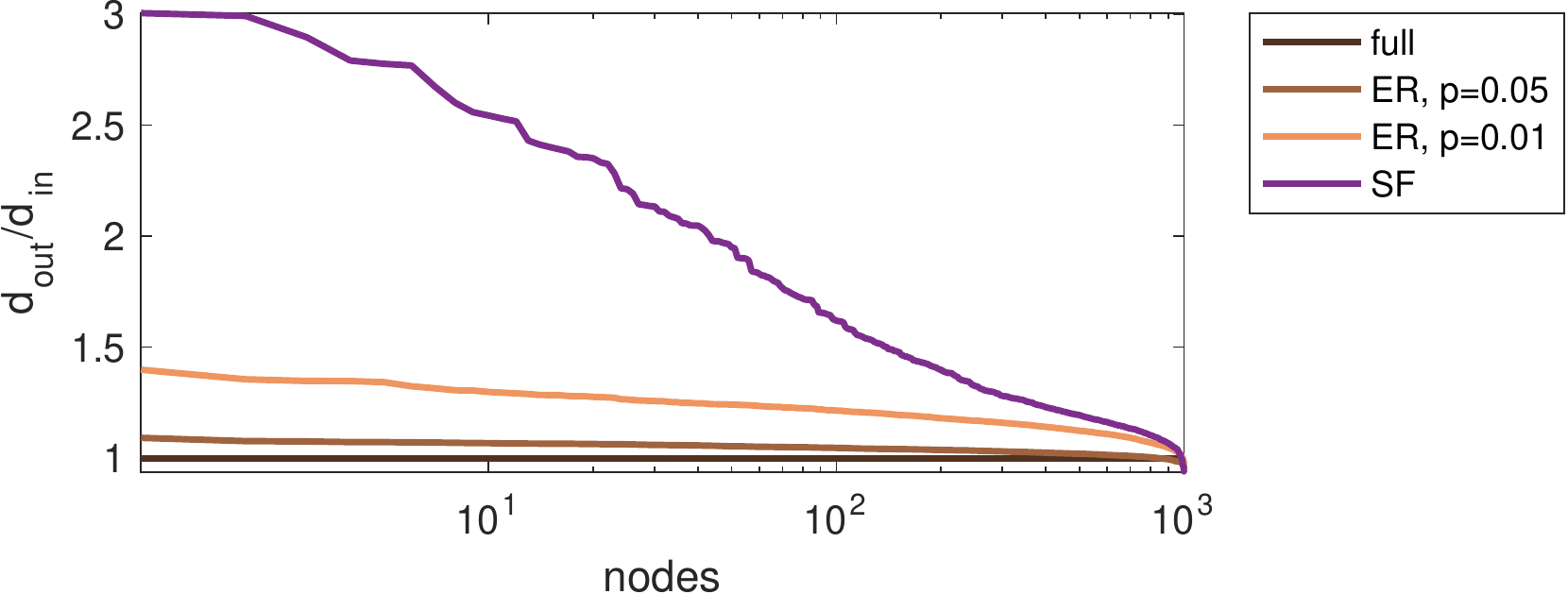}}
\caption[]{\small Degree distributions of the various networks used in this study. 
(a): Average indegree and outdegree of the ER networks (with $ p=0.05 $ and $ p=0.01 $) and SF networks.
The SF networks are generated using the algorithm of \cite{Bollobas:2003:Directed}, with indegree exponent $ \gamma_{\rm in} =3.14 $ and outdegree exponent $ \gamma_{\rm out} =2.87 $. 
To avoid problems with controllability, extra edges are added randomly among different strongly connected components, until strong connectivity is achieved on the entire graph. 
The difference in the in/out exponent is still clearly visible.
(b): Histogram of outdegree/indegree ratio for the various networks. The SF networks reflect our choice of $  \gamma_{\rm out} <  \gamma_{\rm in} $. }
\label{fig:driver_node_placement4}
\end{center}
\end{figure}

\begin{figure}[h]
\begin{center}
\subfigure[]{
\includegraphics[width=9.5cm]{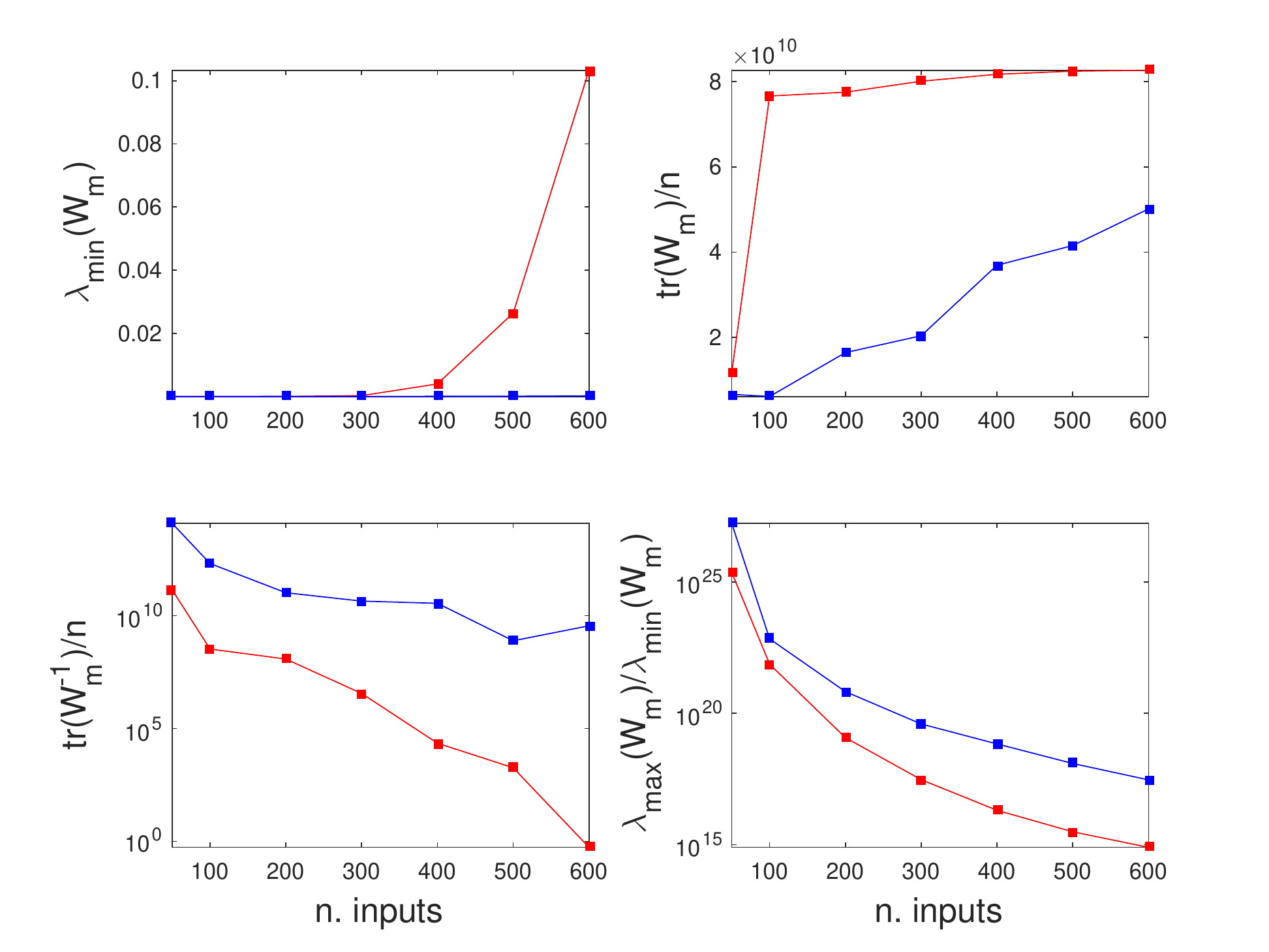}} 
\\
$\;$ 
\vspace{0.5cm}

\subfigure[]{
\includegraphics[width=8.5cm]{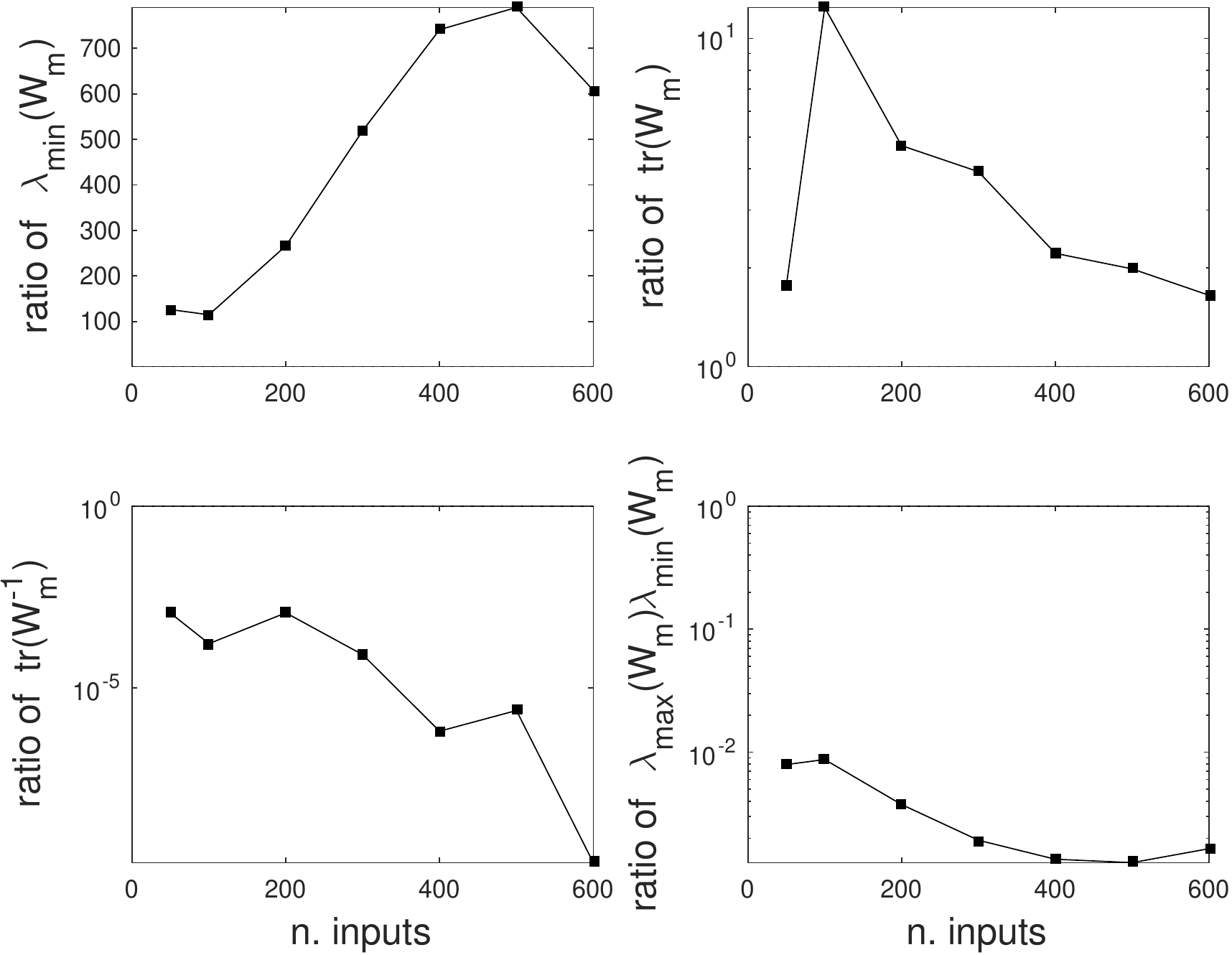}}
\caption[]{\small Driver node placement for SF networks with $ \gamma_{\rm in} =3.14 $ and $ \gamma_{\rm out} =2.87 $.
(a): Comparison between the value of the various measures of control energy obtained for driver node placement strategies based on $ r_w = w_{\rm out}/w_{\rm in}$ (red) and the same measure for random driver node assignments (blue). 
As the ratios in (b) show, the improvement in all measures is normally of several orders of magnitude. 
Also the condition number of $ W_m $ (i.e., $\lambda_{\max} (W_m)/\lambda_{\min} (W_m) $) improves substantially.
}
\label{fig:driver_node_placement2_SF}
\end{center}
\end{figure}

%
%

\begin{figure}[h]
\begin{center}
\subfigure[]{
\includegraphics[width=4cm]{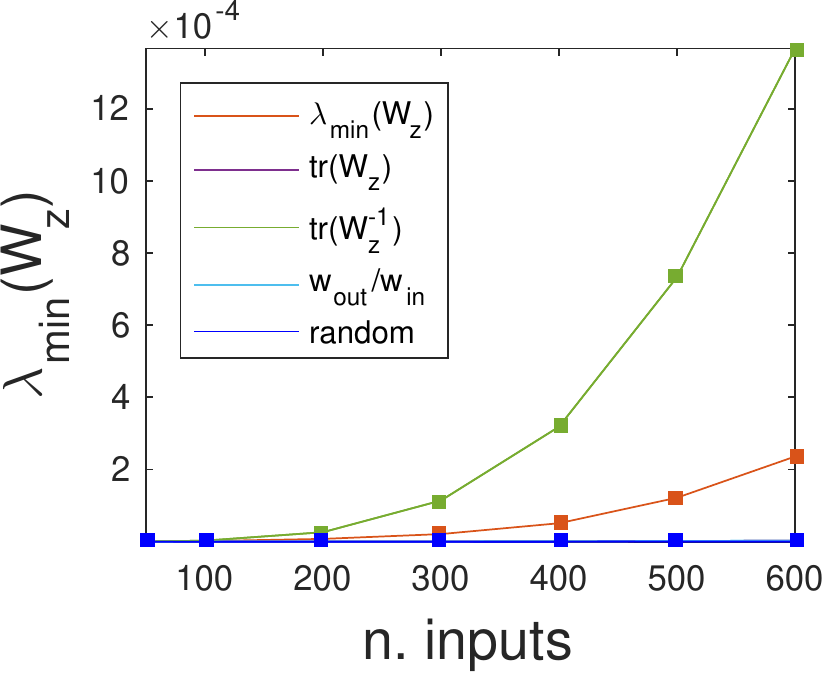}$\;\;\;$
\includegraphics[width=4cm]{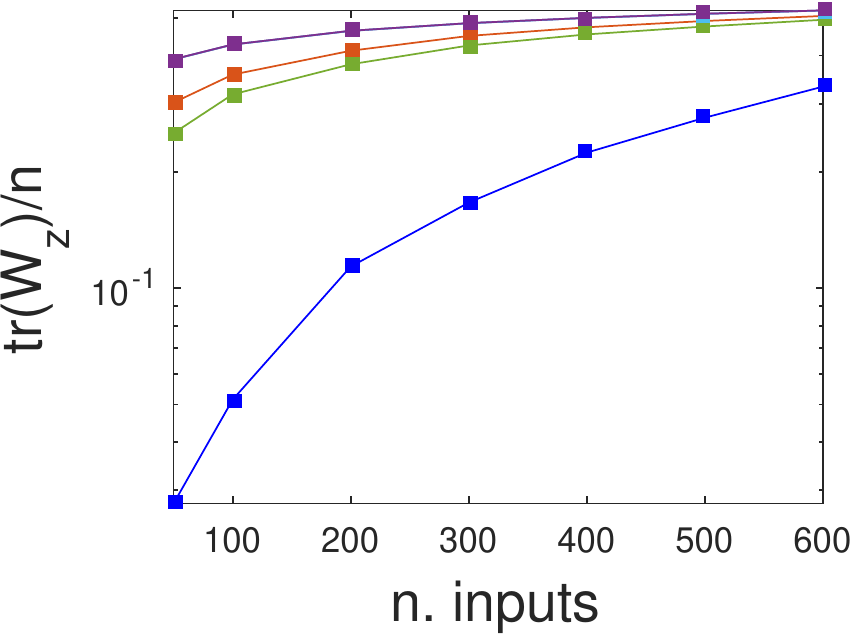}$\;\;\;$
\includegraphics[width=4cm]{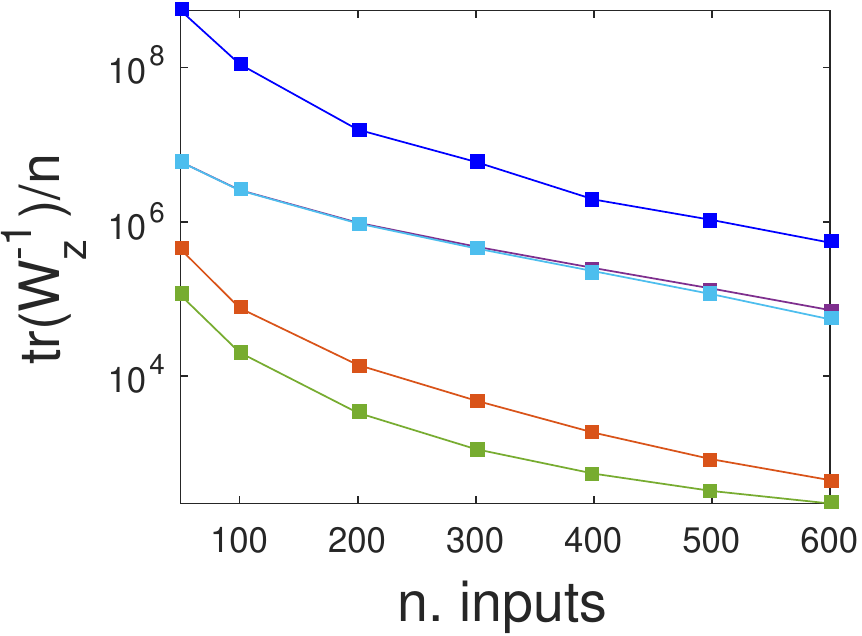}} 
\subfigure[]{
\includegraphics[width=9cm]{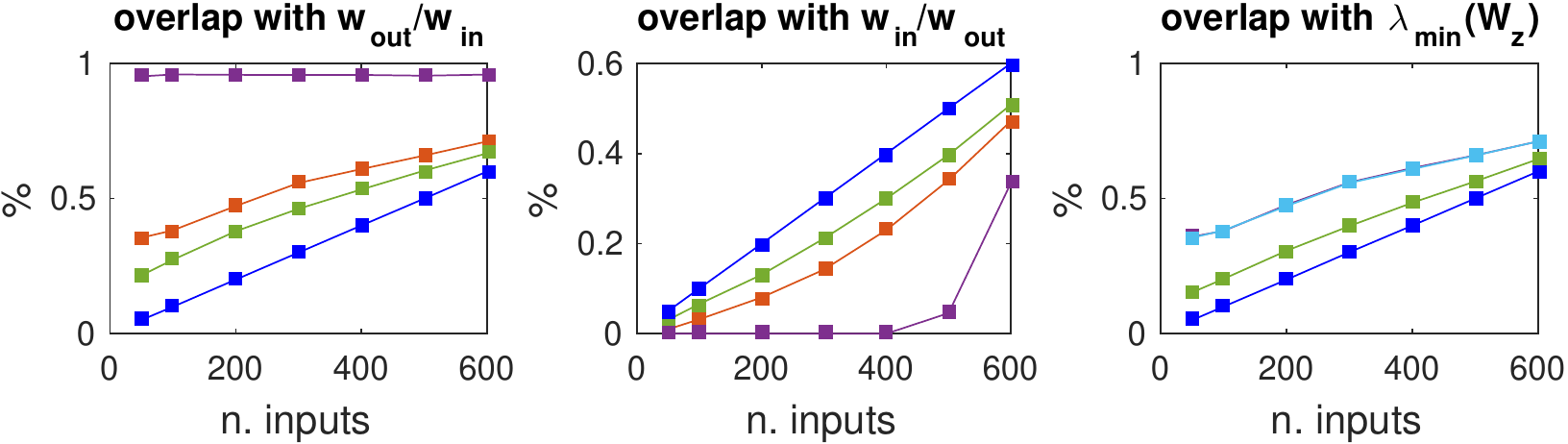}}
\caption[]{\small Driver node placement for a network of $ n=1000$ coupled harmonic oscillators. The figure is the analogous of Fig.~\ref{fig:driver_node_placement_oscill}, but now the coupling matrix $K$ is fully connected. (a): Shown are means over 50 realizations (with 100 edge weight samples taken for each realization). Red: driver node placement based on $ \lambda_{\min}(W_{z})$.  Violet: placement based on $ {\rm tr}(W_{z})$. Green: placement based on $ {\rm tr}(W^{-1}_{z})$. Cyan: placement based on $ w_{\rm out}/w_{\rm in} $. Blue: random input assignment. 
All driver node placement strategies still beat a random assignment, but with worse performances with respect to Fig.~\ref{fig:driver_node_placement_oscill}. 
Of the four measures,  $ \lambda_{\min}(W_{z})$ and $ {\rm tr}(W^{-1}_{z})$ tend to behave similarly and so do $ w_{\rm out}/w_{\rm in} $ and $ {\rm tr}(W_z)$ (in the mid plot they completely overlap, and both give the true optimum). (b): Overlap in the node ranking of the different driver node placement strategies. Color code is the same as in (a). The only highly significant overlap is still between $ w_{\rm out}/w_{\rm in} $ and $ {\rm tr}(W_z)$ ($ > 90\%$), while $ \lambda_{\min}(W_{z})$ and $ {\rm tr}(W^{-1}_{z})$ correspond to different node ranking patterns. None of the strategies orders nodes according to $ w_{\rm in}/w_{\rm out} $, as expected.}
\label{fig:driver_node_placement_oscill_full}
\end{center}
\end{figure}

\begin{figure}[h]
\begin{center}
\subfigure[]{
\includegraphics[width=4cm]{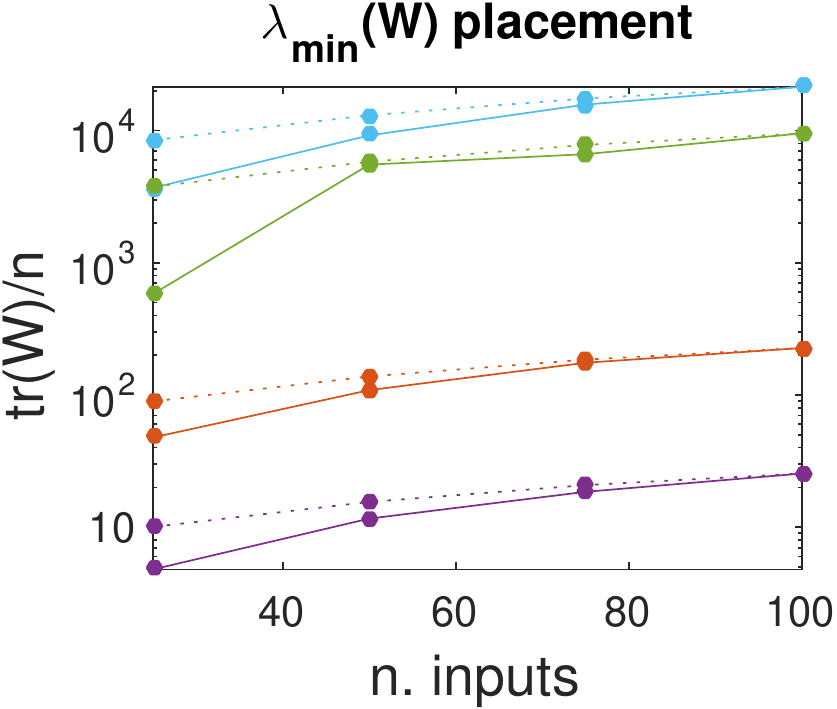}$\;\;\;$
\includegraphics[width=4cm]{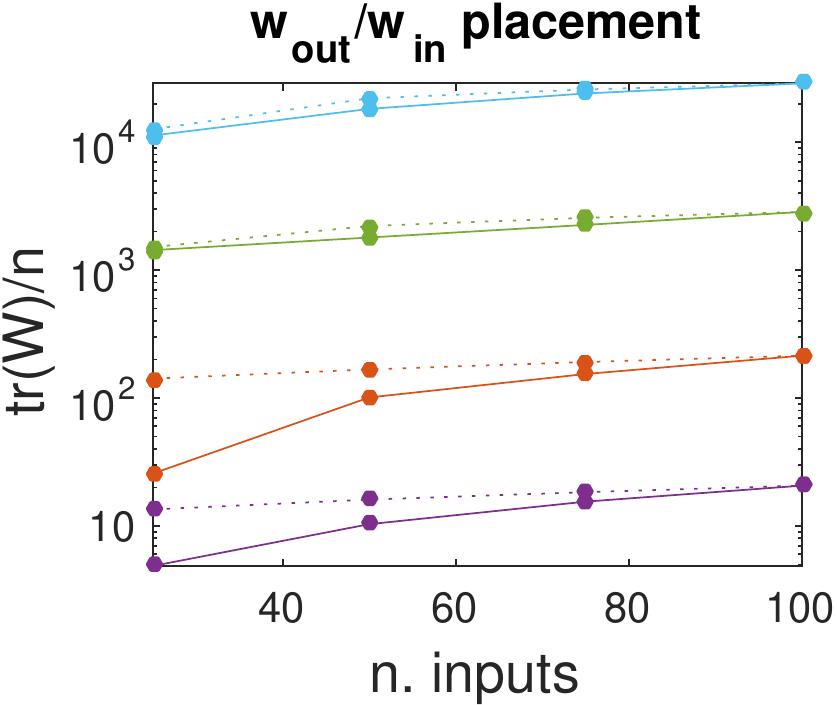}$\;\;\;$
\includegraphics[width=4cm]{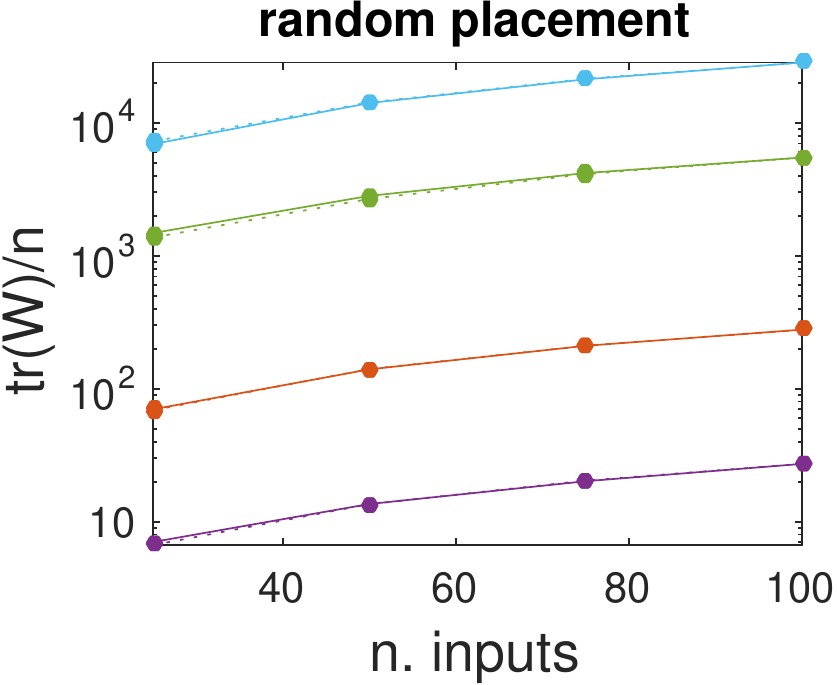}
} 
\subfigure[]{
\includegraphics[width=4cm]{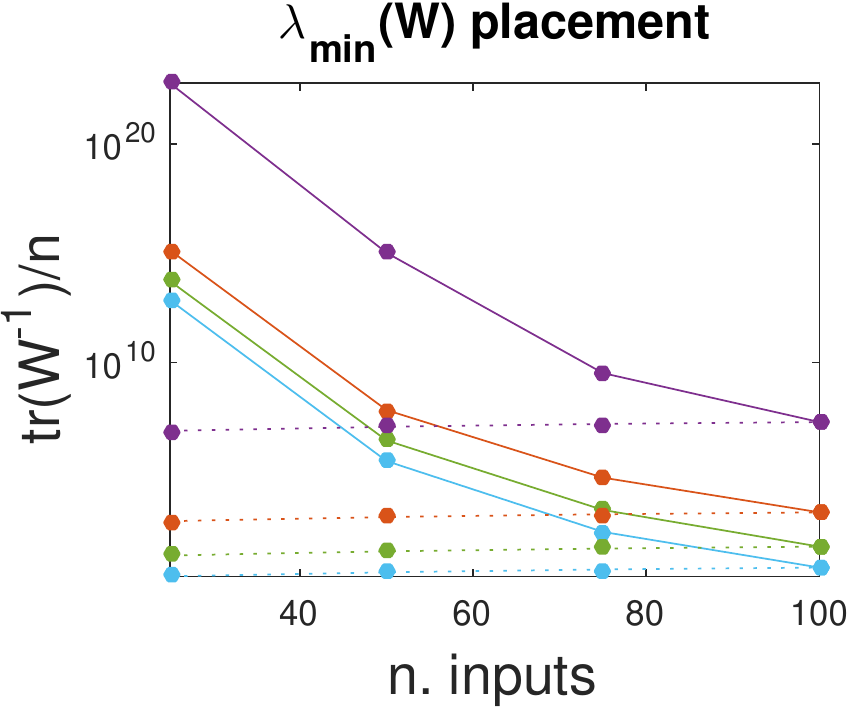}$\;\;\;$
\includegraphics[width=4cm]{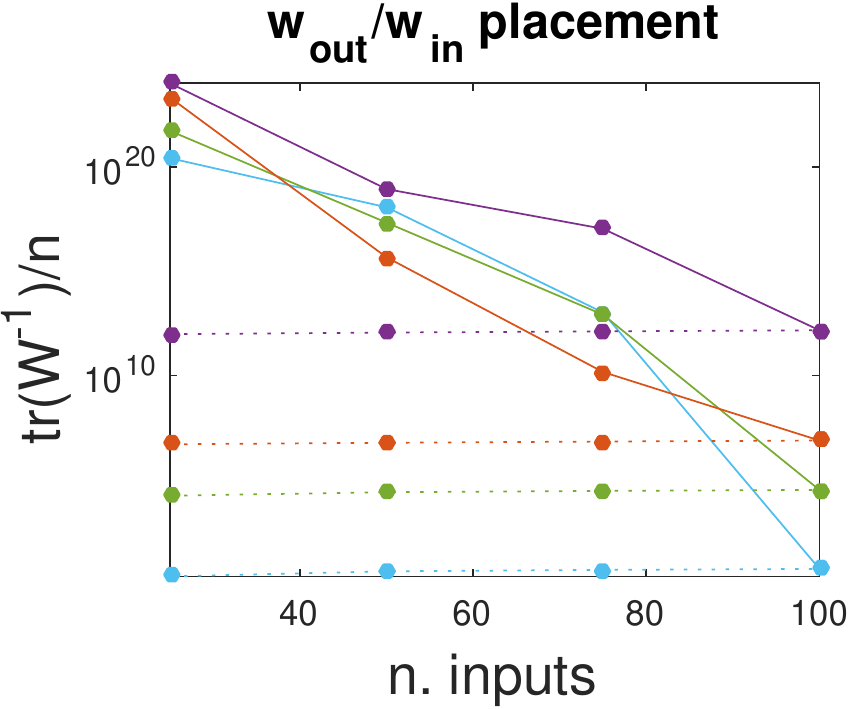}$\;\;\;$
\includegraphics[width=4cm]{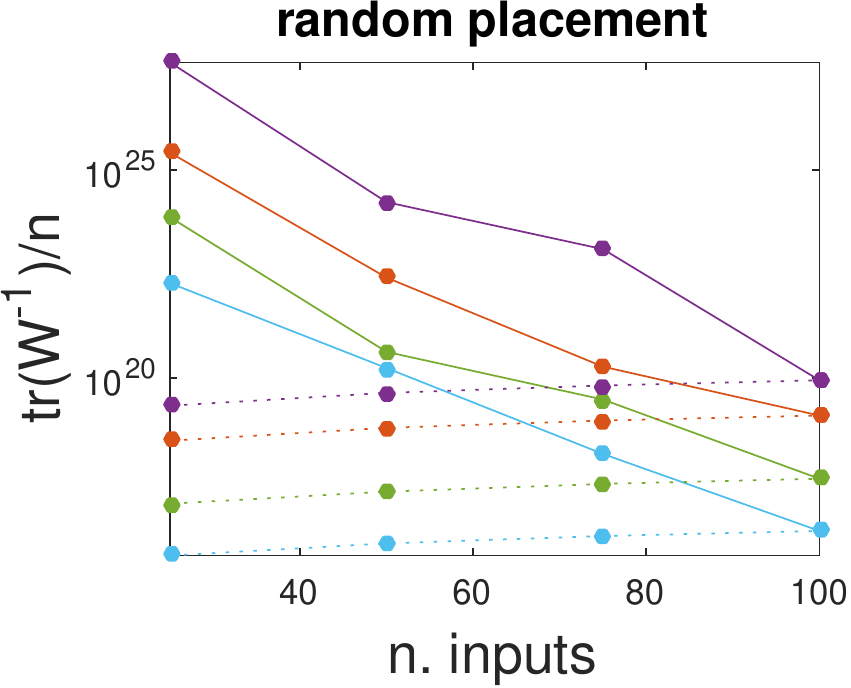}
} 

\caption[]{\small Minimum energy control of power grids with varying damping coefficients. North EU power grid.
This Figure complements Fig.~\ref{fig:driver_node_placement_dampedoscill}(b) of the paper. 
(a): Control energy for the metric $ {\rm tr}(W_r) $ when the driver nodes are placed according to $ \lambda_{\min} (W_r) $ (left panel), $ w_{\rm out}/w_{\rm in} $ (mid panel), or randomly (right panel).    
The color code is a function of the damping coefficients, with the same convention as in Fig.~\ref{fig:driver_node_placement_dampedoscill}(a) of the paper. 
The values of $ {\rm tr}(W_r ) $ are shown in solid lines, while in dotted lines the values of $  {\rm tr}(W_z ) $ are shown (suitably normalized to eliminate the explicit dependence from $ t_f$).
Values are averages over 100 realizations.
(b): Control energy for the metric $ {\rm tr}(W_r^{-1}) $, with the same conventions as in (a).
}
\label{fig:driver_node_placement_dampedoscill2}
\end{center}
\end{figure}

\begin{figure}[h]
\begin{center}
\subfigure[]{
\includegraphics[angle=0, trim=0cm 0cm 0cm 0cm, clip=true, width=4cm]{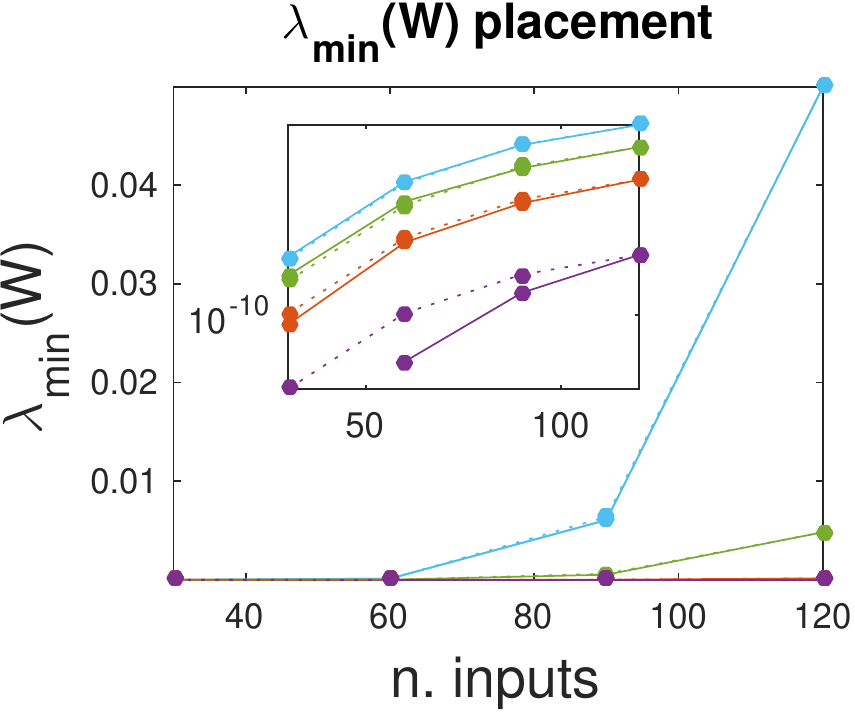}
\includegraphics[angle=0, trim=0cm 0cm 0cm 0cm, clip=true, width=4cm]{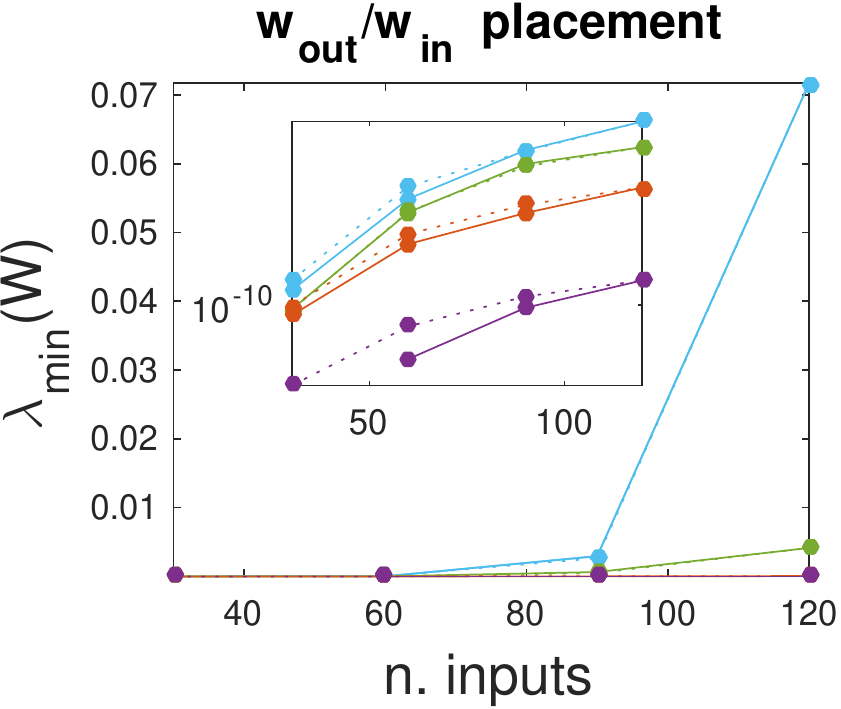}
\includegraphics[angle=0, trim=0cm 0cm 0cm 0cm, clip=true, width=4cm]{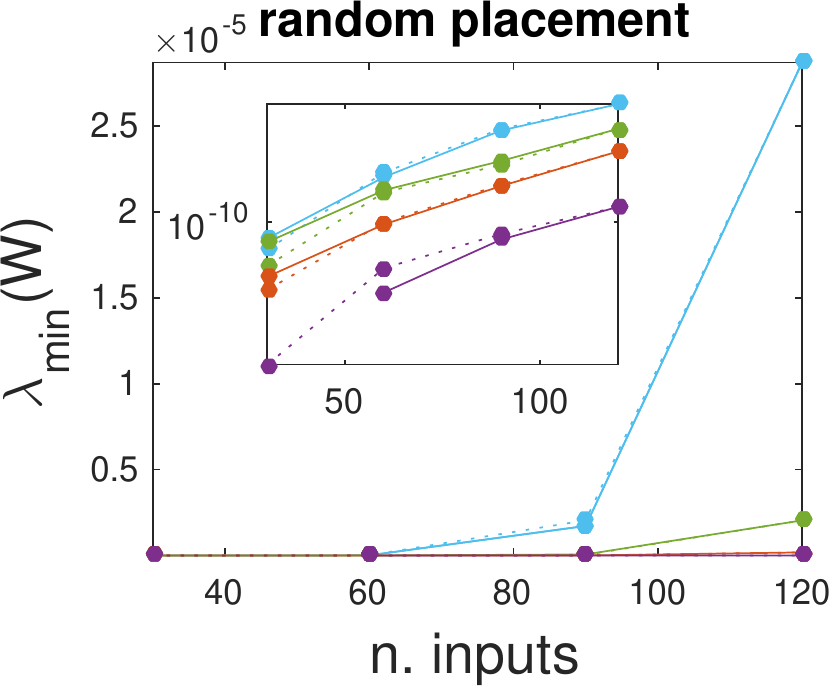}}
\subfigure[]{
\includegraphics[width=4cm]{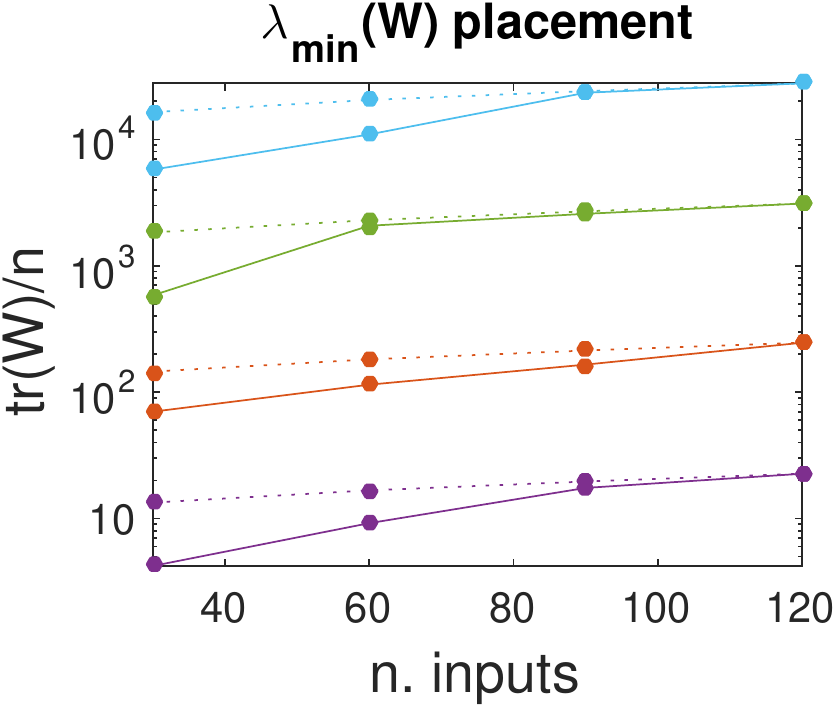}$\;\;\;$
\includegraphics[width=4cm]{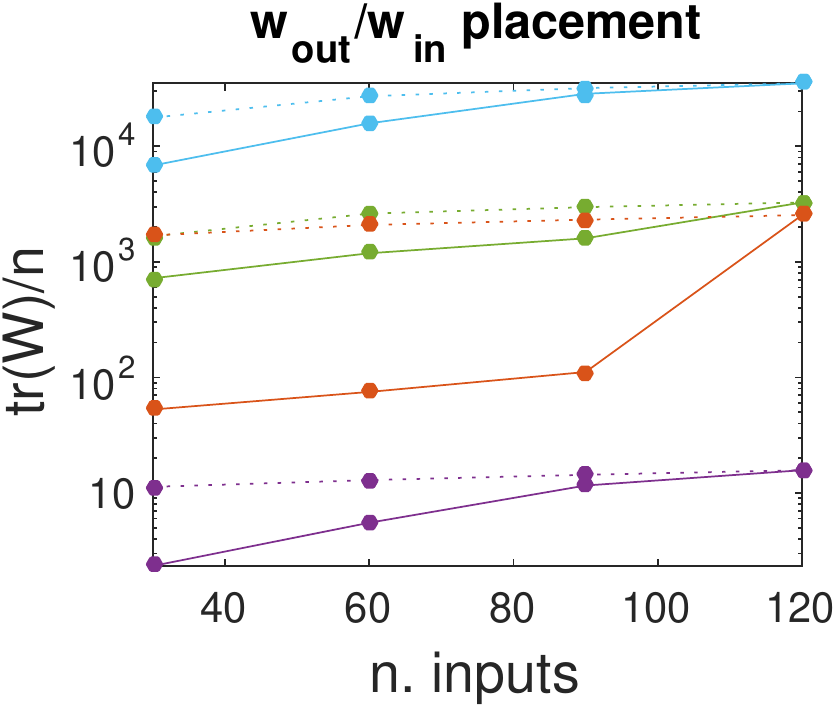}$\;\;\;$
\includegraphics[width=4cm]{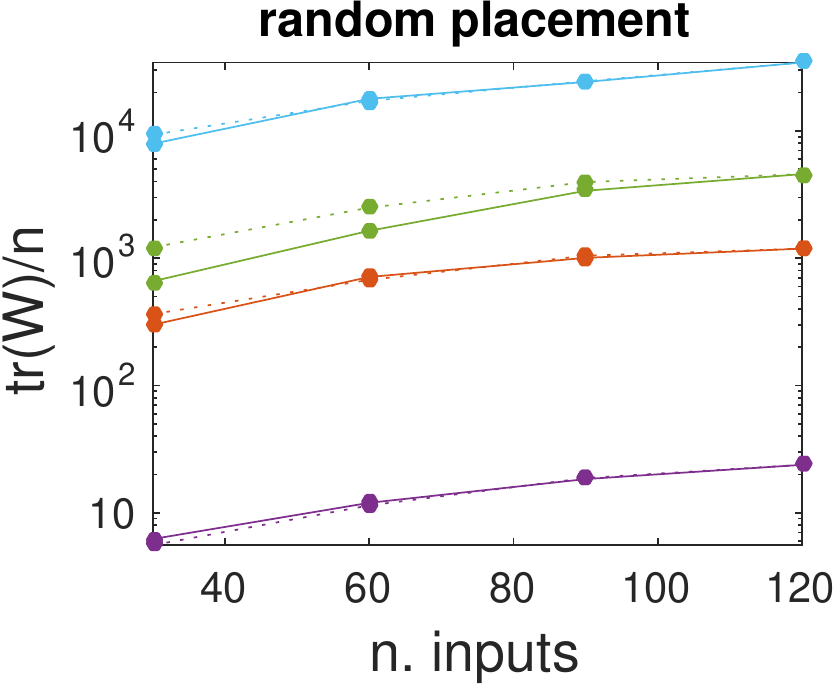}
} 
\subfigure[]{
\includegraphics[width=4cm]{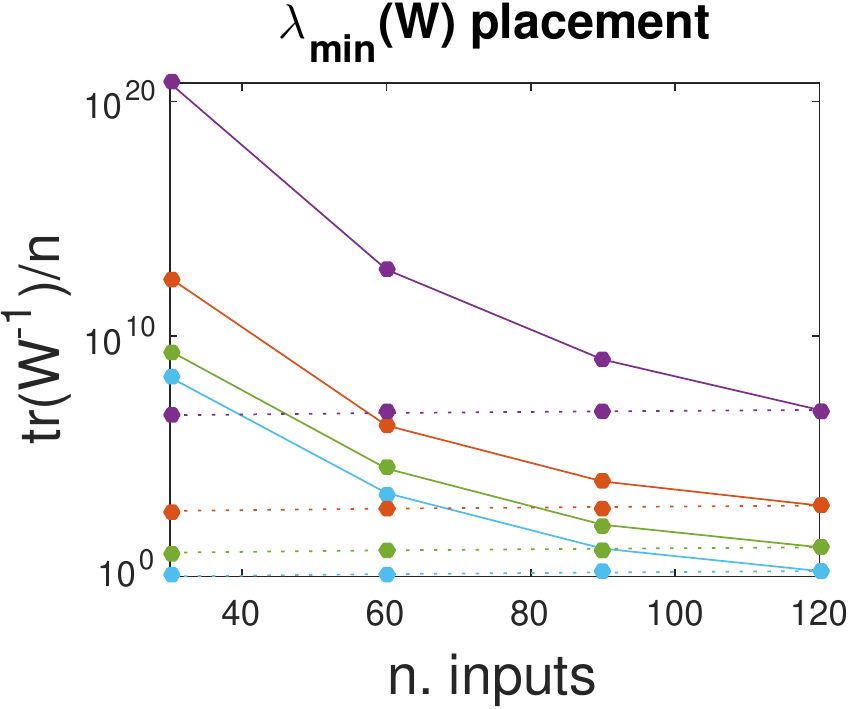} $\;\;$
\includegraphics[width=4cm]{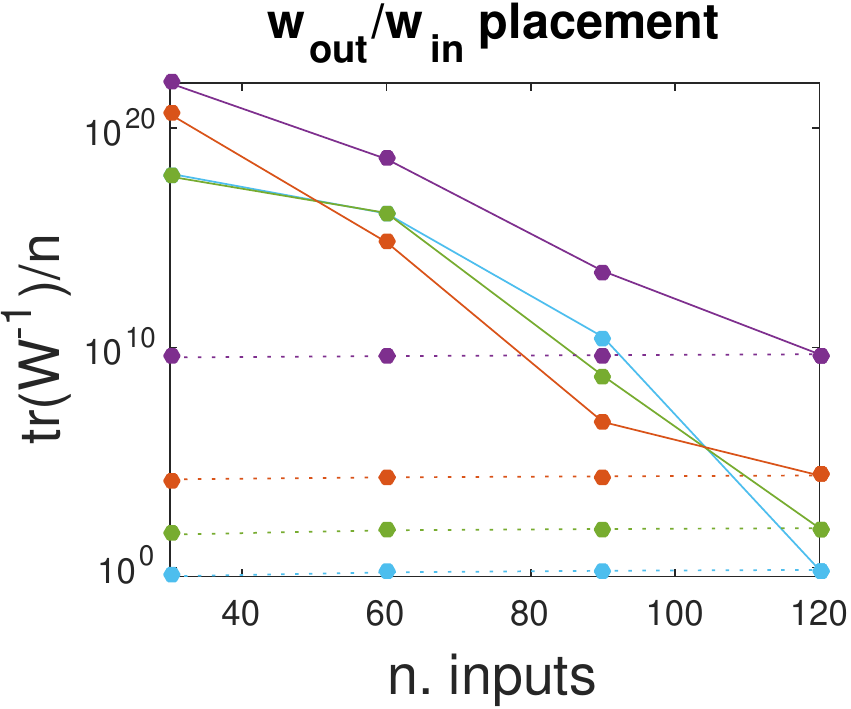}$\;\;$
\includegraphics[width=4cm]{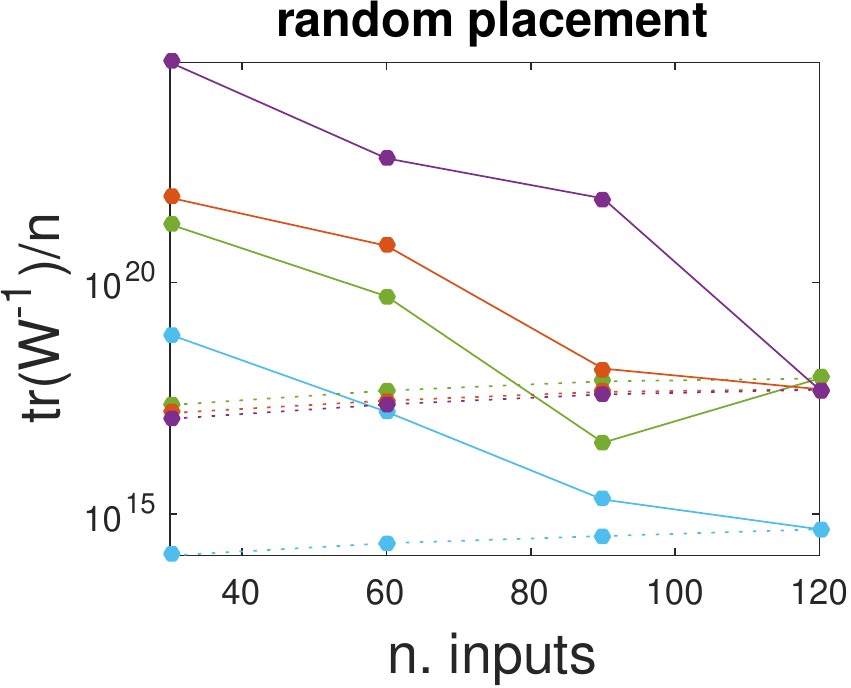}
} 

\caption[]{\small Minimum energy control of power grids with varying damping coefficients. IEEE 300 bus test power network.
(a): Control energy for the metric $ \lambda_{\min} (W_r ) $ when the driver nodes are placed according to $ \lambda_{\min} (W_r) $ (left panel), $ w_{\rm out}/w_{\rm in} $ (mid panel), or randomly (right panel). 
The color code is a function of the damping coefficients, using the same convention as in Fig.~\ref{fig:driver_node_placement_dampedoscill}(a) of the paper. 
The values of $ \lambda_{\min} (W_r ) $ are shown in solid lines, while in dotted lines the values of $  \lambda_{\min} (W_z ) $ are shown (suitably normalized to eliminate the explicit dependence from $ t_f$).
Values are averages over 100 realizations.
Data are missing when the Gramian $ W_r $ is too close to singular in too many trials.
(b): Control energy for the metric $ {\rm tr}(W_r) $ when the driver nodes are placed according to $ \lambda_{\min} (W_r) $ (left panel), $ w_{\rm out}/w_{\rm in} $ (mid panel), or randomly (right panel).    
The values of $ {\rm tr}(W_r ) $ are shown in solid lines, while in dotted lines the values of $  {\rm tr}(W_z ) $ are shown.
(c): Control energy for the metric $ {\rm tr}(W_r^{-1}) $, with the same conventions as in (a) and (b).
}
\label{fig:driver_node_placement_dampedoscill_ieee300}
\end{center}
\end{figure}

\begin{figure}[h]
\begin{center}
\subfigure[]{
\includegraphics[angle=0, trim=0cm 0cm 0cm 0cm, clip=true, width=4cm]{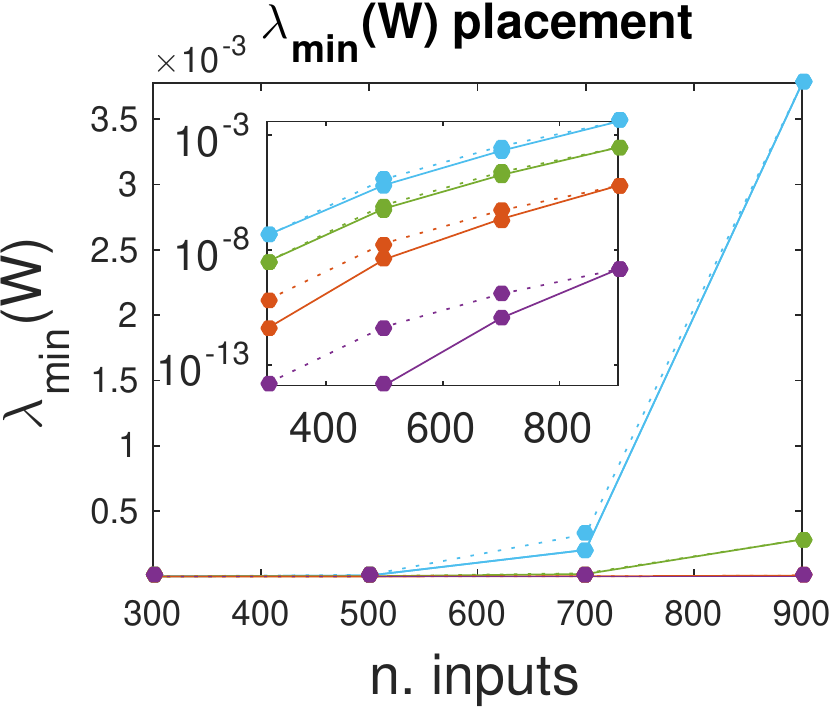}
\includegraphics[angle=0, trim=0cm 0cm 0cm 0cm, clip=true, width=4cm]{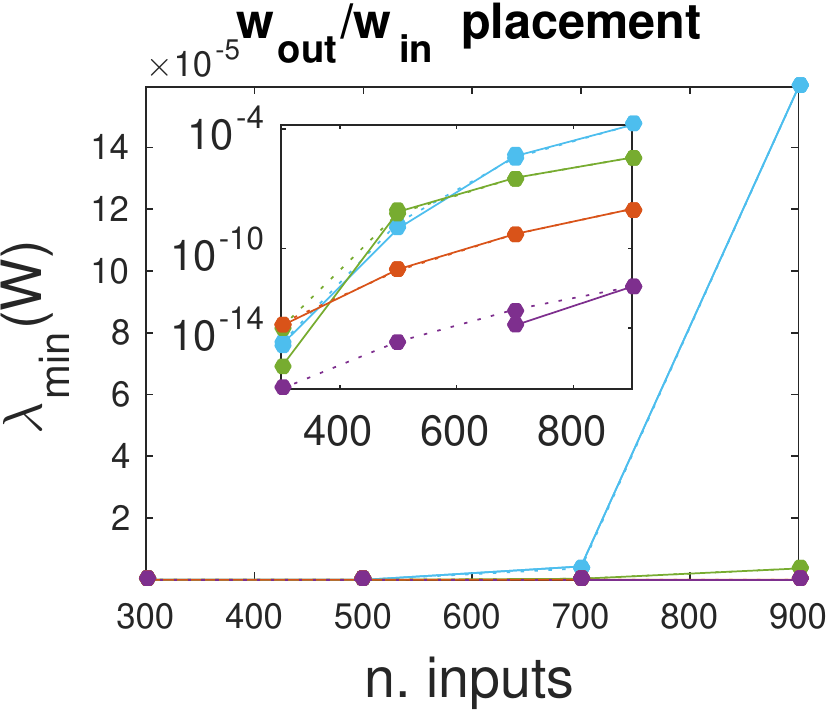}
\includegraphics[angle=0, trim=0cm 0cm 0cm 0cm, clip=true, width=4cm]{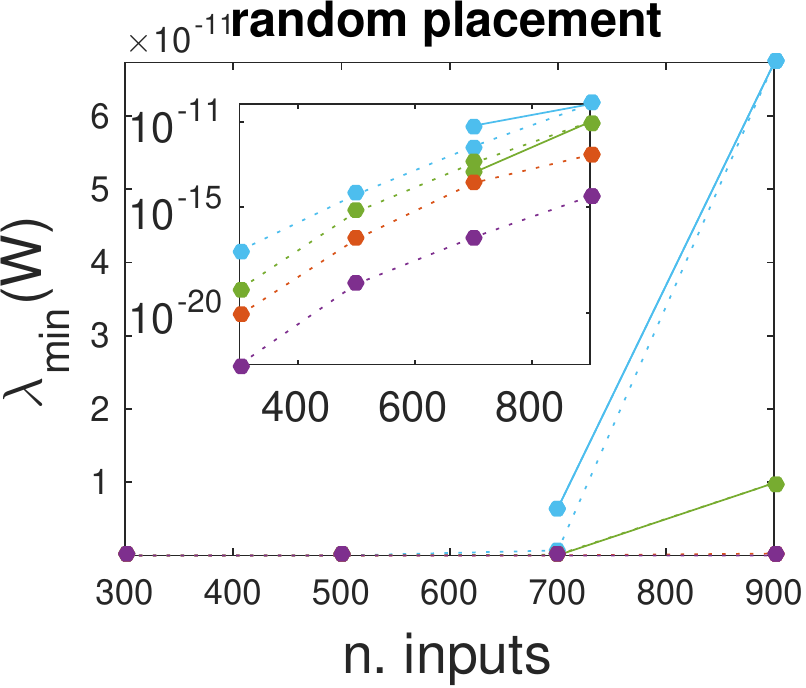}}
\subfigure[]{
\includegraphics[width=4cm]{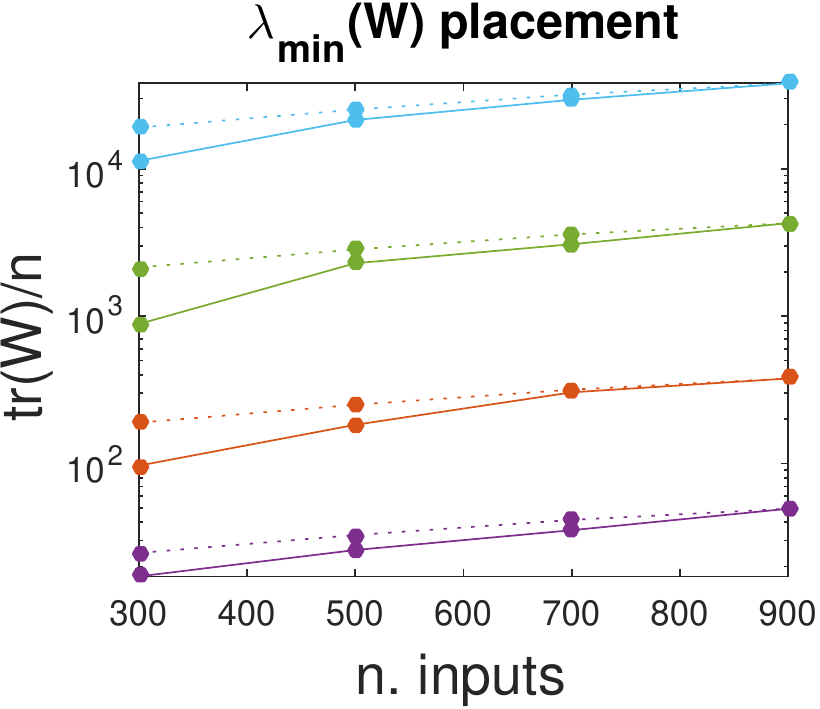}$\;\;\;$
\includegraphics[width=4cm]{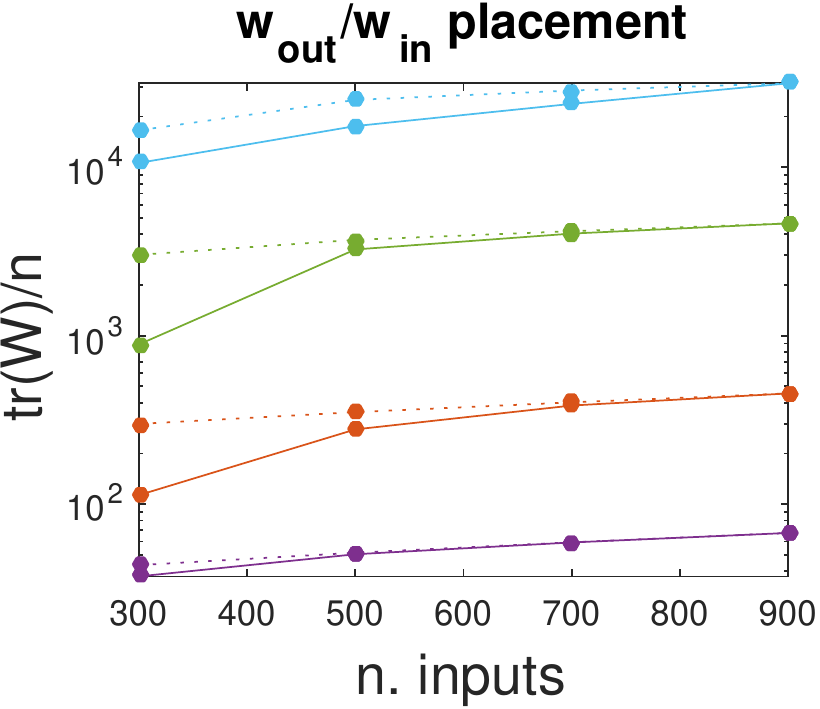}$\;\;\;$
\includegraphics[width=4cm]{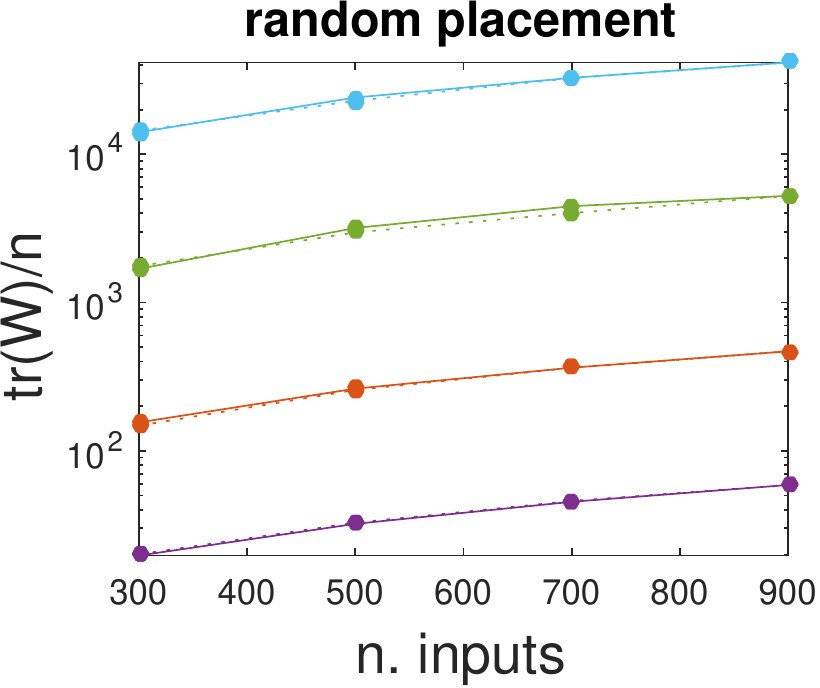}
} 
\subfigure[]{
\includegraphics[width=4cm]{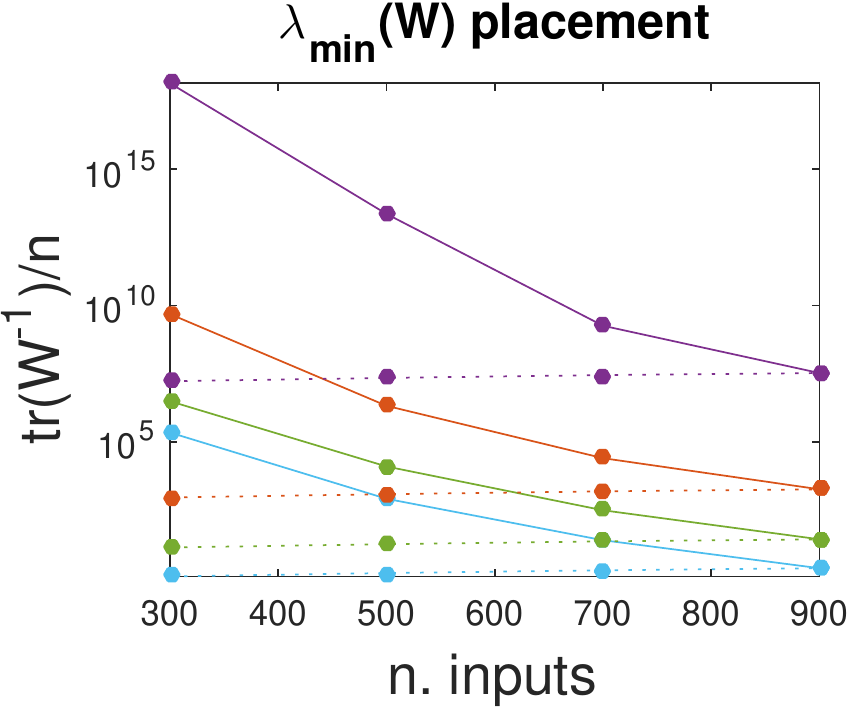}$\;\;$
\includegraphics[width=4cm]{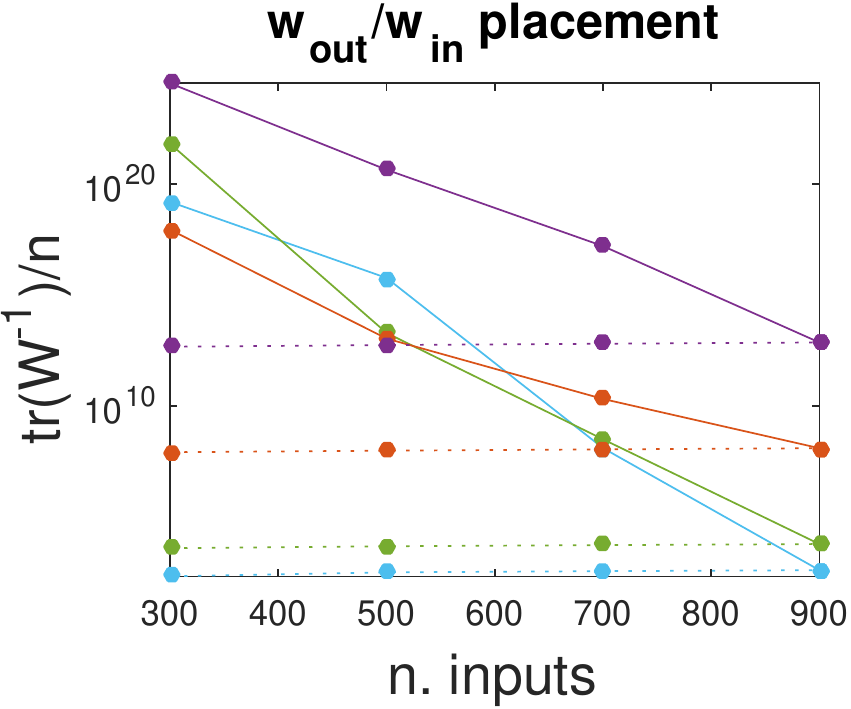}$\;\;$
\includegraphics[width=4cm]{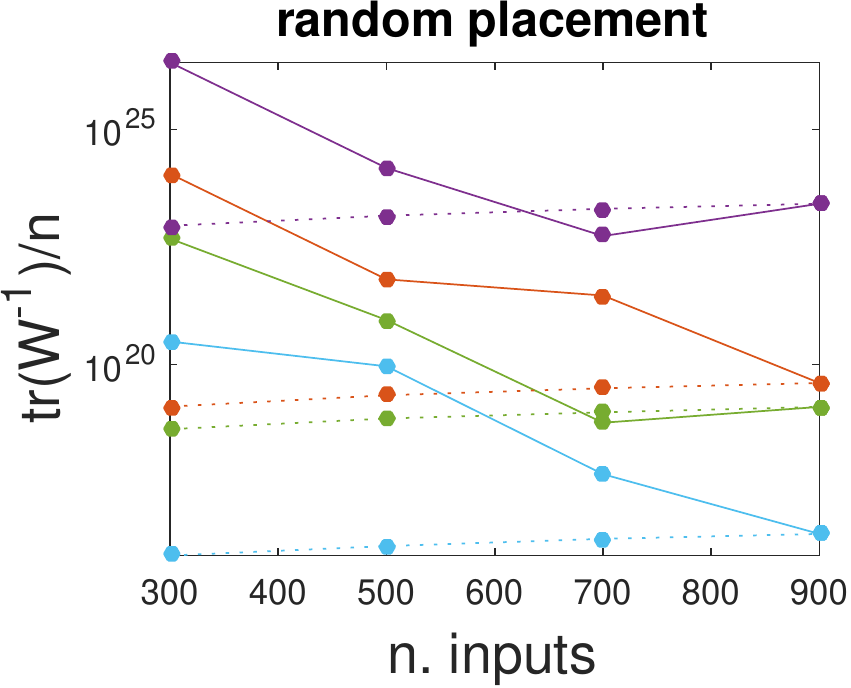}
} 

\caption[]{\small Minimum energy control of power grids with varying damping coefficients. French high/mid voltage power grid.
(a): Control energy for the metric $ \lambda_{\min} (W_r ) $ when the driver nodes are placed according to $ \lambda_{\min} (W_r) $ (left panel), $ w_{\rm out}/w_{\rm in} $ (mid panel), or randomly (right panel). 
The color code is a function of the damping coefficients, using the same convention as in Fig.~\ref{fig:driver_node_placement_dampedoscill}(a) of the paper. 
The values of $ \lambda_{\min} (W_r ) $ are shown in solid lines, while in dotted lines the values of $  \lambda_{\min} (W_z ) $ are shown (suitably normalized to eliminate the explicit dependence from $ t_f$).
Data are missing when the Gramian $ W_r $ is too close to singular (mostly when driver nodes are chosen randomly, right column).
(b): Control energy for the metric $ {\rm tr}(W_r) $ when the driver nodes are placed according to $ \lambda_{\min} (W_r) $ (left panel), $ w_{\rm out}/w_{\rm in} $ (mid panel), or randomly (right panel).    
The values of $ {\rm tr}(W_r ) $ are shown in solid lines, while in dotted lines the values of $  {\rm tr}(W_z ) $ are shown.
(c): Control energy for the metric $ {\rm tr}(W_r^{-1}) $, with the same conventions as in (a) and (b). }
\label{fig:driver_node_placement_dampedoscill_French}
\end{center}
\end{figure}

\begin{figure}[h]
\begin{center}
\subfigure[]{
\includegraphics[angle=0, trim=0cm 0cm 0cm 0cm, clip=true, width=4cm]{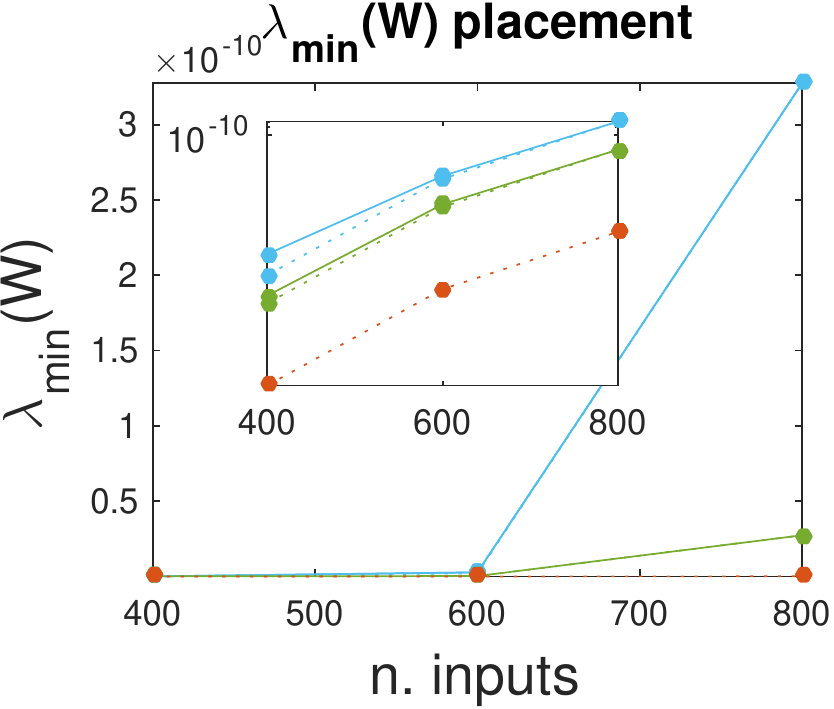}
\includegraphics[angle=0, trim=0cm 0cm 0cm 0cm, clip=true, width=4cm]{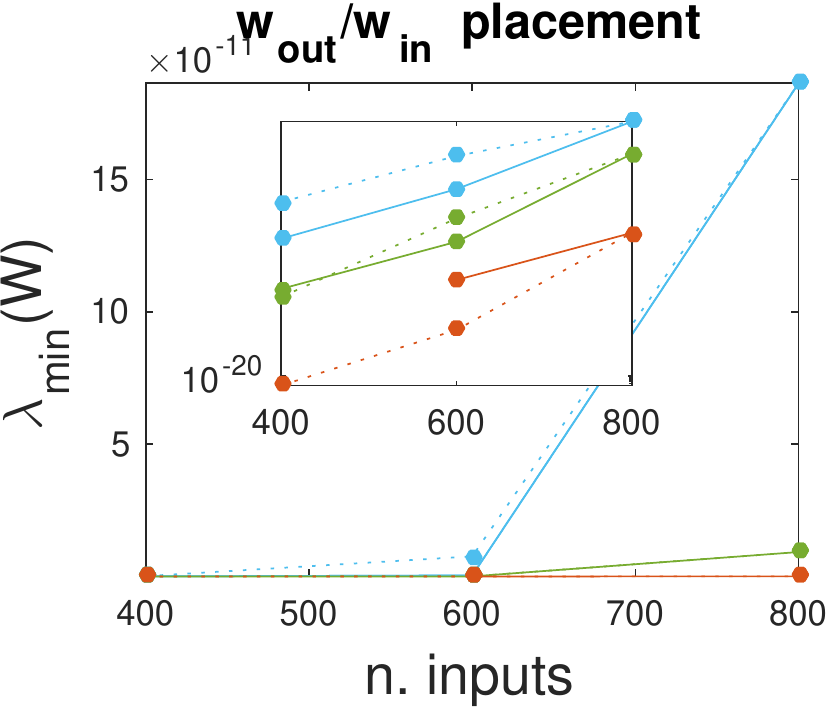}
\includegraphics[angle=0, trim=0cm 0cm 0cm 0cm, clip=true, width=4cm]{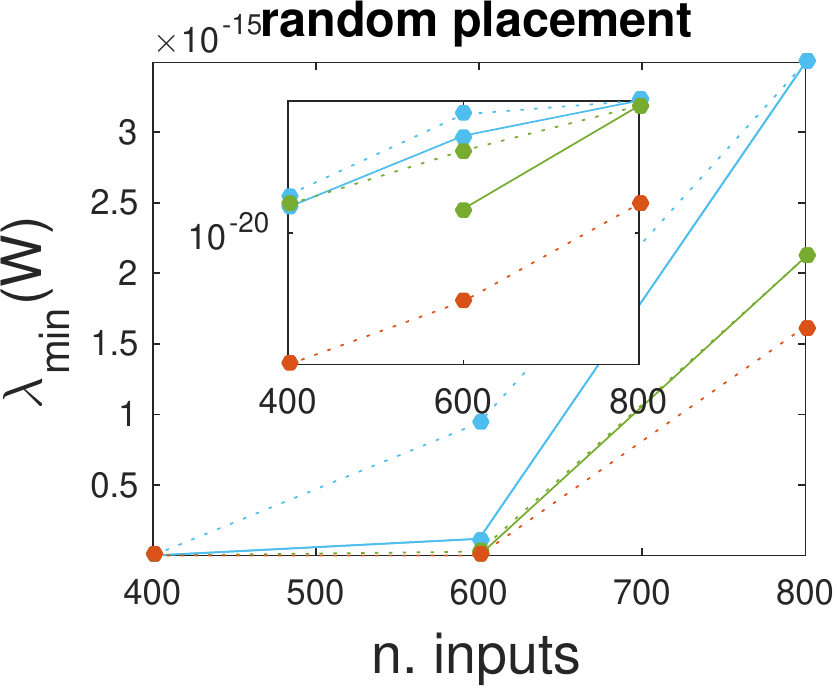}}
\subfigure[]{
\includegraphics[width=4cm]{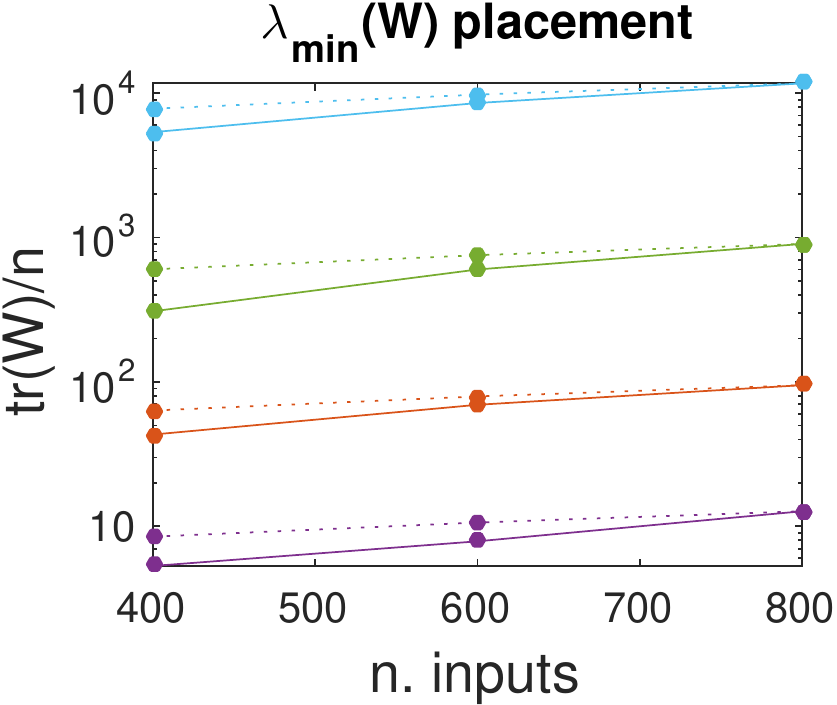}$\;\;$
\includegraphics[width=4cm]{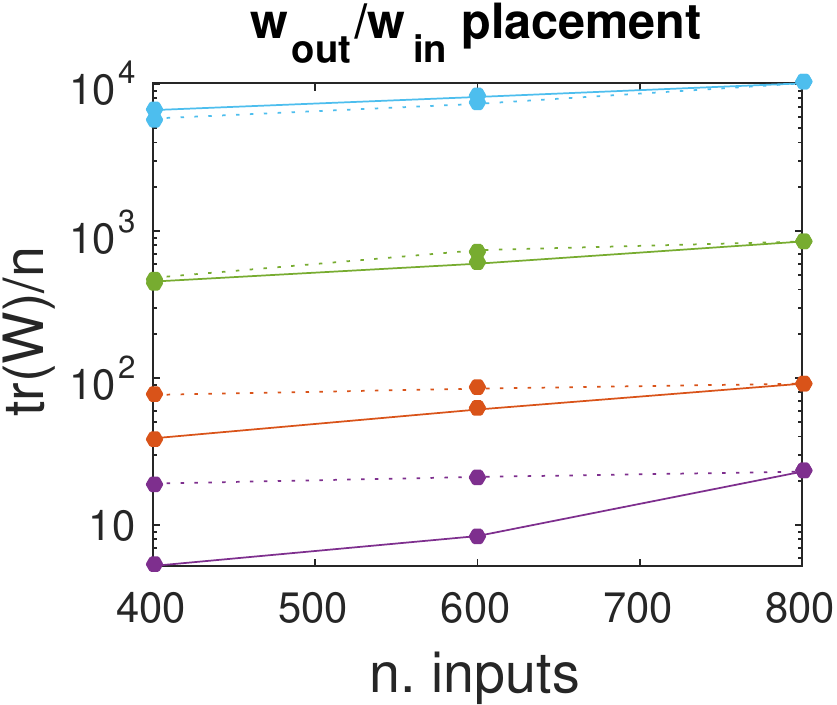}$\;\;$
\includegraphics[width=4cm]{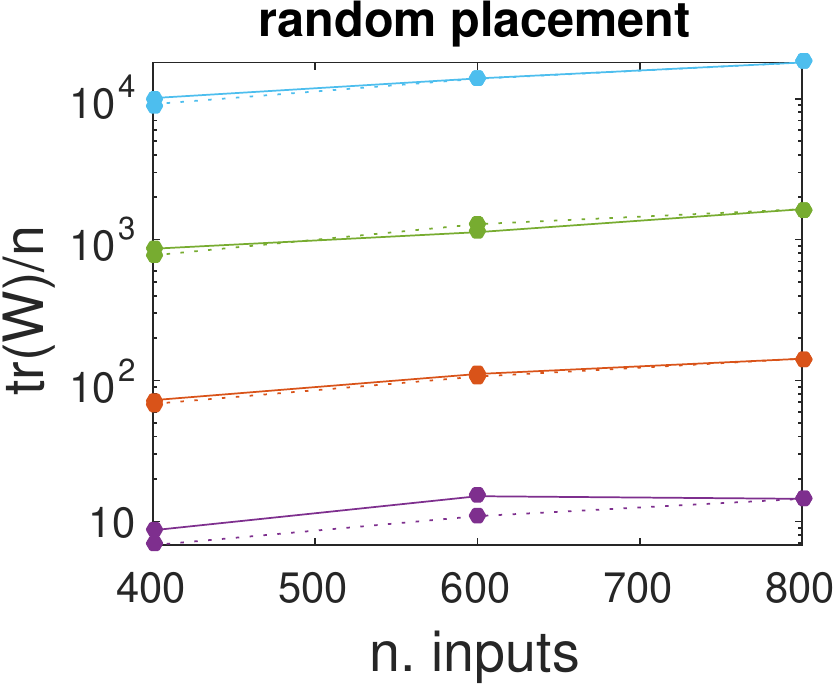}
} 
\subfigure[]{
\includegraphics[width=4cm]{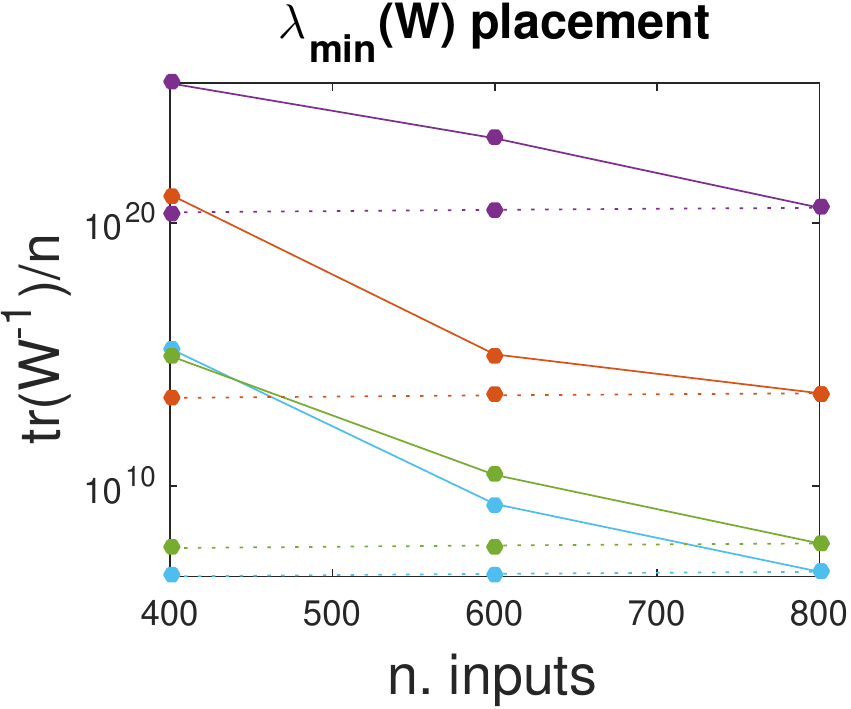}$\;\;$
\includegraphics[width=4cm]{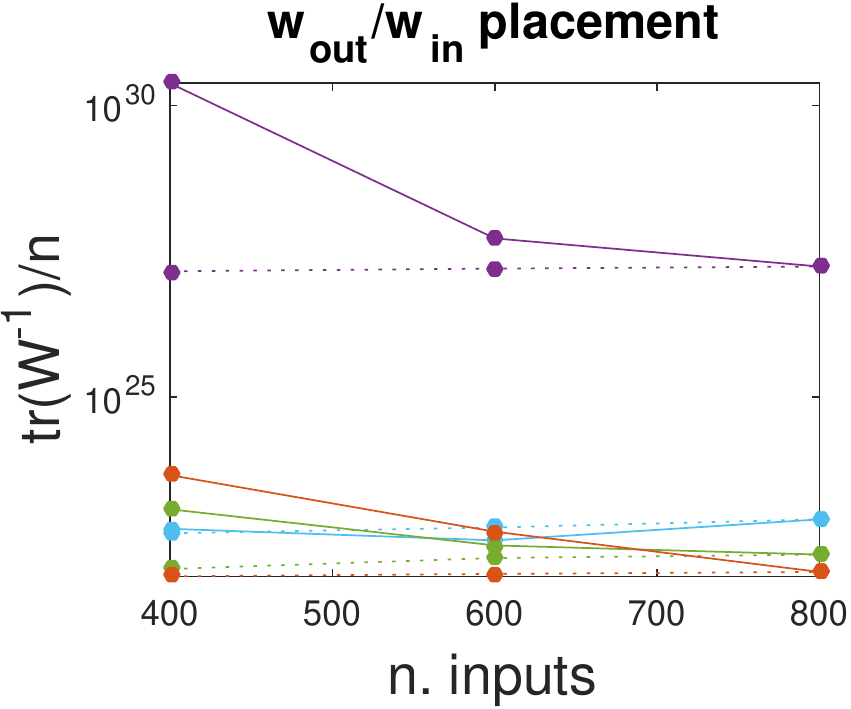}$\;\;$
\includegraphics[width=4cm]{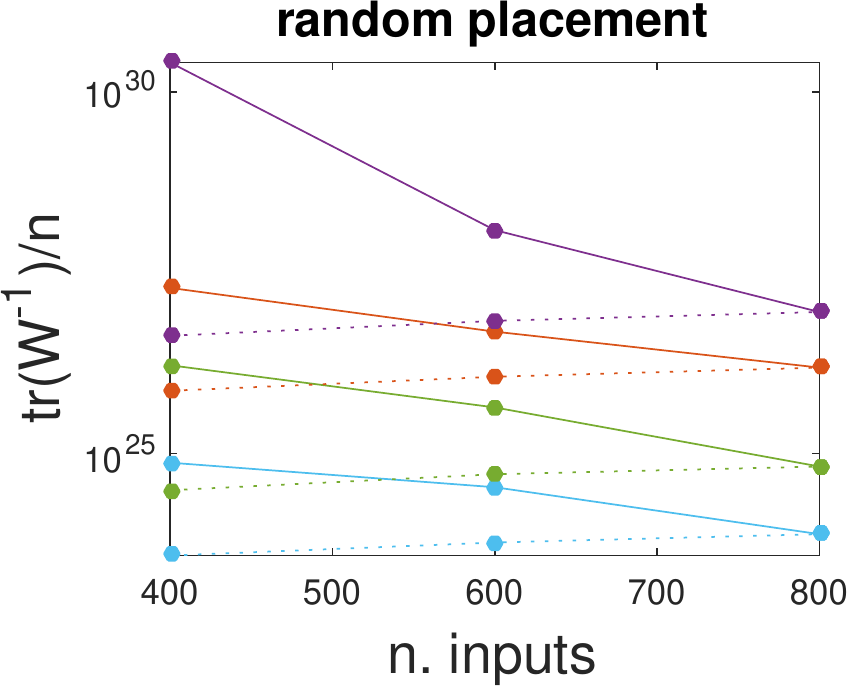}
} 

\caption[]{\small Minimum energy control of power grids with varying damping coefficients. USA  power grid.
(a): Control energy for the metric $ \lambda_{\min} (W_r ) $ when the driver nodes are placed according to $ \lambda_{\min} (W_r) $ (left panel), $ w_{\rm out}/w_{\rm in} $ (mid panel), or randomly (right panel). 
The color code is a function of the damping coefficients, using the same convention as in Fig.~\ref{fig:driver_node_placement_dampedoscill}(a) of the paper. 
The values of $ \lambda_{\min} (W_r ) $ are shown in solid lines, while in dotted lines the values of $  \lambda_{\min} (W_z ) $ are shown (suitably normalized to eliminate the explicit dependence from $ t_f$).
Data are missing when the Gramian $ W_r $ is numerically too close to singular in too many trials.
(b): Control energy for the metric $ {\rm tr}(W_r) $ when the driver nodes are placed according to $ \lambda_{\min} (W_r) $ (left panel), $ w_{\rm out}/w_{\rm in} $ (mid panel), or randomly (right panel).    
The values of $ {\rm tr}(W_r ) $ are shown in solid lines, while in dotted lines the values of $  {\rm tr}(W_z ) $ are shown.
(c): Control energy for the metric $ {\rm tr}(W_r^{-1}) $, with the same conventions as in (a) and (b). }
\label{fig:driver_node_placement_dampedoscill_USA}
\end{center}
\end{figure}

\end{document}